\documentclass[%
preprint,
showpacs,preprintnumbers,
nofootinbib,
 aps,
prd,
floatfix,
titlepage,
longbibliography]{revtex4-2}

\RequirePackage{etex}
\usepackage[a4paper, total={6.5in, 9in}]{geometry}
\usepackage{macros}
\usepackage{graphicx}
\usepackage{dcolumn}
\usepackage{bm}
\usepackage{fancyhdr}
\usepackage{simplewick} 
\usepackage{subfigure} 
\usepackage{color} 
\usepackage{listings}
\usepackage{grffile}
\usepackage{comment}
\usepackage{etex}
\reserveinserts{18}
\usepackage{morefloats}
\usepackage{slashed}
\usepackage{float}
\usepackage{amsmath,amssymb}
\usepackage{comment}

\newcommand{\csw}{{\textrm{c}}_{\textrm{sw}}}

\begin{document}
\title{
Nonperturbative renormalization of the quark chromoelectric dipole moment with the gradient Flow:
Power divergences}
\author{Jangho Kim}
\email{jkim@th.physik.uni-frankfurt.de}
\affiliation{%
Institut f\"ur Theoretische Physik, Goethe-Universit\"at Frankfurt am Main, Max-von-Laue-Str. 1, 60438 Frankfurt am Main, Germany
}%
\author{Thomas Luu}
 \email{t.luu@fz-juelich.de}
\affiliation{
 Institute for Advanced Simulation (IAS-4) \& JARA-HPC\\
 Forschungszentrum J\"ulich, Wilhelm-Johnen-Stra{\ss}e, 52428 J\"ulich, Germany
}%
\author{Matthew D. Rizik}
\email{rizik@nscl.msu.edu}
\affiliation{%
Facility for Rare Isotope Beams \& Department of Physics and Astronomy,
Michigan State University,
640 South Shaw Lane
East Lansing, MI 48824, USA
}%
\author{Andrea Shindler}
\email{shindler@frib.msu.edu}
\affiliation{%
Facility for Rare Isotope Beams \& Department of Physics and Astronomy,
Michigan State University,
640 South Shaw Lane
East Lansing, MI 48824, USA
}%
\collaboration{SymLat Collaboration}\noaffiliation

\noaffiliation
\vspace{-1.0cm}
\begin{figure}[h!]
\flushright{ \includegraphics[scale=0.1]{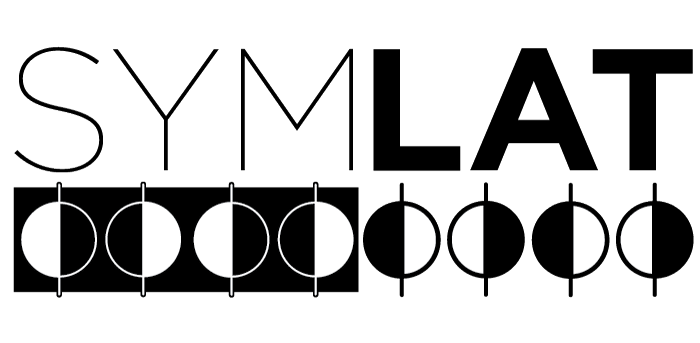}}
\end{figure}

\begin{abstract}
The  $CP$-violating quark chromoelectric dipole moment (qCEDM) operator, contributing to the electric dipole moment (EDM), 
mixes under renormalization and particularly on the lattice with the pseudoscalar density.
The mixing coefficient is power-divergent with the inverse lattice spacing squared, $1/a^2$,
regardless of the lattice action used. 
We use the gradient flow to define a multiplicatively renormalized qCEDM operator 
and study its behavior at small flow time.
We determine nonperturbatively the linearly divergent coefficient with the flow time, $1/t$, up to subleading logarithmic corrections,
and compare it with the 1-loop perturbative expansion 
in the bare and renormalized strong coupling.
We also discuss the O($a$) improvement of the qCEDM defined at positive flow time.

\end{abstract}
\maketitle
\thispagestyle{fancy}

\section{Introduction}
\label{sec:intro}

The Standard Model (SM) of particle physics is described by a Lagrangian density of
dimension $d=4$, with fundamental fields associated with 
all the elementary particles already experimentally observed.
Physics beyond the Standard Model (BSM) at the hadronic scale can be described
by higher dimensional ($d>4$) operators, with coefficients usually suppressed,
up to logarithmic corrections, by powers of the BSM energy scale
$1/\Lambda_\BSM^{d-4}$.

The baryon asymmetry in the universe can be determined by comparing the
abundances of the light elements (D, ${}^3$He, ${}^4$He) determined experimentally
with the prediction of standard big-bang nucleosynthesis~\cite{Fields:2019pfx, Tanabashi:2018oca}.
A completely independent determination of the baryon asymmetry can be obtained 
from the cosmic microwave background (CMB)~\cite{Jungman:1995bz, Planck:2018vyg} giving 
perfectly consistent results~\cite{Zyla:2020zbs}. 
The amount of $CP$-violation stemming from the CKM (Cabibbo-Kobayashi-Maskawa) matrix~\cite{Shaposhnikov:1987tw, Gavela:1993ts, Huet:1994jb}
is impossible to reconcile with the amount of $CP$-violation needed to explain the observed baryon asymmetry.
This lead to the conclusion that new sources of $CP$-violation are needed.

$CP$-violation can be investigated by studying the electric dipole moment (EDM)
of particles with nonzero spin. 
The current bound on the neutron EDM (nEDM)~\cite{Abel:2020gbr}
is several orders of magnitude larger than
the value of the nEDM that would be induced by the CKM matrix in the weak sector~\cite{Seng:2014lea}.  
An experimental signal of a nonzero EDM would thus provide strong evidence for BSM physics.  

To provide possible clues on the type of BSM physics it is crucial to 
determine separately every $CP$-violating contribution to a nonzero EDM.
Using lattice QCD it is possible, at least in principle, to determine 
the renormalized matrix elements of the $CP$-violating operators.
An important contribution to the EDM comes from the $CP$-violating chromoelectric quark operator
\be
\Oc^{ij}(x)= \psibar_i(x)\gamma_5\sigma_{\mu\nu}F_{\mu\nu}(x)\psi_j(x)\,,
\ee
where $\psi_i$ and $\psibar_i$ are fermion fields with flavor $i=1, \ldots,
N_f$, and $F_{\mu\nu}(x) = F_{\mu\nu}^a(x) T^a$ is the gluon field tensor with
$a=1, \ldots, 8$.
The operator is known as the quark chromoelectric dipole moment (qCEDM) operator.

The determination of the renormalized qCEDM using lattice QCD is particularly difficult
because the $\Oc$ operator mixes under renormalization with 
the lower-dimensional pseudoscalar density
and this poses a challenge in extracting the physical matrix element. 
In this paper we advocate the use of the gradient 
flow~\cite{Luscher:2010iy, Luscher:2011bx, Luscher:2013cpa, Luscher:2013vga}
to define the quark chromo-EDM.

The qCEDM defined at nonvanishing flow time, $t$, is multiplicatively
renormalizable and to determine the physical matrix element one needs to study
its behavior at small flow time.
The local operators contributing to the short flow time expansion have the same
symmetry transformation properties of the qCEDM. 
Of particular importance is the leading behavior at small flow time stemming from the
pseudoscalar density, which scales as $1/t$ and is thus linearly divergent with the flow time.

Power divergences have been tackled in several ways, for example by imposing
renormalization conditions on hadronic matrix elements at finite lattice
spacing~(see e.g.~\cite{Constantinou:2015ela}).
We use the gradient flow to probe the short distance behavior of the quark chromoelectric dipole moment and
determine nonperturbatively the mixing coefficient of the qCEDM operator with the lowest dimensional 
operator -- in this case the pseudoscalar density. 

Here the application of the gradient flow is potentially advantageous 
for several reasons. We can perform the continuum 
limit at each stage of the calculation
at finite flow time $t$, thus the classification of the operators 
contributing to the renormalization of the qCEDM can be done using
continuum symmetries. 
Additionally the analysis of cutoff effects is simplified~\cite{Luscher:2013cpa}
and, by using chiral symmetry transformation at finite flow time~\cite{Shindler:2013bia},
we can determine the leading dimension $6$ operators contributing to the O($a$) cutoff effects.

If the lattice QCD action breaks chiral symmetry, the qCEDM mixes with lower
dimensional operators with different chiral transformation properties, such as
the topological charge density. 
The definition of the qCEDM at nonzero flow time allows us to perform the
continuum limit at fixed flow time, thus forbidding the contribution of such
operators to the small flow time behavior. 
The \emph{taming} of induced power divergences at finite flow 
time $t$ is the main advantage of the gradient flow and thus 
motivates our use of it with regards to the $\Oc$ operator.  

In this paper we determine the linearly divergent mixing coefficient 
of the qCEDM with the pseudoscalar density as function 
of the renormalized and bare strong coupling.
We note that the particular dependence on the bare coupling is specific 
to the lattice action chosen in this paper, which is the Iwasaki gauge action~\cite{Iwasaki:1985we}
with O($a$) improved Wilson fermions~\cite{Sheikholeslami:1985ij, Luscher:1996sc, Luscher:1996ug}.

The paper is organized as follows: we introduce the main 
definitions and discuss the power divergences in the context of flowed 
fields in Sec.~\ref{sec:sfte}.
In Sec.~\ref{sec:Oaimpro} we define the lattice correlation functions we analyze  
and discuss the O($a$) improvement of the qCEDM at finite flow time $t$.
Our analysis and the ensuing results are presented in Sec.~\ref{sec:analysis}
in terms of the renormalized coupling and in Sec.~\ref{sec:analysis2}
in terms of the bare coupling. In Sec.~\ref{sec:Oaflow} we show results
for the specific O($a$) contributions to the flowed correlation functions.
We finally recapitulate our findings and provide an outlook on future calculations in 
Sec.~\ref{sec:summary}.
We defer to Appendices~\ref{app:fermion_flow} and~\ref{app:data} for more details of our numerical analysis
and in Appendix~\ref{app:PT} we discuss the perturbative calculation performed
using the same scheme adopted in this paper for the nonperturbative determination
of the mixing with the pseudoscalar density.

\section{Quark chromoelectric dipole moment at short flow time}
\label{sec:sfte}

In this section we discuss the definition of the qCEDM with flowed
fermion and gauge fields, its behavior at small flow time $t$ and 
the definition of the lattice correlation functions.
We assume that the reader is familiar with the gradient flow (GF) formalism for both 
for gauge and fermion fields~\cite{Luscher:2010iy,Luscher:2011bx,Luscher:2013vga}. 
Otherwise, for detailed references on the gradient flow and the use of the gradient flow 
to determine $CP$-violating operators contributing to the EDM
we recommend the reader to consult~\cite{Shindler:2013bia,Shindler:2015aqa,Dragos:2019oxn,Rizik:2020naq,
Suzuki:2013gza,Makino:2014taa,Endo:2015iea,Hieda:2016lly}, 
and the recent review~\cite{Shindler:2021bcx}.
Our numerical implementation of the GF for fermions using 
a 4th-order Runge-Kutta integration scheme follows closely that of~\cite{Luscher:2010iy}. 

We denote the fermion and antifermion fields with flavor $i$ 
at nonzero flow time $t$ as $\chi_i(x;t)$, and
$\chibar_i(x;t)$, respectively, and the gluon fields $B_\mu(x;t)$. 
We define the qCEDM operator at nonzero flow time
as
\be
\Oc^{ij}(x;t) = \chibar_i(x;t)\gamma_5\sigma_{\mu\nu}G_{\mu\nu}(x;t)\chi_j(x;t)\,,
\label{eq:OCt}
\ee
where $\sigma_{\mu\nu} = \frac{i}{2}\left[\gamma_\mu, \gamma_\nu\right]$ and $G_{\mu\nu}(x;t)$
is the gauge field tensor with flowed field $B_\mu(x;t)$.

At vanishing flow time, $t=0$, the renormalization of the qCEDM is 
nontrivial, and it mixes divergently with several operators
of the same dimension and less, $d \le 5$.
The list of operators that can mix with the qCEDM has
been analyzed in Ref.~\cite{Bhattacharya:2015rsa} in the context of the RI-MOM renormalization
of the qCEDM and classified based on their engineering dimension.
The proliferation of local operators contributing to the mixing 
motivates in large measure our use of the gradient flow to define
the nonperturbatively renormalized qCEDM operator. 
We note that in Ref.~\cite{Jenkins:2017dyc} the qCEDM has been renormalized in the 
$\MSbar$ scheme using a background gauge 
to ensure that only gauge invariant divergences contribute.  
Our method, described below in Sec.~\ref{ssec:sfte}, 
is based on the short flow time expansion~\cite{Luscher:2013vga} and,
as it will become clear in the next section, 
is based on gauge invariant correlation functions. 

Taking into account gauge invariance the list of operators reduces to
\be
P^{ij},\quad\partial^2 P^{ij},\quad\frac{e(q_i + q_j)}{2} \psibar_i \gamma_5 \sigma_{\mu\nu} 
F_{\mu\nu}^{\textrm{em}}\psi_j\,,
\ee
where $F_{\mu\nu}^{\textrm{em}}$ is the electromagnetic field tensor 
and $q_i$ is the quark $i$ charge in units 
of the proton charge.
Other operators, proportional to the quark mass, can contribute at small flow
time, but we choose a massless scheme and our results are extrapolated to zero
quark mass.

In this work we focus only on the power divergence
of the qCEDM and the only contribution to the power divergence comes from the 
pseudoscalar density
\be
P^{ij}(x) = \psibar_i(x)\gamma_5\psi_j(x)\,.
\ee
If we perform our renormalization using the lattice as a regulator, the mixing
pattern will depend on the choice of the lattice action and in particular
whether the action preserves (a remnant of) chiral symmetry or not. 

The renormalization of the qCEDM presented in Ref.~\cite{Bhattacharya:2015rsa}
assumes that the lattice action breaks chiral symmetry, thus the classification
of operators includes the ones with opposite chirality with respect to the
qCEDM, such as the topological charge density.  For example, if our calculation were to use the RI-MOM scheme and
nonperturbatively O($a$) improved Wilson
fermions~\cite{Sheikholeslami:1985ij,Luscher:1996ug}, operators of dimension
$d=4$, with opposite chirality with respect to the qCEDM operator, would
contribute to a linear power divergence in the lattice spacing, $\sim 1/a$.
These additional operators have been also classified in
Ref.~\cite{Bhattacharya:2015rsa}.

As our strategy to study the power divergences 
is based on the use of the GF,
we can perform the continuum limit before
studying the short distance behavior of the qCEDM.
This allows to classify operators using the symmetries 
of the continuum theory, such as chiral symmetry.

Operators with opposite chirality do contribute 
to O($a$) cutoff effects of the qCEDM correlation functions under the GF, however.
Such contributions can be analyzed using a Symanzik effective theory 
at finite flow time~\cite{Luscher:2013cpa}
and then using a generalization of chiral symmetry at 
finite flow time~\cite{Shindler:2013bia}.
We present our analysis of this effect relevant for the qCEDM in 
Sec.~\ref{sec:Oaflow}.

\subsection{Short flow time expansion}
\label{ssec:sfte}

To study the power divergent mixing of the qCEDM 
operator with the pseudoscalar density $P^{ij}$, we probe 
the short distance behavior of the qCEDM 
with the GF and by means of the short flow time expansion (SFTE).
The SFTE, as the name suggests, is an operator product expansion for $t \rightarrow 0$.  Its application is
valid for renormalized fields and can be used to define renormalized operators via nonperturbative subtraction
at nonvanishing flow time~\cite{Luscher:2013vga}.
We first perform the SFTE in the continuum assuming that any correlation
function calculated on a lattice with spacing $a$ has been extrapolated
to the continuum at a fixed physical value of the flow time $t$.
We do not specify at the moment the renormalization scheme we use
to renormalize the operators. The SFTE reads
\be
\label{eq:c_CP}
\left[\Oc^{ij}\right]_R(x;t) \stackrel{t \rightarrow 0}{\sim} \cCP(t) P_R^{ij}(x)\,,
\ee
where $\cCP \sim 1/t$ and we have neglected any additional contribution coming from higher
dimensional operators. 
The final goal is to determine nonperturbatively the coefficient $\cCP$  
so as to subtract the power divergence and thus renormalize our qCEDM matrix element of interest. 

The expansion coefficient $\cCP= \cCP^{(1)}g^2 + O(g^4)$
can be also computed in perturbation theory and its value can provide us with the behavior
of $\cCP$ at small coupling. In Ref.~\cite{Rizik:2020naq} we have calculated $\cCP$ at one-loop order
in perturbation theory in an off-shell scheme with two external quarks obtaining
\be
\cCP= \cCP^{(1)}\gbar^2 + O(\gbar^4)\,, \qquad c_{CP}^{(1)}=\frac{1}{2\pi^2t}\,,
\label{eq:cP_PT}
\ee
where $\gbar$ denotes the strong coupling renormalized at a scale $\mu=(8t)^{-1/2}$.
We have repeated the perturbative calculation of $\cCP$, using 
the same gauge invariant correlation functions that we use in our  
nonperturbative lattice determination of $\cCP$ described 
in Secs.~(\ref{ssec:lattice_corr}, \ref{sec:analysis}, \ref{sec:analysis2}). 
Details on the perturbative calculation can be found in Appendix~\ref{app:PT}.
The final result for the linear divergence is the same 
as the one obtained in Ref.~\cite{Rizik:2020naq} given by Eq.~\eqref{eq:cP_PT}.
This result suggests a form of universality for the leading order
divergence. We comment in more details 
on this result in Appendix~\ref{app:PT}.

\subsection{Lattice correlators}
\label{ssec:lattice_corr}

We have a certain amount of freedom when choosing how to compute the expansion
coefficient nonperturbatively of a given correlator.
Being an operator relation, the SFTE can be applied to different correlation
functions, the choice of which is only dictated by theoretical or numerical
convenience.
The specific choice of the external state allows one to select, to a certain
extent, which operators that contribute to the SFTE we want to study.
We use gauge invariant correlation functions
that are numerically unproblematic and allows us to select the contribution coming from the 
pseudoscalar density to the SFTE shown in Eq.~\eqref{eq:c_CP}.
\footnote{The determination of $\cCP$ using a gauge 
invariant two-point function is more problematic in 
perturbation theory though (see Appendix~\ref{app:PT}).}

We consider a four-dimensional, hypercubic Euclidean lattice with spacing $a$ and we define
the two-point function
\be
\Gamma_{CP}(x_4;t) = a^3\sum_{\bx}\left\langle\Oc^{ij}(x_4,\bx;t) P^{ji}(0,\bzero;0)\right\rangle\,,
\qquad 
\label{eq:COP}
\ee
where the probe is represented by the pseudoscalar density at $t=0$.
To define the lattice correlation function we need to specify the 
form of the field tensor
\be 
G_{\mu\nu}(x) = \frac{1}{8 i a^2}\left[Q_{\mu\nu}(x) - Q_{\nu\mu}(x)\right]\,,
\label{eq:clover}
\ee 
where $Q_{\mu\nu}(x)$ is the ``clover" term, i.e.
the sum of the $4$ plaquettes in the $\mu\nu$ plane around the point $x$.

Inserting the SFTE~\eqref{eq:c_CP} in Eq.~\eqref{eq:COP} and thereby neglecting higher dimensional
operators, we obtain
\be
\left[\Gamma_{CP}(x_4;t)\right]_R = \cCP\left[\Gamma_{PP}(x_4)\right]_R\,,
\label{eq:Gamma_sfte}
\ee
where
\be
\Gamma_{PP}(x_4) = \Gamma_{PP}(x_4;t=0) = 
a^3\sum_{\bx}\left\langle P^{ij}(x_4,\bx;t=0) P^{ji}(0,\bzero)\right\rangle\,,
\label{eq:CPP}
\ee 
is the usual pseudoscalar two-point function.
The SFTE suggests that the following dimensionless renormalized ratio 
\be
\left[R_P(x_4;t)\right]_R = t \frac{\left[\Gamma_{CP}(x_4;t)\right]_R}{\left[\Gamma_{PP}(x_4)\right]_R}\,,
\label{eq:RP}
\ee
tends to $ t\cdot\cCP$ for small enough values of $t$.
Before detailing the method, we want to comment on this particular choice 
to determine $\cCP$. We could have chosen any interpolator with the quantum numbers
of the qCEDM, but we find the pseudoscalar density very convenient to work with.
Another important aspect is the choice to keep $\Oc$ and $P$ at nonzero physical 
distance $x_4 > 0$. This choice is done to avoid possible spurious
contact terms at $x_4 \sim 0$. 
We also notice that the denominator of $R_P$ will have additional contact terms if $x_4=0$.

In determining $\cCP$ it is advantageous to study the correlation functions
in Eq.~\eqref{eq:RP} for large Euclidean times $x_4 \gg \sqrt{8t}$.
Performing a spectral decomposition of the $2$ correlation functions~\eqref{eq:COP}
and~\eqref{eq:CPP} and retaining only the contributions of the ground state we obtain
\be
\Gamma_{CP}(x_4 ;t) = \frac{1}{2 m_\pi}\left\langle 0|\Oc^{ij}(t)|\pi\right\rangle
\left\langle \pi|P^{ij}|0\right\rangle \e^{-m_\pi x_4}\,,
\ee
\be
\Gamma_{PP}(x_4) = \frac{1}{2 m_\pi}\left\langle 0|P^{ij}|\pi\right\rangle
\left\langle \pi|P^{ij}|0\right\rangle \e^{-m_\pi x_4}\,,
\ee
where $\left.|\pi\right\rangle$ is the pion state. 
The bare ratio $R_P$ thus has a spectral decomposition for large $x_4$ 
\be
R_P(t) = t \frac{\left\langle 0|\left[\Oc\right]_R^{ij}(t)|\pi\right\rangle}
{\left\langle 0|P_R^{ij}|\pi\right\rangle}\,.
\ee

The renormalization of the pseudoscalar density and any flowed operators 
is well understood \cite{Luscher:2010iy, Luscher:2011bx, Luscher:2013cpa}
\be
P_R^{ij} = Z_P P^{ij}\,, \qquad \left[\Oc\right]_R^{ij}(t) = Z_\chi \Oc^{ij}(t)\,,
\ee
where $Z_\chi^{1/2}$ is the renormalization factor of the flowed fermion field.
If we make explicit the renormalization constants in Eq.~\eqref{eq:Gamma_sfte} we then obtain
\be
    Z_{\chi} \Gamma_{CP}(x_4 ;t) = \cCP Z_P \Gamma_{PP}(x_4)\,.
\label{eq:sfte_Z}
\ee

To determine $\cCP$ in the continuum limit 
we need an independent determination of $Z_P$ and $Z_\chi$ 
(or the use of the ringed fermion fields~\cite{Makino:2014taa}).
In this first study of the qCEDM we calculate $2$ different quantities.
The first is the nonperturbative renormalization that connects the qCEDM at
finite flow time with the pseudoscalar density, also  at finite flow time.
This quantity, labeled $\Delta(\gbar^2)$ below in Eq.~\eqref{eq:cCP_cP},
is finite and represents the leading dependence of the coefficient of the linear divergence
on the renormalized coupling.
The second quantity we calculate is the bare 
expansion coefficient of the linear divergence divided by $Z_\chi$,
labeled $c_\chi$ below in Eq.~\eqref{eq:c_chi}. 

To calculate the finite renormalization, instead of using the ratio $R_P(x_4;t)$ in Eq.~\eqref{eq:RP}, we consider the ratio
\be
\left[\Rbar_P(x_4;t)\right]_R = 
t \frac{\left[\Gamma_{CP}(x_4;t)\right]_R}{\left[\Gamma_{PP}(x_4;t)\right]_R}\,,
\label{eq:RPbar}
\ee
where the denominator now contains a flowed pseudoscalar density
\be 
\Gamma_{PP}(x_4;t) = a^3\sum_{\bx}\left\langle P^{ij}(x_4,\bx;t) P^{ji}(0,\bzero)\right\rangle\,,
\label{eq:CPPt}
\ee 
where
\be
P^{ij}(x;t) = \chibar_i(x;t)\gamma_5\chi_j(x;t)\,.
\ee
The reason for this definition is that now we can perform the continuum limit 
of $\Rbar_P(x_4;t)$ without any knowledge of renormalization factors since
the ratio in Eq.~\eqref{eq:RPbar} is scheme independent and free
of renormalization ambiguities.
To determine the expansion coefficient $\cCP$ one still needs to determine
the expansion coefficient, $\cP$, of the pseudoscalar density, $P^{ij}(x;t)$
\be 
P^{ij}(x;t) = \cP(t) P^{ij}(x) +\mcO(t)\,.
\label{eq:cP}
\ee 
We do this via the relation
\be 
\cCP (t) = \frac{1}{t}\Delta(\gbar^2) \cP(t) +\mcO(t)\,,
\label{eq:cCP_cP}
\ee 
where $\Delta(\gbar^2)$ is the nonperturbative finite renormalization
we determine in this work:
\be 
\Delta(\gbar^2) = \left[\Rbar_P(x_4;t)\right]_R \,, \qquad x_4 \gg \sqrt{8t}\,.
\label{eq:Delta}
\ee 
In this way we determine nonperturbatively 
the power-divergent coefficient up to subleading logarithmic
corrections described by of the expansion coefficient of the pseudoscalar density.
Using the gradient flow shows 
that we can determine the power-divergent coefficient 
in the continuum limit, and once we determine the expansion coefficient
$\cP$ of the pseudoscalar density it is possible to estimate also
the subleading logarithmic contribution to the power divergence. 
Operating with the RG operator $\mu\frac{d}{d\mu}$ 
on the SFTE of the pseudoscalar density~\eqref{eq:cP}, 
it is possible to determine the flow time dependence of the expansion coefficient $\cP$
given the anomalous dimension of the pseudoscalar density, of the flowed fermion field
and the beta function~\cite{Luscher:2013vga}. We leave the determination of $\cP$
for a future work.

The second quantity we determine
is the bare expansion coefficient 
\be
c_\chi \equiv \frac{t}{Z_\chi} \cCP =  \frac{1}{Z_P} R_P(x_4;t)\,,
\label{eq:c_chi}
\ee
and make use of the determination of $Z_P(g_0^2)$ in Ref.~\cite{Aoki:2010wm}
(the values of $Z_P$ are also listed for completeness in
Table~\ref{tab:latpar1}).
With this definition we can study the dependence of $c_\chi$ on the bare
coupling $g_0$ leaving for future calculations the determination of $Z_\chi$ or
the use of "ringed" fermion fields~\cite{Makino:2014taa}.
Employing a Pad\'e approximant in combination with perturbation theory, we determine the
dependence on the bare coupling of $c_\chi$ for our choice of the lattice
action.
In Sec.~\ref{sec:analysis} we present the numerical details for the 
determination of $\Delta(\gbar^2)$, while in 
Sec.~\ref{sec:analysis2} we show results for $c_\chi$.

\section{O($a$) improvement of the Quark Chromoelectric Dipole Moment operator}
\label{sec:Oaimpro}

The theoretical analysis of cutoff effects in 
lattice field theory follows the Symanzik 
description~\cite{Symanzik:1983dc,Symanzik:1983gh,Luscher:1996sc,Luscher:1996ug}
in terms of effective action and operators close to the continuum limit.
To study the cutoff effects of correlation functions involving fields
at positive flow time $t$, it is better to rely on the quantum theoretical description
based on the $4+1$ dimensional field theory~\cite{Luscher:2013cpa,Shindler:2013bia,Ramos:2015baa}.
The extra dimension is represented by the flow time $t$ and the GF 
equations are imposed augmenting the theory with an additional Grassmanian field 
$\lambda$ that acts as a Lagrange multiplier. With this field theoretical description 
it is possible to write the corresponding Symanzik effective action by
adding higher dimensional operators to the continuum theory. We recall that 
the higher dimensional operators are constrained by the symmetries of the lattice theory.

We consider here nonperturbatively clover-improved Wilson fermions and
gauge invariant correlation functions of fields, 
at nonzero physical distance with each other.
With this prescription we can use the equations of motion
to reduce the number of higher dimensional operators and thus avoid spurious
O($a$) terms stemming from contact terms.

Following Ref.~\cite{Luscher:2013cpa} the lattice action
is improved by adding the following $d=5$ operators
\be
\mcO_1(x) = \frac{i}{4}\sum_{i=1}^{N_f}\psibar_i(x) \sigma_{\mu\nu} F_{\mu\nu}(x) \psi_i(x)\,,
\ee
\be
\mcO_2(x) = \sum_{i=1}^{N_f}\lambdabar_i(x) \lambda_i(x)\,,
\ee
where $\lambda(x)$ is the Lagrange multiplier that, once integrated over in the Functional Integral,
enforces the fermion fields, defined at positive flow time, to be solutions of the 
GF equations. We discuss in Appendix~\ref{app:fermion_flow}
how we numerically contract the Lagrange multiplier with the fermion fields.
Here we just recall the basic properties. 
The field $\lambda$ has energy dimension $\frac{5}{2}$ in $d=4$ spacetime
dimensions and the chiral transformation properties of this field are described
in Ref.~\cite{Shindler:2013bia}, as well as those of other higher dimensional
operators.
The operator $\mcO_2$ breaks chiral symmetry and is present as an O($a$)
operator in the Symanzik effective action.  The O($a$) operators are all
evaluated at $t=0$, i.e. they are boundary terms for the $4+1$ dimensional
theory.
This is a consequence of the invariance of the action under chiral symmetry
in the bulk of the $4+1$ dimensional theory~\cite{Shindler:2013bia}.
While the term proportional to $\mcO_1$ is the usual clover term, with a
tunable $\csw$ parameter in the lattice action, the term proportional to
$\mcO_2$ has a tunable $\cfl$ coefficient that only contributes to contractions
between fermion fields at nonvanishing flow time~\cite{Luscher:2013cpa}.
In this work we will never consider
fermion contractions between two flowed fermion fields, thus from now on we assume that 
$\cfl$ is not needed to remove O($a$) effects from the lattice correlators we compute.

The effective action is not sufficient to describe correlation 
functions in the Symanzik effective theory.
We also need to consider O($a$) terms from the local operators.
The list of higher dimensional operators, parametrizing O($a$) cutoff effects, 
for the fermion fields and the pseudoscalar and scalar densities is given in
Ref.~\cite{Luscher:2013cpa}.
Here we just list for completeness the form
of the renormalized O($a$) improved operators using the same notation as in~\cite{Luscher:2013cpa}.
For the fields at positive flow time the renormalization is dictated always
by the same field renormalization factor $Z_\chi^{1/2}$.   
Chiral symmetry implies that only O($am$) contributions be present,
\be
\chi_R(x,t) = Z_\chi^{1/2}\left(1 + a~\frac{b_\chi}{2} m_q + a~\frac{\overline{b}_\chi}{2} \Tr M \right) \chi(x,t)\,,
\ee
\be
P^{ij}_R(x,t) = Z_\chi\left(1 + a~b_\chi m_{q,ij} + a~\overline{b}_\chi \Tr M \right) P^{ij}(x,t)\,,
\ee
where $m_q$ is the subtracted bare quark mass and $m_{q,ij}$ is the average of the subtracted quark masses,
$m_{q,ij} = \frac{1}{2}(m_{q,i} + m_{q,j})$. 
To determine the subtracted quark mass
it is possible to use the PCAC mass, 
\be
m_{q,ij} = \frac{\sum_\bx \left\langle \partial_0 A_0^{ij}(x_4,\bx;0) P^{ji}(0;0) \right\rangle }
{\sum_\bx \left\langle P^{ij}(x_4,\bx;0) P^{ji}(0;0)\right \rangle}\,.
\label{eq:PCAC}
\ee
The correlator on the right-hand-side (rhs) of Eq.~\eqref{eq:PCAC} should be independent of $x_4$ up to cutoff effects.

With respect to the standard treatment of O($a$) cutoff effects, the fields at
the boundary $t=0$ show additional O($a$) terms, if these are then contracted
with fields at positive flow time
\be
\left[A_\mu^{ij}(x)\right]_I = A_\mu^{ij}(x) + a~c_A \partial_\mu P^{ij}(x) + a~\wcA \wA_\mu^{ij}(x)\,,
\ee
\be
\left[P^{ij}(x)\right]_I = P^{ij}(x) + a~\wcP \wP^{ij}(x)\,.
\label{eq:P_impr}
\ee
The two additional O($a$) terms are proportional to 
\be
\wA_\mu^{ij}(x) = \lambdabar_i(x)\gamma_\mu \gamma_5 \psi_j(x) +
\psibar_i(x)\gamma_\mu \gamma_5 \lambda_j(x)\,,
\label{eq:Atilde}
\ee
\be
\wP^{ij}(x) = \lambdabar_i(x)\gamma_\mu \gamma_5 \psi_j(x) +
\psibar_i(x)\gamma_\mu \gamma_5 \lambda_j(x)\,.
\label{eq:Ptilde}
\ee
The presence of the Lagrange multipliers confirms that these O($a$) terms contribute
only when contracted with flowed local operators.
According to Ref.~\cite{Luscher:2013cpa} at tree level perturbation theory
we have
\be
\cfl = -\wcA = - \wcP = \frac{1}{2}\,, \qquad b_\chi = 1\,.
\ee

To determine the O($a$) terms for the qCEDM operator we consider its chiral
symmetry properties and construct the higher dimensional operator following
Ref.~\cite{Shindler:2013bia}. 
Any operator of the form $\chibar_i(t) \Gamma(t) \chi_j(t)$ at $t>0$, where
$\Gamma(t)$ is either constant or it contains, as in this case, the flowed
gauge field, renormalizes with $Z_\chi$ and shows only O($am$) cutoff effects
\be
\left[\Oc^{ij}(x,t)\right]_R = Z_\chi \left[\Oc^{ij}(x,t)\right]_I\,, \qquad \left[\Oc^{ij}(x,t)\right]_I =  
\left(1 + a~b_\chi m_{q,ij} + a~\overline{b}_\chi \Tr M \right)\Oc^{ij}(x,t)\,,
\label{eq:Oc_impr}
\ee
where again this form is constrained by the chiral symmetry at $t>0$.
In addition to the O($a$) terms shown in~\eqref{eq:Oc_impr}, we have to add an
additional O($a$) improvement term because the pseudoscalar density is
contracted with a field at nonvanishing flow time, which introduces the term
proportional to $\wP^{ij}$.
The final form of the renormalized and improved correlation function is then 
\be
\left[\Gamma_{CP}(x_4;t)\right]_R = Z_\chi Z_P \left[\Gamma_{CP}(x_4;t)\right]_I\,,
\ee
\be
\left[\Gamma_{CP}(x_4;t)\right]_I = \left(1 + a~b_\chi m_{q,ij} + 
a~\overline{b}_\chi \Tr M \right)\Gamma_{CP}(x_4;t)+ a~\wcP
\wGamma_{CP}(x_4;t)\,,
\label{eq:Gamma_CP_improved}
\ee
where
\be
\wGamma_{CP}(x_4;t) = a^3\sum_{\bx}\left\langle\Oc^{ij}(x_4,\bx;t)
\wP^{ji}(0,\bzero;0)\right\rangle\,.
\label{eq:wC_OP}
\ee

To improve the denominator in the ratio of Eq.~\eqref{eq:RPbar} we define
\be 
\left[\Gamma_{PP}(x_4;t)\right]_R = Z_\chi Z_P \left[\Gamma_{PP}(x_4;t)\right]_I\,,
\ee
\be
\left[\Gamma_{PP}(x_4;t)\right]_I = \left(1 + a~b_\chi m_{q,ij} + 
a~\overline{b}_\chi \Tr M \right)\Gamma_{PP}(x_4;t)+ a~\wcP
\wGamma_{PP}(x_4;t)\,,
\label{eq:Gamma_PPt_improved}
\ee
where
\be
\wGamma_{PP}(x_4;t) = a^3\sum_{\bx}\left\langle P^{ij}(x_4,\bx;t)
\wP^{ji}(0,\bzero;0)\right\rangle\,.
\label{eq:wC_PP}
\ee
The renormalized and improved ratio $\Rbar_P$ then reads
\be 
\left[\Rbar_P(x_4;t)\right]_R = 
t \frac{\left[\Gamma_{CP}(x_4;t)\right]_I}{\left[\Gamma_{PP}(x_4;t)\right]_I}\,.
\label{eq:RPbar2}
\ee
We show in Sec.~\ref{sec:analysis2} 
(see Figs.~\ref{fig:C_PP_tilde_A3},\ref{fig:C_OP_tilde})
that the O($a$) terms proportional 
to $\wGamma_{CP}(x_4;t)$ and $\wGamma_{PP}(x_4;t)$ 
vanish for $x_4$ slightly larger than zero for any value of the flow time we consider.

To improve the ratio in Eq.~\eqref{eq:RP} we also need to improve the denominator
\be
\left[\Gamma_{PP}(x_4)\right]_I = \left(1 + a~b_P m_{q,ij} + 
a~\overline{b}_P \Tr M \right)\Gamma_{PP}(x_4)\,,
\label{eq:Gamma_PP_improved}
\ee
where the tree-level value is $b_p^{(0)}=1$.
All the coefficients coming from sea quark effects on fermion correlation
functions, such as the parameters $\overline{b}_X = O(g^4)$, are neglected. 
The tree level value of $b_P$ coincides with the value of $b_\chi$\footnote{The value of $b_\chi$, to the best of our knowledge, is
unknown beyond tree level.} thereby simplifying the ratio~\eqref{eq:RP}.
Thus we are left with O($am$) discretization errors of O($g^2$). 
The O($am$) terms of the pseudoscalar density probing the qCEDM 
in Eqs.~(\ref{eq:Gamma_CP_improved}, \ref{eq:Gamma_PP_improved})
simplify once we determine the ratios in Eqs.~(\ref{eq:RP}, \ref{eq:RPbar}).

In the next section we evaluate the correlation functions in Eqs.~(\ref{eq:wC_OP}, \ref{eq:wC_PP})
and show that it contributes only at short distances when $x_4 \sim \sqrt{8t}$.
To conclude, at nonzero physical distances $x_4 \gg \sqrt{8t}$, for all
practical purposes the ratio in Eq.~\eqref{eq:RPbar} is renormalization group
invariant and automatically O($a$) improved, after we improve the action, while
the ratio in Eq.~\eqref{eq:RP} is O($a$) improved,  up to O($a m g^2$).
\section{Numerical Analysis}
\label{sec:analysis}

\begin{table}[t]
		  \caption{Summary of the lattice bare parameters for the ensembles used.
		  $N_G$ is the number of gauge configurations and $Z_P$ is the value of the renormalization constant determined
		  in Ref.~\cite{Aoki:2010wm}.
			\label{tab:latpar1}}
		\begin{tabular}{|c|c|c|c|c|c|c|c|c|c|c|c|c|}
		  \hline
		  Designation& $\beta$ & $\kappa_l$ & $\kappa_s$ & L/a & T/a & $\csw$ & $N_G$ & $a$ [fm] & $m_\pi$ [MeV] & $m_{N}$ [GeV] & $Z_P$ \\
		  \hline\hline                                                                
		  M$_1$ & 1.90 & 0.13700 & 0.1364 & 32 & 64 & 1.715 & 399 & 0.0907(13) & 699.0(3) & 1.585(2) & 0.49605 \\
		  \hline\hline                                                                                              
		  M$_2$ & 1.90 & 0.13727 & 0.1364 & 32 & 64 & 1.715 & 400 & 0.0907(13) & 567.6(3) & 1.415(3) & 0.49605 \\
		  \hline                                                                                                
		  M$_3$ & 1.90 & 0.13754 & 0.1364 & 32 & 64 & 1.715 & 450 & 0.0907(13) & 409.7(7) & 1.219(4) & 0.49605 \\
		  \hline\hline                                                                                            
		  A$_1$ & 1.83 & 0.13825 & 0.1371 & 16 & 32 & 1.761 & 800 & 0.1095(25) & 710(1)   & 1.65(1)  & 0.44601 \\
		  \hline                                                                                                
		  A$_2$ & 1.90 & 0.13700 & 0.1364 & 20 & 40 & 1.715 & 790 & 0.0936(33) & 676.3(7) & 1.549(6) & 0.49605 \\
		  \hline                                                                                                
		  A$_3$ & 2.05 & 0.13560 & 0.1351 & 28 & 56 & 1.628 & 650 & 0.0684(41) & 660.4(7) & 1.492(5) & 0.51155 \\
		  \hline
		\end{tabular}
\end{table}

To determine the ratios $R_P$ and $\Rbar_P$ in Eqs.~(\ref{eq:RP}, \ref{eq:RPbar}) 
we need to calculate the two-point functions in Eqs.~(\ref{eq:COP}, \ref{eq:CPP})
and~\eqref{eq:CPPt}.
We use publicly available~\cite{BECKETT20111208} lattice gauge configurations generated with 
the Iwasaki gauge action~\cite{Iwasaki:1985we} and nonperturbatively clover-improved 
fermions~\cite{Sheikholeslami:1985ij, Luscher:1996ug}.
Details on the generation of these ensembles can be found in Refs.~\cite{Aoki:2008sm, Aoki:2010wm}.
The improvement coefficient $\csw$ for this choice of lattice action has been determined in
Ref.~\cite{Aoki:2005et}. The bare parameters of our ensembles are listed in Table~\ref{tab:latpar1}
together with some basic quantities and the pseudoscalar renomalization constant $Z_P$.
In this work, as illustration, we use the values of $Z_P$ determined using the 
Schr\"odinger Functional scheme in Ref.~\cite{Aoki:2010wm} at some low-energy scale. 
The exact value of the renormalization scale is not relevant for the method we use
to determine the linear divergent term. 

The computation of the fermion correlation functions requires, beside the
standard quark propagators, the calculation of the quark fields contractions
between fields at zero and nonzero flow time $t$. 
The flow time dependence is calculated taking the standard quark propagator 
as initial condition of the GF equation.
We always place the flowed field at the ``sink": in this way we do 
not have to repeat the inversion of the lattice Dirac operator 
for each value of the flow time $t$.

Recall that in Sec.~\ref{sec:sfte} we discussed 
$2$ different strategies to study the behavior at small flow time $t$ 
of the qCEDM operator~\eqref{eq:OCt}. 
The first strategy was based on the determination of the ratio
$\Rbar_P$ defined in Eq.~\eqref{eq:RPbar}, or in other words the determination
of the finite renormalization connecting the qCEDM and the pseudoscalar density
at finite flow time.  We now discuss this strategy in more detail.

\subsection{Finite renormalization}

\begin{figure}
    \centering
    \includegraphics[width=0.49\textwidth]{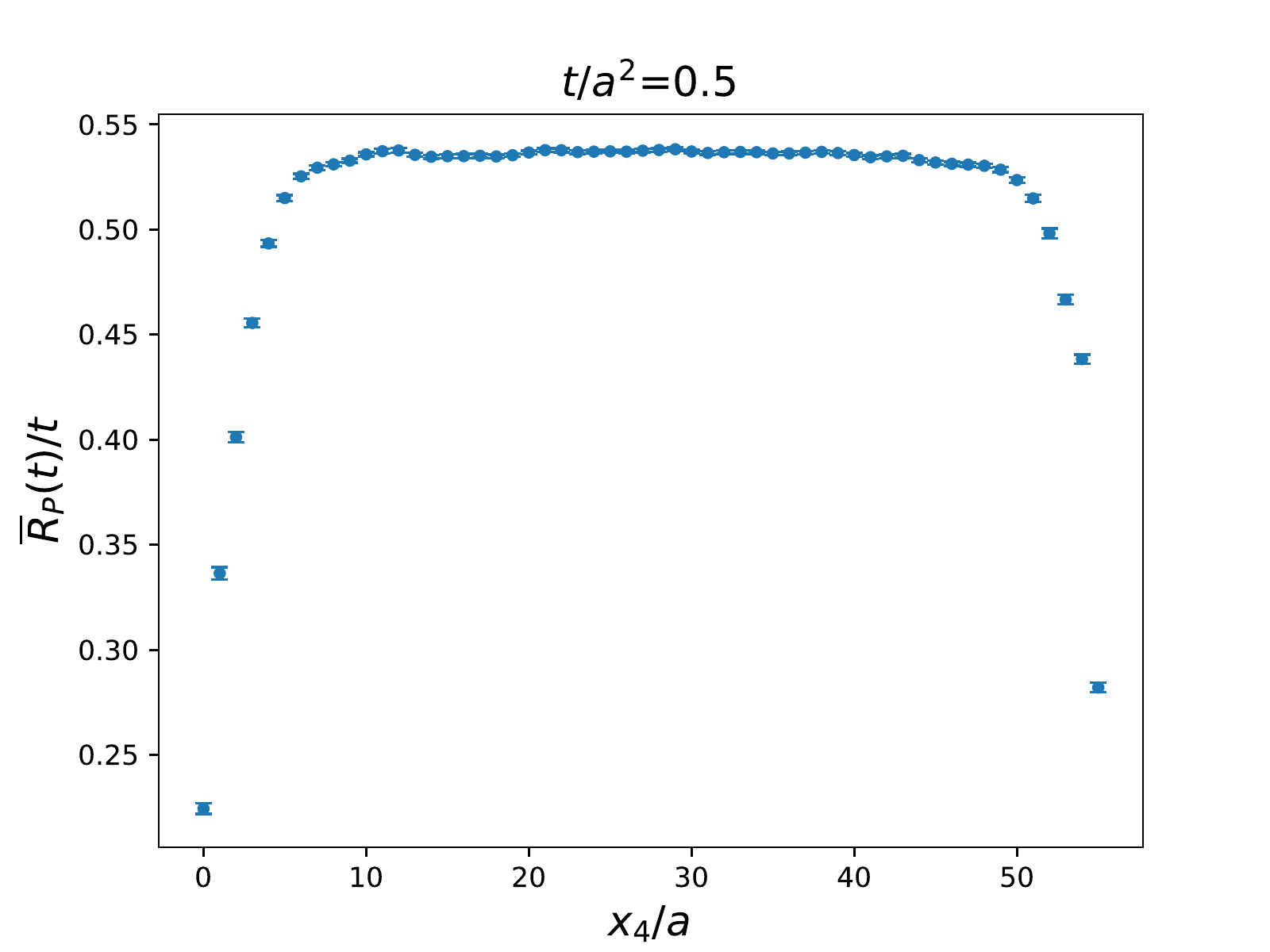}
    \includegraphics[width=0.49\textwidth]{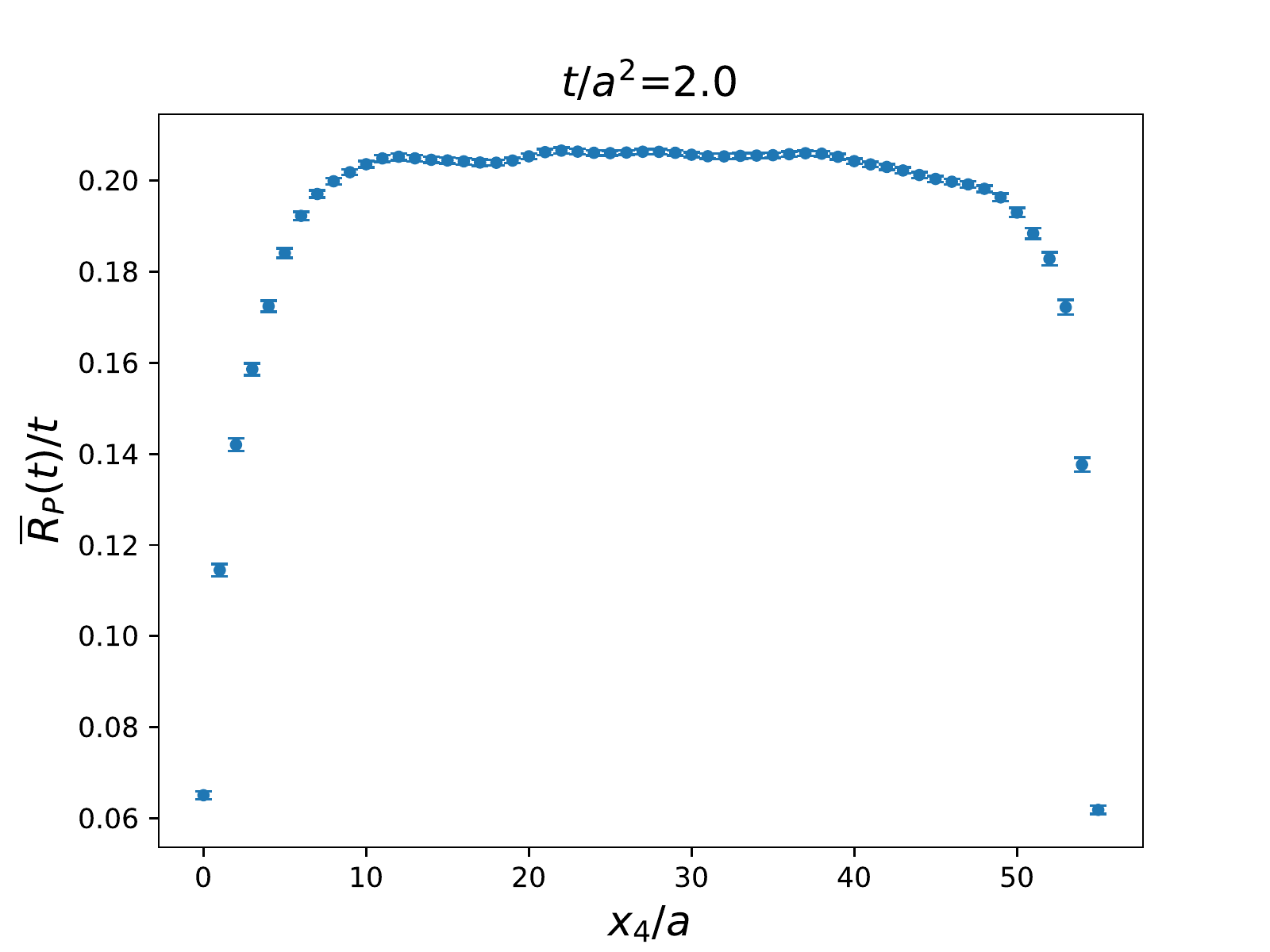}
		\caption{$\Rbar_P(t)/t$ defined in Eq.~\eqref{eq:RPbar} and determine on the ensemble $A_3$ 
		(see Table~\ref{tab:latpar1}) for values of the flow time $t/a^2 = 0.5, 2.0$, 
		corresponding to a flow time radius $r_f = \sqrt{8t} = 2a, 4a$. }
    \label{fig:Rbar_plateau}
\end{figure}

To determine the ratio $\Rbar_P$ in Eq.~\eqref{eq:RPbar} we need to calculate 
the two-point functions in Eqs.~\eqref{eq:COP} and~\eqref{eq:CPPt}.
The spectral decomposition of Eq.~\eqref{eq:RPbar} is straightforward 
and if we retain only the ground state contribution we obtain
\be
\Rbar_P(t) = t \frac{\left\langle 0|\left[\Oc\right]_R^{ij}(t)|\pi\right\rangle}
{\left\langle 0|P_R^{ij}(t)|\pi\right\rangle}\,,
\label{eq:RP_hadron}
\ee
where we assume that $x_4 \gg \sqrt{8t}$ in order to make sure that the state propagating
are pion states. It is easy to check numerically if this condition is satisfied.
In Fig.~\ref{fig:Rbar_plateau} we show examples of the Euclidean time, $x_4$, dependence of 
$\Rbar_P(t)/t$. The quark propagators are determined using a pointlike source.
We have studied the influence of using smeared sources and we found no 
obvious advantage in the determination of the plateau in Eq.~\eqref{eq:RP_hadron}.
We discuss the impact of smeared sources in Sec.~\ref{sec:analysis2}.

\begin{figure}
	\centering
	\includegraphics[width=0.49\textwidth]{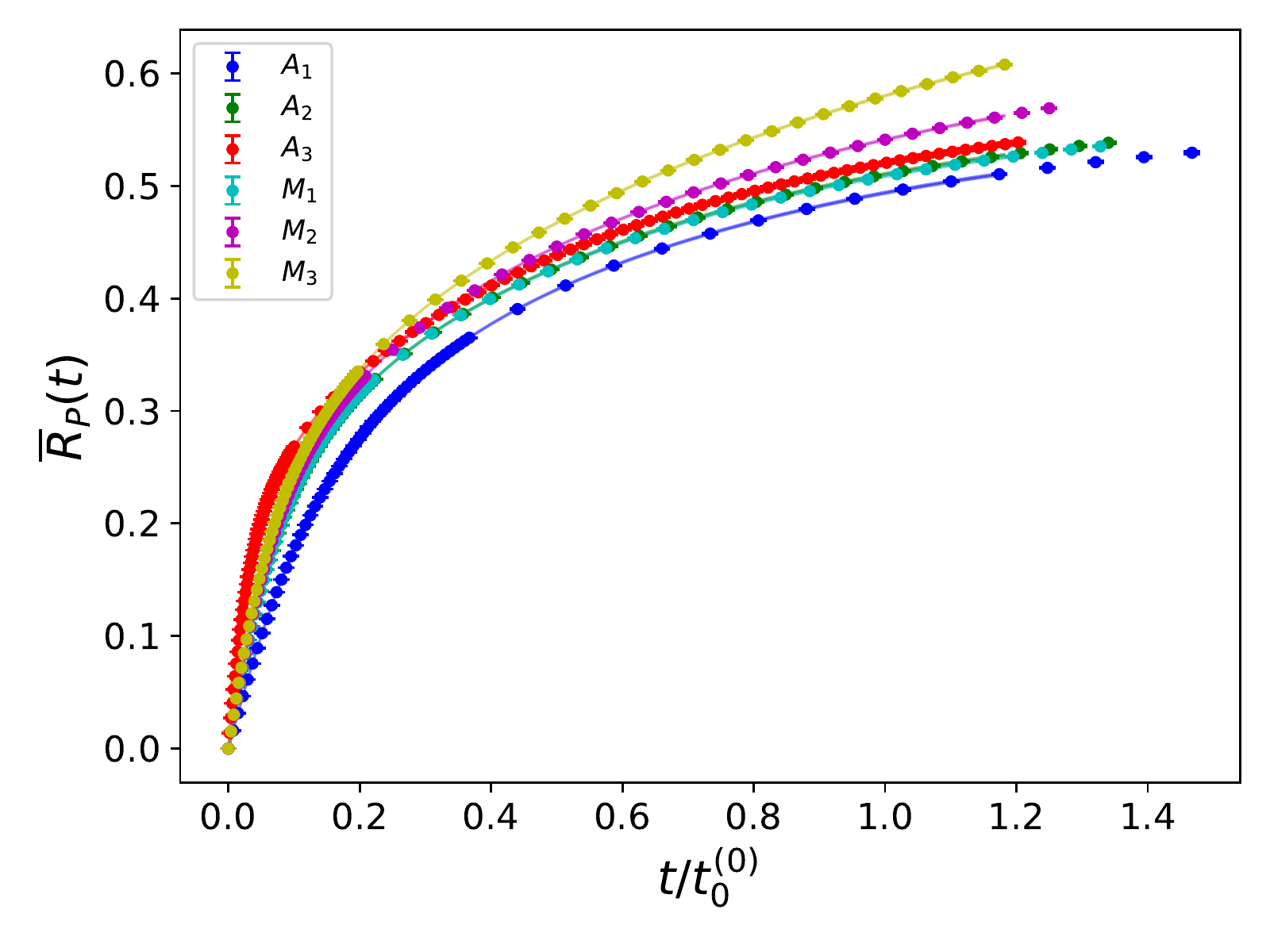}
	\caption{Flow time dependence of the ratio $\Rbar_P$ for all our ensembles.
	For the x-axis we choose the standard definition of $t_0/a^2$, i.e.
	$t_0=t_0^{(0)}$ (see main text).} \label{fig:RP_bart}
\end{figure}

In Fig.~\ref{fig:RP_bart} we show the flow time dependence, 
in unit of $t/t_0$, of $\Rbar_P$ for all our ensembles in Table~\ref{tab:latpar1}.
The data allow a smooth interpolation with a cubic spline 
and we can perform a robust chiral and continuum limit extrapolation with a global fit
at fixed values of the flow time.

To assess the uncertainty due to the continuum limit, we have calculated
the values of $t_0$ for each ensemble,
removing tree-level cutoff effects, at different orders in
$a^2$~\cite{Fodor:2014cpa}.
\begin{figure}
	\centering
	\includegraphics[width=0.49\textwidth]{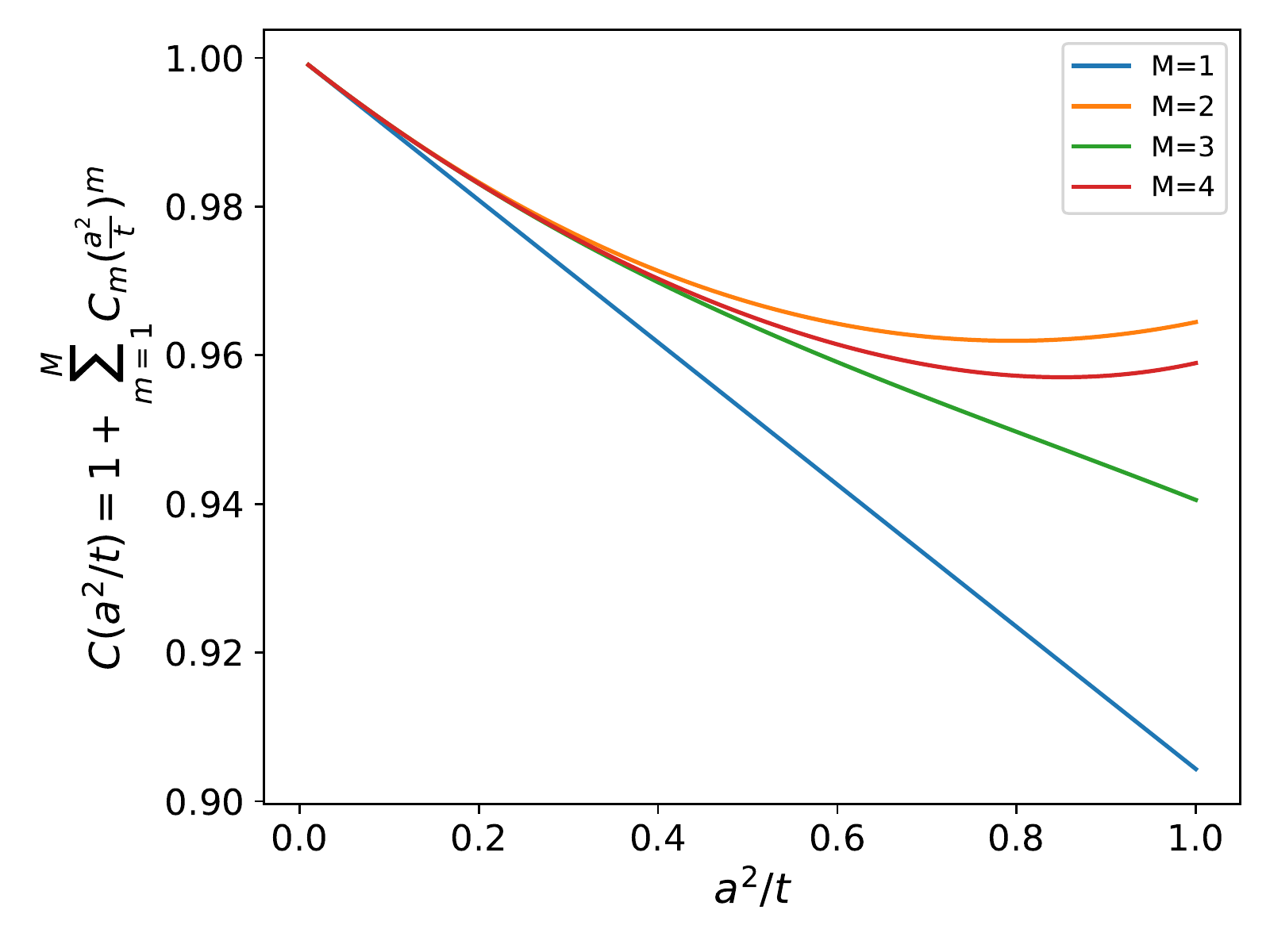}
	\caption{Tree-level lattice artifacts of the energy density defined in Eq.~\eqref{eq:Coft} for different
	orders $(\frac{a^2}{t})^M$.
	\label{fig:C_imp}}
\end{figure}
\begin{table}
	\caption{\label{tab:t0_imp} Here we present the $t_0^{(M)}/a^2$ values with improvements in Ref.~\cite{Fodor:2014cpa}, where $M$ is the order of the improvements in $C(a^2/t)=1+\sum^{M}_{m=1} C_{m} (\frac{a^{2}}{t})^{m}$.}
  \centering\begin{tabular}{|c|c|c|c|c|c|}
    \hline
		Designation & $t_0^{(0)}/a^2$ & $t_0^{(1)}/a^2$ & $t_0^{(2)}/a^2$ & $t_0^{(3)}/a^2$ & $t_0^{(4)}/a^2$ \\
    \hline
    \hline
		$M_1$       & 2.2586(12) & 2.1344(12) & 2.1723(12) & 2.1655(12) & 2.1680(12)\\
    \hline
		$M_2$       & 2.3993(12) & 2.2739(11) & 2.3094(11) & 2.3033(11) & 2.3054(11)\\
    \hline
		$M_3$       & 2.5371(15) & 2.4088(15) & 2.4435(15) & 2.4379(15) & 2.4397(15)\\
    \hline\hline
		$A_1$       & 1.3627(15) & 1.2397(15) & 1.3028(14) & 1.2849(15) & 1.2951(14)\\
    \hline
		$A_2$       & 2.2378(24) & 2.1145(23) & 2.1526(23) & 2.1457(23) & 2.1482(23)\\
    \hline
		$A_3$       & 4.9879(65) & 4.8652(64) & 4.8815(64) & 4.8802(64) & 4.8804(64)\\
    \hline
  \end{tabular}
\end{table}
The value of $t_0/a^2$ determined on a given lattice 
is usually defined as the value of the flow time satisfying
\be 
t_0^2\left\langle E(t_0) \right\rangle_{\textrm{lat}} = 0.3\,, 
\label{eq:t0}
\ee 
where $E = \frac{1}{4}G_{\mu\nu}^aG_{\mu\nu}^a$ and the index ``lat" reminds us that
the expectation value is evaluated on the lattice.
The discretization effects depend on the form of the discretization of $3$
different aspects of the calculation: the lattice action, the lattice GF
equation and the lattice definition of the observable which is the energy in this case.
By evaluating the energy lattice expectation value at tree-level it is possible to
remove tree-level cutoff effects.
The tree-level ratio
\be 
C(t) = \frac{\left\langle E(t) \right\rangle^{(0)}_{\textrm{lat}}}{\left\langle E(t) \right\rangle_{\textrm{c}}^{(0)}}\,,
\label{eq:Coft}
\ee 
between the lattice and the continuum expectation values is $1$ in the continuum limit,
and dimensionless. At tree-level of perturbation theory, in a pure gauge calculation,
the only scale available, beside the lattice spacing $a$, is the flow time $t$, thus $C=C(a^2/t)$
and expanding in powers of $a^2/t$ one has
\be 
C(t) = 1 + \sum_{m=1}^M C_m \left(\frac{a^2}{t}\right)^m\,.
\ee 
The values of the coefficients $C_m$, for a wide selection of different lattice
actions, GF equations and energy definitions, are given in
Ref.~\cite{Fodor:2014cpa}.
Our choice corresponds to the Iwasaki gauge action, a lattice GF equation using
the ``plaquette" discretization for the field tensor,
and an energy defined using the so called ``clover" definition [see Eq.~\eqref{eq:clover}]. 
We can have different determinations of $t_0/a^2$
\be 
\left(t_0^{(M)}\right)^2\left\langle E(t_0^{(M)}) \right\rangle_{\textrm{lat}}\times \frac{1}{C(t_0^{(M)})}=0.3\,,
\label{eq:t0M}
\ee 
depending on the order in $a^2/t$ to which we evaluate $C(t)$. 
In Table~\ref{tab:t0_imp} we list all the values of $t_0/a^2$ 
we have determined for all the ensembles
and in Fig.~\ref{fig:C_imp} we show the ratio $C(t)$ 
for given orders $M$ and for our lattice setup.
The complete Symanzik analysis of the O($a^2$) for flowed gauge observable
can be found in Ref.~\cite{Ramos:2015baa}.

We can now perform the chiral and continuum extrapolation of the renormalized coupling
defined as
\be 
\gbar^2 = \frac{16 \pi^2}{3}t^2 \left\langle E(t) \right\rangle\,,
\label{eq:g2}
\ee
at fixed values of $t/t_0$ and using different definitions of $t_0/a^2$.
We parametrize our data with 
\be 
\left[\gbar^2\right]_{\textrm{fit}} = A_g(t) + B_g(t)a^2 + C_g(t) m_q + D_g(t)m_q^2\,,
\label{eq:gbar_fit}
\ee 
where $X_g(t)$, $X=A, B, C, D$, are the fit parameters.
We keep the freedom to have both cutoff effects and 
the quark mass term depending on the flow time, but we do not have 
sufficient ensembles to be sensitive to mass dependent cutoff effects.
The quark mass is determined with the PCAC relation~\eqref{eq:PCAC}.
\begin{figure}
	\centering
	\includegraphics[width=0.49\textwidth]{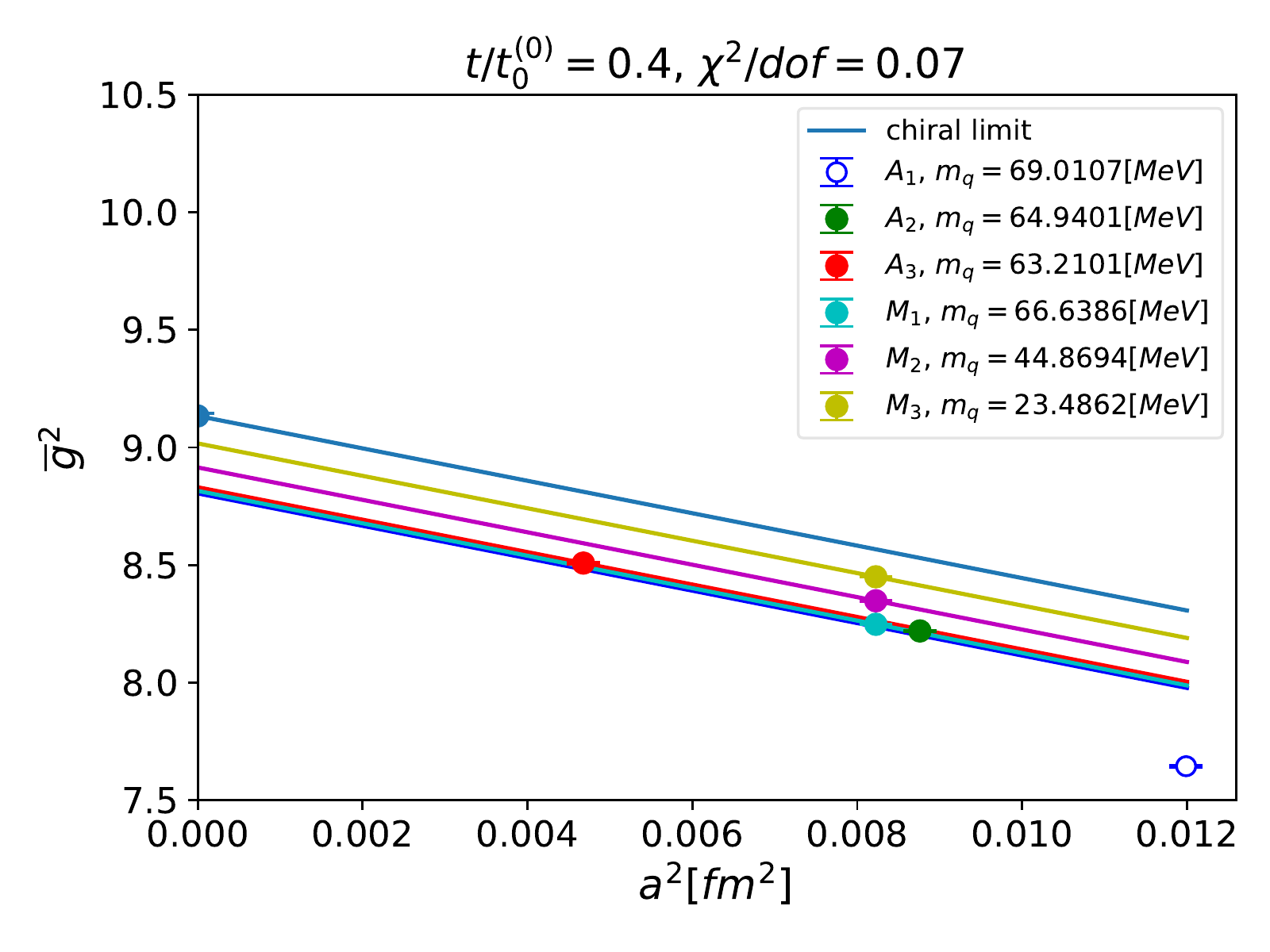}
	\includegraphics[width=0.49\textwidth]{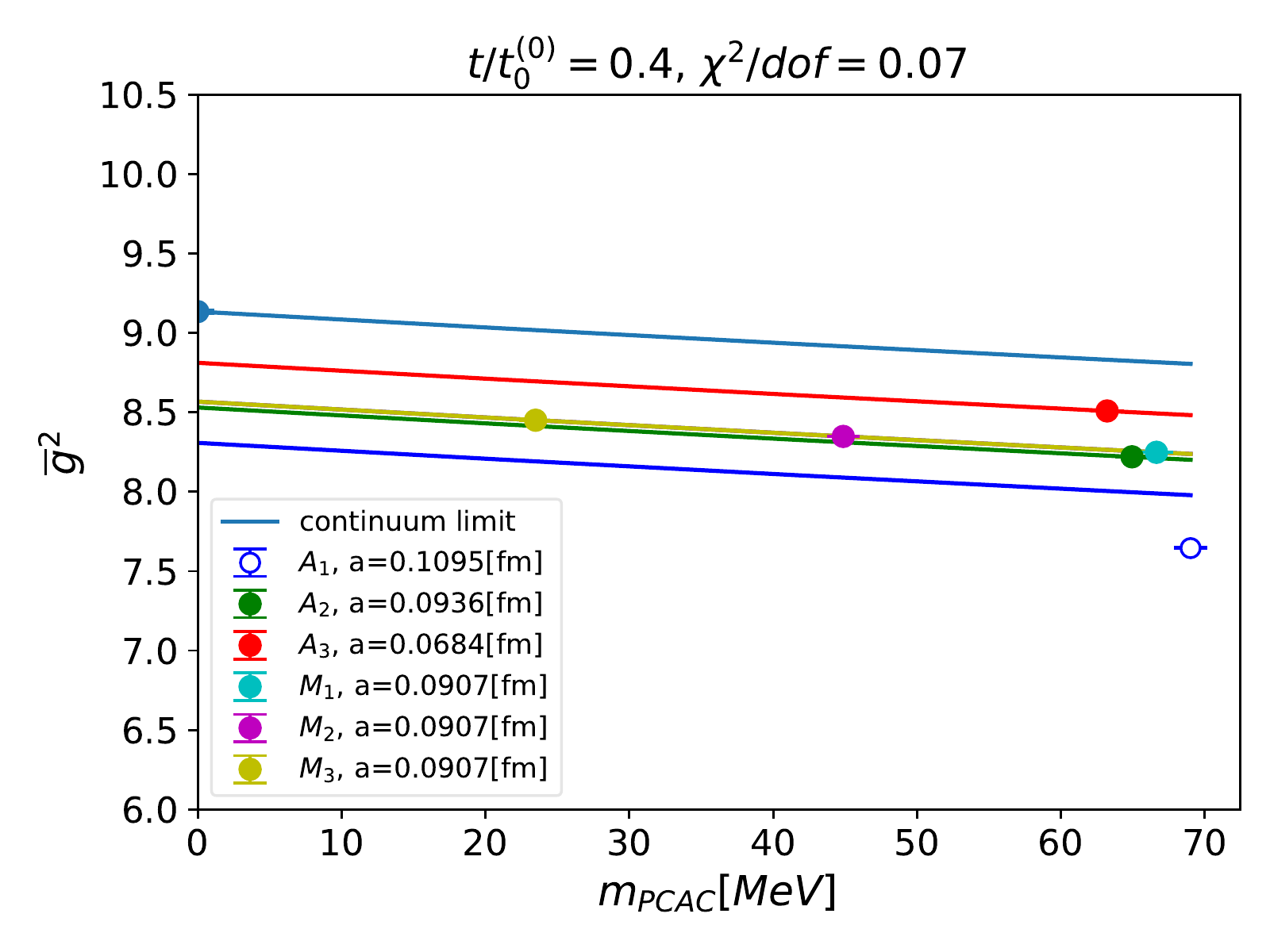}
	\caption{Simultaneous fit for chiral and continuum extrapolation of renormalized 
	coupling $\gbar^2$ at $t/t_0=0.4$ choosing the standard definition of $t_0$
	with $M=0$. See main text for more details. 
	The empty point at the coarser lattice spacing is excluded from the 
	continuum and chiral extrapolation.}
	\label{fig:g2_cl}
\end{figure}

\begin{figure}
	\centering
	\includegraphics[width=0.49\textwidth]{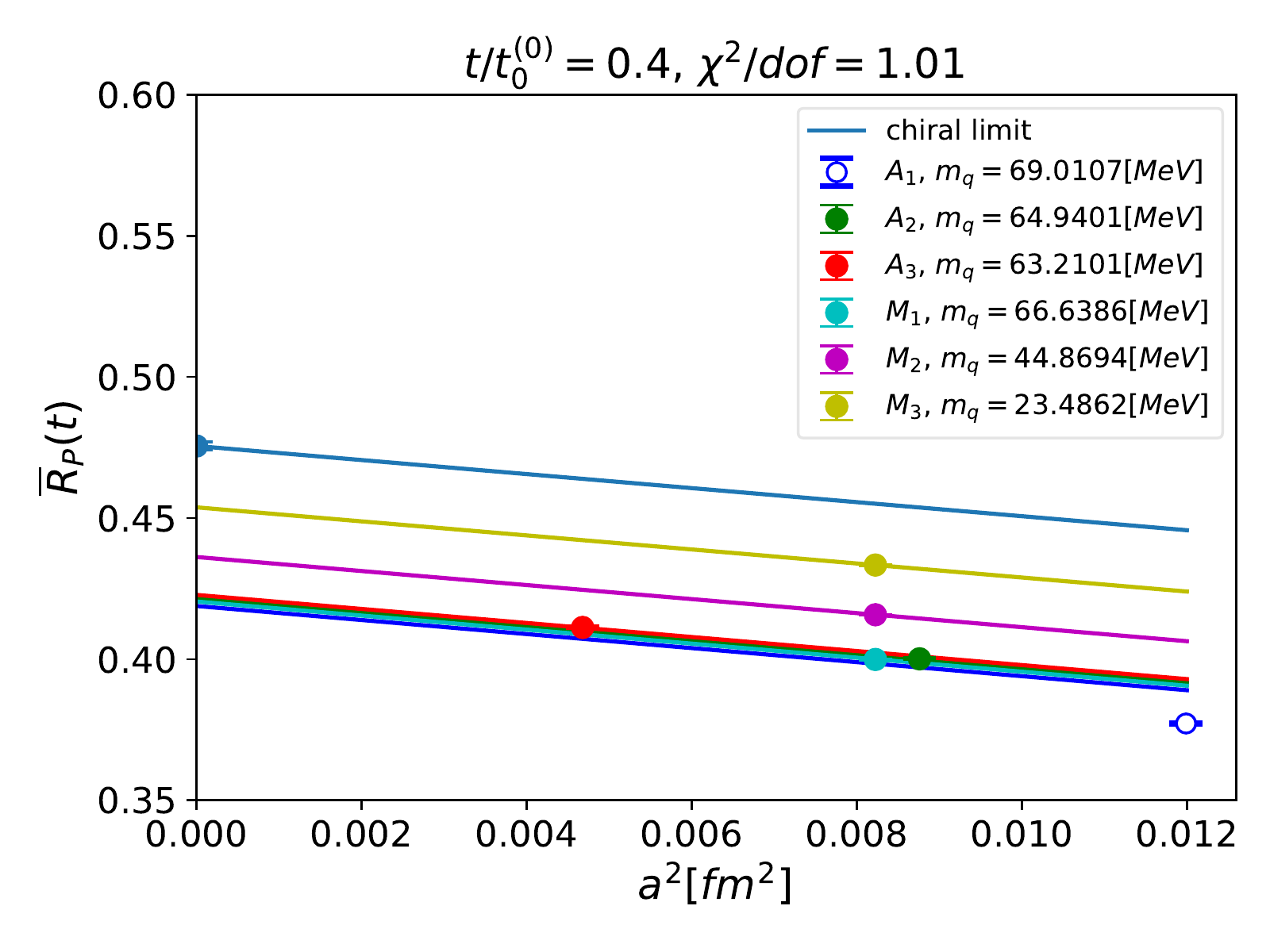}
	\includegraphics[width=0.49\textwidth]{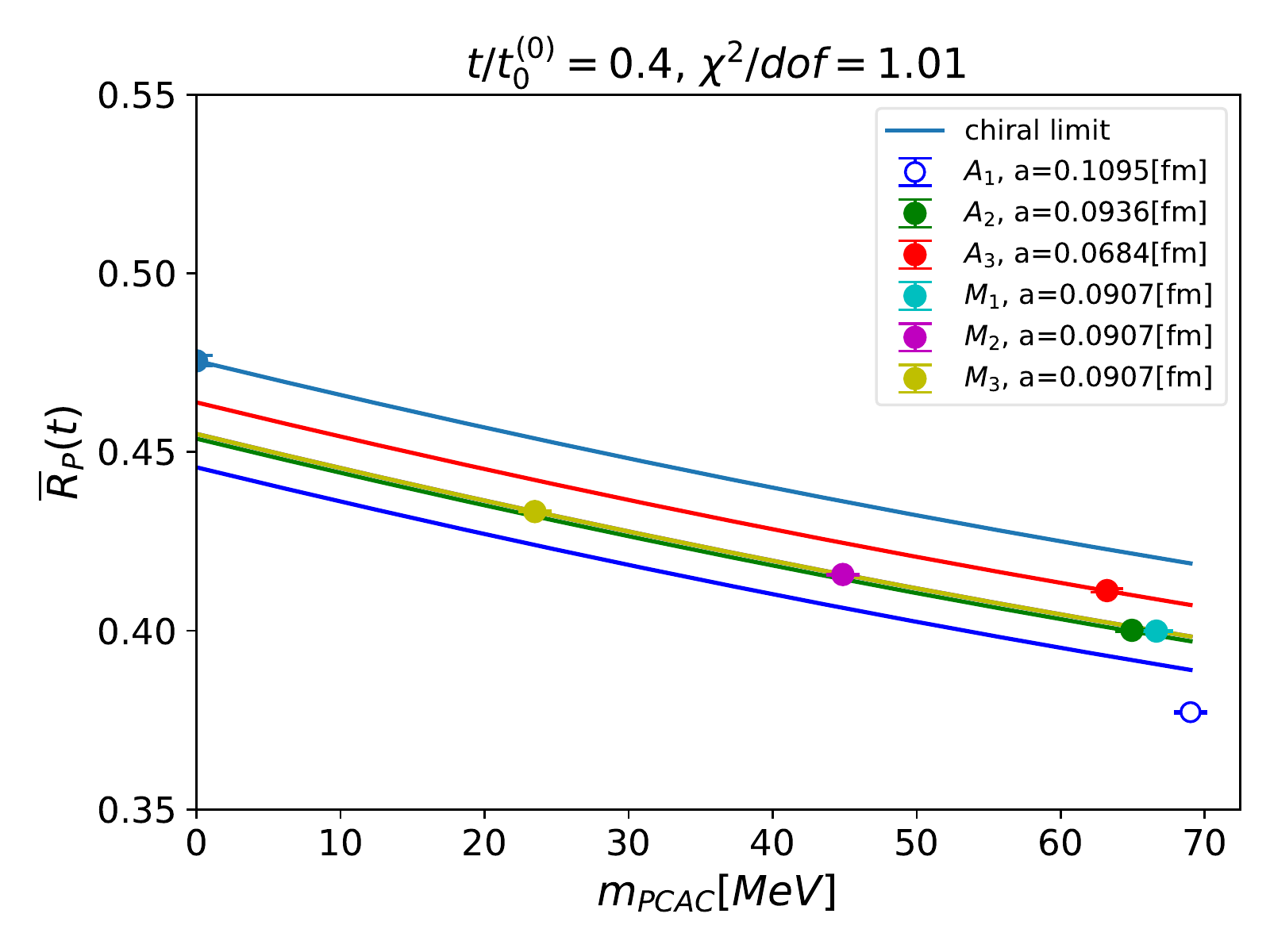}
	\caption{Global fit of the ratio $\Rbar_P$ for the
	definition of $t_0$, with $M=0$. See main text for more details. 
	The empty point at the coarser lattice spacing is excluded from the 
	continuum and chiral extrapolation.}
\label{fig:Rbar_cl}
\end{figure}

In Fig.~\ref{fig:g2_cl} we show the 
results of our global fits for the coupling $\gbar^2$ 
at fixed values of $t/t_0 =0.4$. 
In the left plot we show the 
continuum extrapolation for the standard definition of $t_0$, i.e.
with $C(t)=1$, while on the right plot we show the chiral extrapolation.
The variation of the results obtained for different choices of $t_0/a^2$ represents
our estimate of the systematic uncertainty of the continuum extrapolation.
We can now repeat the same analysis for the ratio $\Rbar_P$ of Eq.~\eqref{eq:RPbar}.
We parametrize our data with the function
\be 
\left[\Rbar_P\right]_{\textrm{fit}} = A_R(t) + B_R(t)a^2 + C_R(t) m_q + D_R(t) m_q^2\,,
\label{eq:Rbar_fit}
\ee 
where $X_R(t)$, $X=A, B, C, D$, are the fit parameters. 
In Fig.~\ref{fig:Rbar_cl} we show respectively the
continuum limit (left plot) and the chiral extrapolation (right plot) 
at fixed $t/t_0=0.4$ using the standard definition of $t_0/a^2$ with $C(t)=1$.

\begin{figure}
	\centering
	\includegraphics[width=0.45\textwidth]{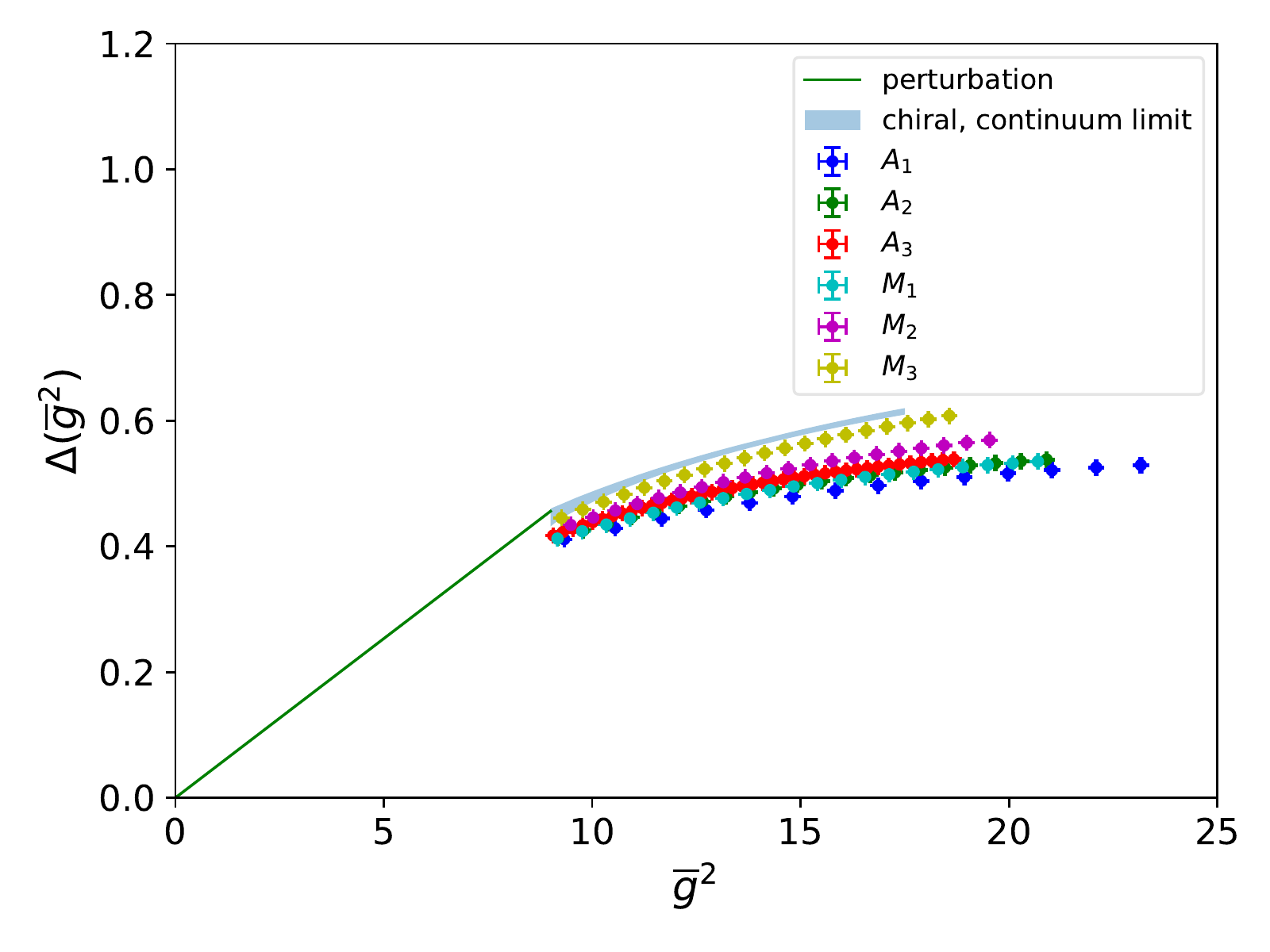}
	\caption{Blue band corresponds to chiral, continuum limit of data. 
	The error is obtained from the maximum difference between $t_0^{(M)}$, $M=0, 1, 2, 3, 4$. 
	For plotting $\gbar^2$ for data points are obtained using $t_0^{(4)}$.}
	\label{fig:Rbar_match}
\end{figure}

Having determined both $\Rbar_P(t)$ and $\gbar(t)$ in the continuum limit, we
can analyze the dependence of $\Delta(\gbar^2)$ as a function of the renormalized
coupling and compare it with perturbation theory.
We have calculated using continuum perturbation theory
at order O($\gbar^2$) the same $\Delta(\gbar^2)$. Details of the calculation
can be found in Appendix~\ref{app:PT}. 
The final result at O($\gbar^2$), 
consistent with our previous determination~\cite{Rizik:2020naq}, 
is given by
\be 
\Delta^{(1)}\gbar^2 = \frac{1}{2\pi^2}\gbar^2\,,
\label{eq:Rbar_PT}
\ee 
and is represented in Fig.~\ref{fig:Rbar_match} by a green straight line.
In Fig.~\ref{fig:Rbar_match} we also show our raw data, obtained with the $M=4$ definition of 
$t_0/a^2$, and the continuum extrapolation. The raw data show statistical uncertainties
in both the y- and x-directions, the latter coming from the uncertainty on 
the determination of $\gbar^2$.
The blue band represents the statistical and systematic uncertainty in the continuum extrapolation
obtained considering the maximal variation of results varying among all the values 
of $M=0,\ldots,4$.
The plot shows the data and our continuum extrapolation
for values above the renormalized coupling $\gbar^2\simeq 9$.
The reason for this choice stems from our ability to control the continuum limit, 
which for smaller values of $\gbar^2$ become increasingly difficult due to larger cutoff effects.

\begin{figure}
	\centering
	\includegraphics[width=0.32\textwidth]{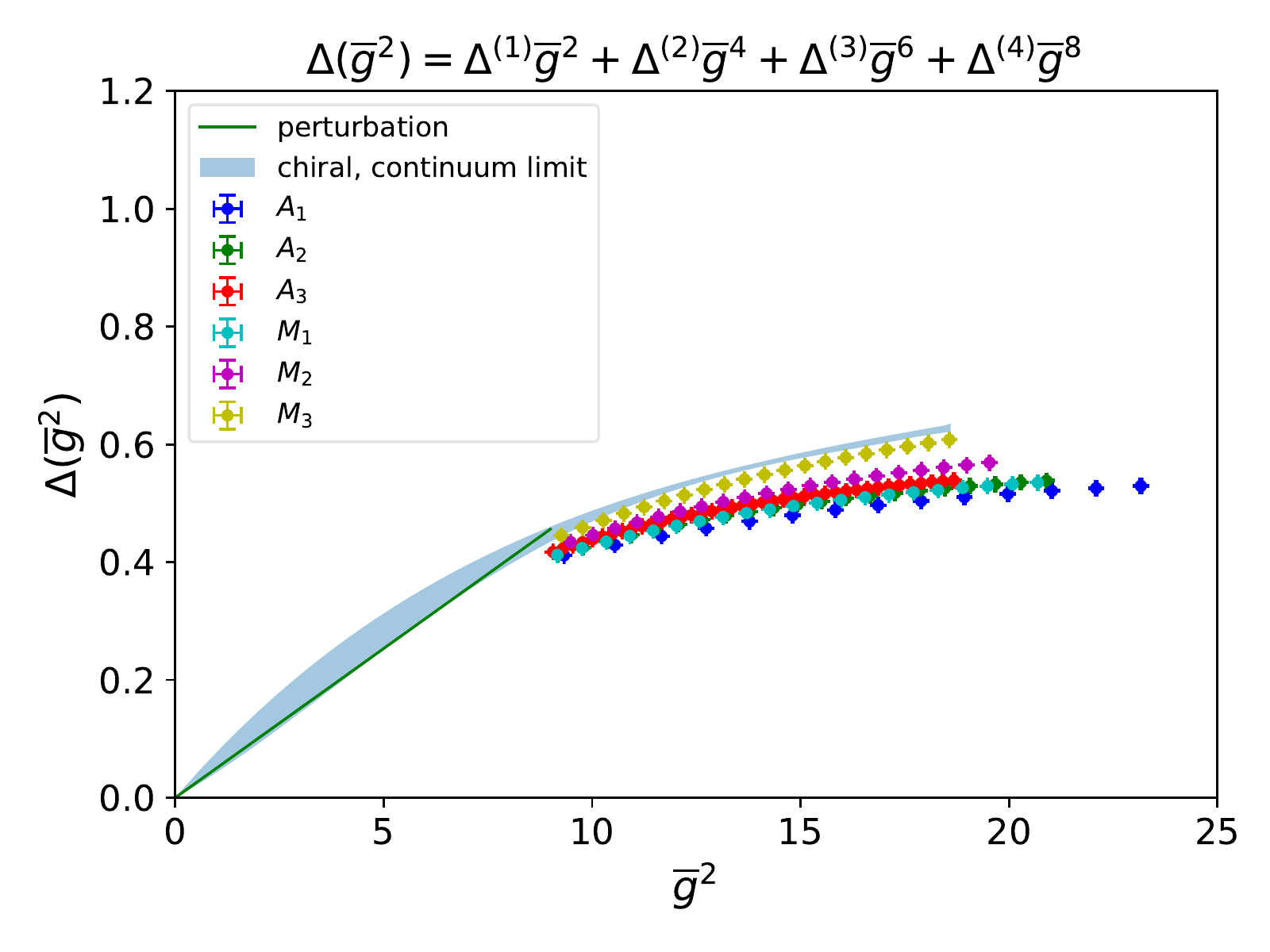}
	\includegraphics[width=0.32\textwidth]{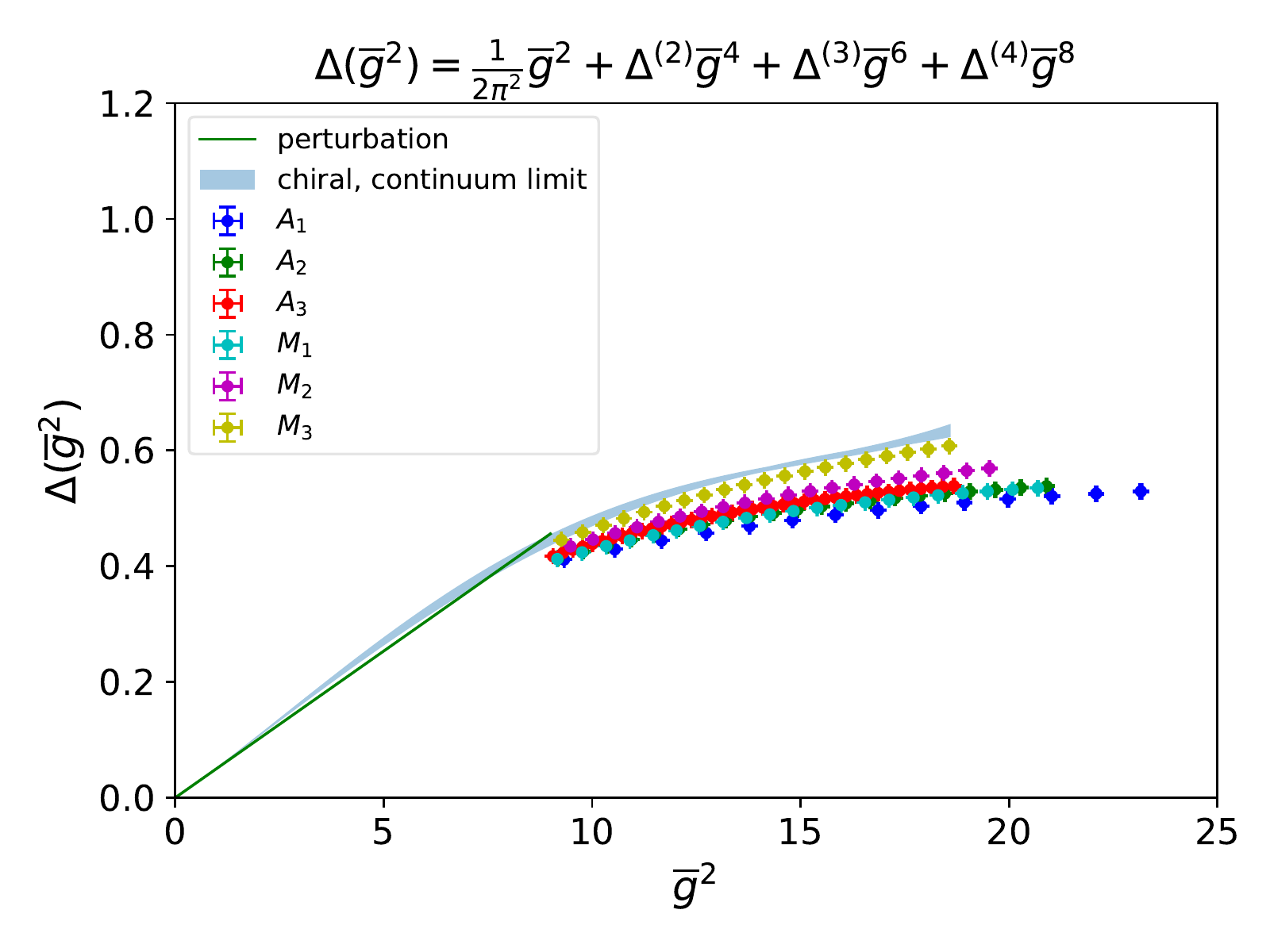}
	\includegraphics[width=0.32\textwidth]{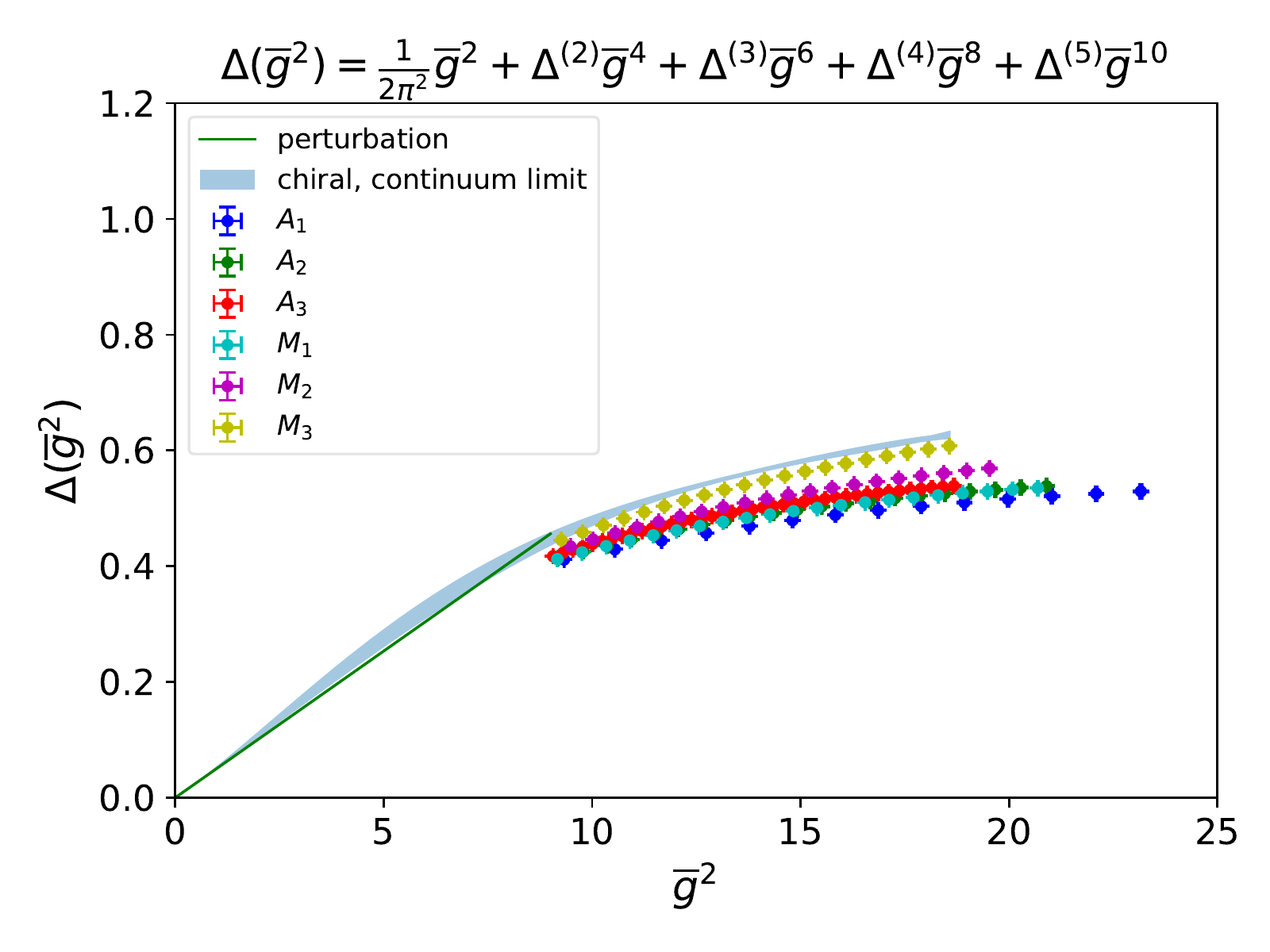}
	\caption{Error bands are obtained from the fitting of the chiral, 
	continuum limit data. 
	The error band is obtained from the maximum difference between the fitting 
	results of extrapolated data using different $t_0^{(M)}$, $M=0, 1, 2, 3, 4$.}
	\label{fig:Rbar_poly}
\end{figure}

\begin{table}
	\caption{\label{tab:g2_fitting} Results for the polynomial fit 
	$\Delta(\gbar^2)$ as a function of $\gbar^2$. The perturbative result is given by 
	$\Delta^{(1)} = \frac{1}{2\pi^2}=0.05066$.
	}
  \centering
	\begin{tabular}{|c|c|c|c|c|c|c|}
  \hline
		\multicolumn{7}{|c|}{$\Delta(\gbar^2)=\Delta^{(1)}\gbar^2+\Delta^{(2)}\gbar^4+\Delta^{(3)}\gbar^6+\Delta^{(4)}\gbar^8$} \\
		\hline
		M & $\Delta^{(1)}$ & $\Delta^{(2)}$  & $\Delta^{(3)}$  & $\Delta^{(4)}$  &  $\chi^2/d.o.f.$ & d.o.f.\\
    \hline
		0 & 0.07255(71) & -0.00279(17) &  0.000027(11)  &  0.00000059(25) & 0.003  & 785 \\
		1 & 0.08156(55) & -0.00444(13) &  0.0001289(85) & -0.00000155(18) & 0.0004 & 815 \\
		2 & {\textbf{0.0491(10)}}  & 0.00191(23)  & -0.000286(16)  &  0.00000749(35) & 0.13   & 874 \\
		3 & 0.07357(66) & -0.00304(16) &  0.000046(10)  &  0.00000012(22) & 0.001  & 833 \\
		4 & 0.0409(11)  & 0.00376(27)  & -0.000420(19)  &  0.00001063(43) & 0.45   & 888 \\
  \hline
		\multicolumn{7}{|c|}{$\Delta(\gbar^2)=\frac{1}{2\pi^2}\gbar^2+\Delta^{(2)}\gbar^4+\Delta^{(3)}\gbar^6+\Delta^{(4)}\gbar^8$} \\
		\hline
		M & $\Delta^{(2)}$  & $\Delta^{(3)}$  & $\Delta^{(4)}$ & $\chi^2/d.o.f.$ & \multicolumn{2}{c|}{d.o.f.}\\
    \hline
		0 & 0.002321(44) & -0.0003604(46) & 0.00001016(14) & 0.40 &\multicolumn{2}{c|}{786} \\
		1 & 0.002611(46) & -0.0003923(46) & 0.00001097(14) & 1.09 &\multicolumn{2}{c|}{816} \\
		2 & 0.001551(38) & -0.0002593(44) & 0.00000683(14) & 0.13 &\multicolumn{2}{c|}{875} \\
		3 & 0.002210(42) & -0.0003444(43) & 0.00000955(13) & 0.59 &\multicolumn{2}{c|}{834} \\
		4 & 0.001489(38) & -0.0002493(45) & 0.00000647(14) & 0.57 &\multicolumn{2}{c|}{889} \\
    \hline
		\multicolumn{7}{|c|}{$\Delta(\gbar^2)=\frac{1}{2\pi^2}\gbar^2+\Delta^{(2)}\gbar^4+\Delta^{(3)}\gbar^6+\Delta^{(4)}\gbar^8+\Delta^{(5)}\gbar^{10}$} \\
		\hline
		M & $\Delta^{(2)}$  & $\Delta^{(3)}$  & $\Delta^{(4)}$ & $\Delta^{(5)}$ & $\chi^2/d.o.f.$ & d.o.f.\\
		\hline
		0 & 0.004075(68) & -0.000765(14) &  0.0000405(10)  & -0.000000743(22) & 0.005 & 785 \\
		1 & 0.005016(62) & -0.000934(12) &  0.00005056(80) & -0.000000941(18) & 0.04  & 815 \\
		2 & 0.001565(86) & -0.000262(20) &  0.0000071(14)  & -0.000000005(30) & 0.13  & 874 \\
		3 & 0.004007(63) & -0.000751(13) &  0.00003944(88) & -0.000000714(19) & 0.01  & 833 \\
		4 & 0.00095(10)  & -0.000126(22) & -0.0000027(16)  &  0.000000221(36) & 0.51  & 888 \\
    \hline
  \end{tabular}
\end{table}
This behavior is expected at such short distances, but perhaps surprisingly,
as we can see from Fig.~\ref{fig:Rbar_match}, we are still able to match the
perturbative results even at these relatively large values of the renormalized
coupling.
We then attempt to parametrize the dependence on the renormalized coupling of $\Delta(\gbar^2)$ over the whole range of renormalized couplings with polynomials of the form
\be 
\Delta(\gbar^2) = \sum_{i=1}^{N}\Delta^{(i)} \gbar^{2i}\,,\qquad N=4, 5\,.
\label{eq:Rbar_poly}
\ee 
In Fig.~\ref{fig:Rbar_poly} we show the results of these polynomial fits: for 
$N=4$ we compare the results leaving the leading
O($\gbar^2$) coefficient as a free fit parameter (left plot), while
in the middle plot with constrain it to our perturbative results.
From the plots and the fit results in Table~\ref{tab:g2_fitting}
we observe that without constraining the fit 
parameter $\Delta^{(1)}$, we obtain results consistent with one-loop perturbation
theory if we consider the uncertainties due to the continuum limit.
We also notice that the value for $M=2$ is perfectly consistent
with the perturbative result.
In the right plot of Fig.~\ref{fig:Rbar_poly} we show the same fit up to the 
order O($\gbar^{10}$). The blue bands represent the uncertainty 
stemming from the uncertainty in the continuum limit.
The fit parameters and their uncertainties are given in Table~\ref{tab:g2_fitting}.
The results indicate a very good description of the 
numerical data combined with the perturbative results and demonstrate a rather fast convergence 
of the polynomial.  There is still some instabilities in the fit parameters, especially when changing
the order of the polynomial (however the $M=2$ case shows very stable results).
As our final result we quote the parametrization with $N=4$.
This curve is universal and renormalization group invariant.
It can be applied to any determination of the expansion coefficient of the pseudoscalar
density and to any corresponding hadronic matrix element.
\subsection{Analysis in terms of the bare coupling}
\label{sec:analysis2}

A second strategy to study the behavior of the qCEDM 
at small flow time is to define an effective expansion coefficient
$c_\chi$ in Eq.~\eqref{eq:c_chi} and determine its value
in terms of the bare coupling.
To estimate $c_\chi$, as discussed in Sec.~\ref{sec:sfte}, we want to determine
the ratio $R_P(t)$ in Eq.~\eqref{eq:RP_hadron}.
The calculation of the correlation functions in~\eqref{eq:COP}
and~\eqref{eq:CPP} requires the same propagators used in the analysis presented
in the previous section.
The only difference is that in the denominator of $R_P$ we do not flow the quark propagators.

\begin{figure}
    \centering
    \includegraphics[width=0.32\textwidth]{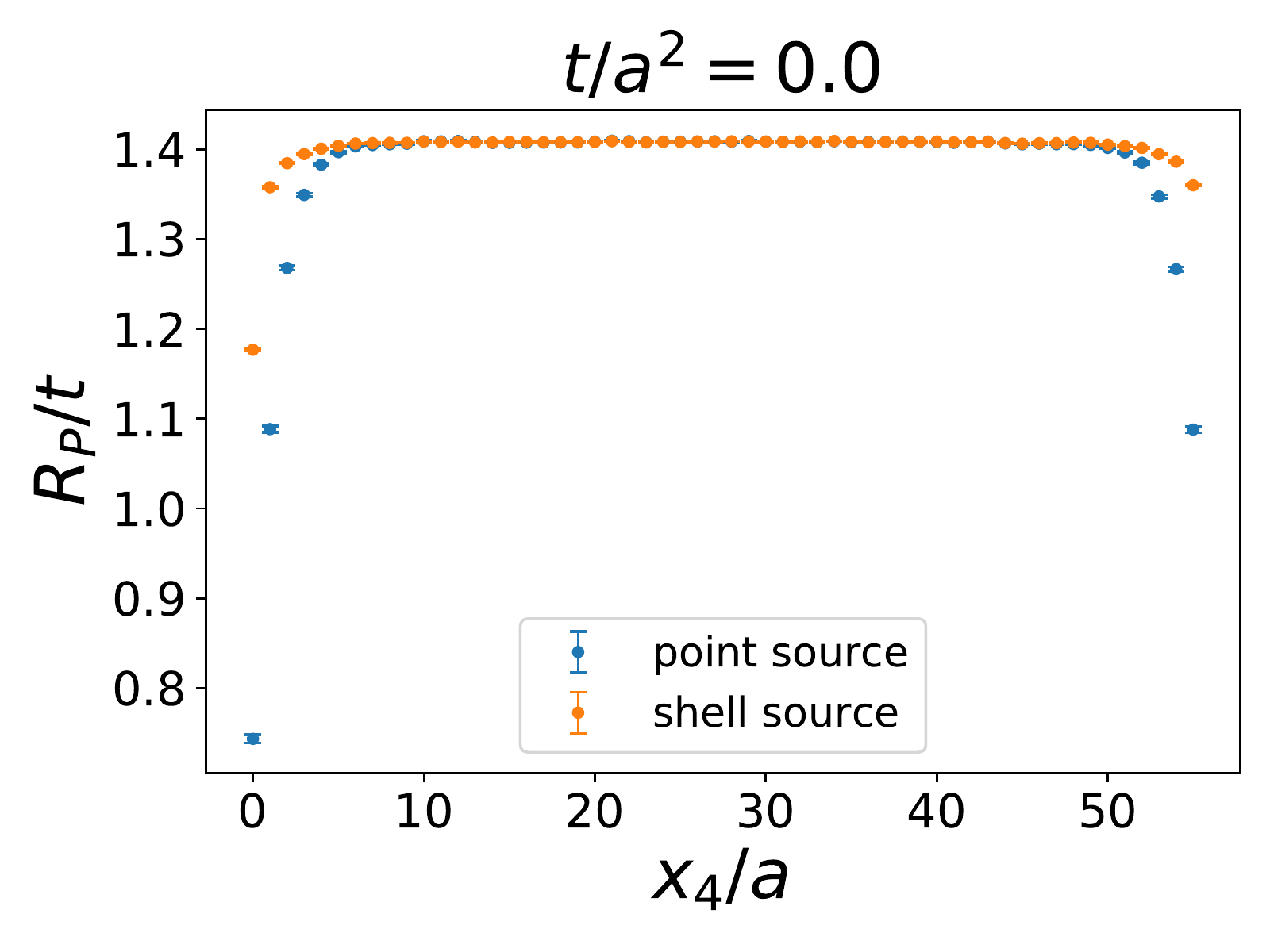}
    \includegraphics[width=0.32\textwidth]{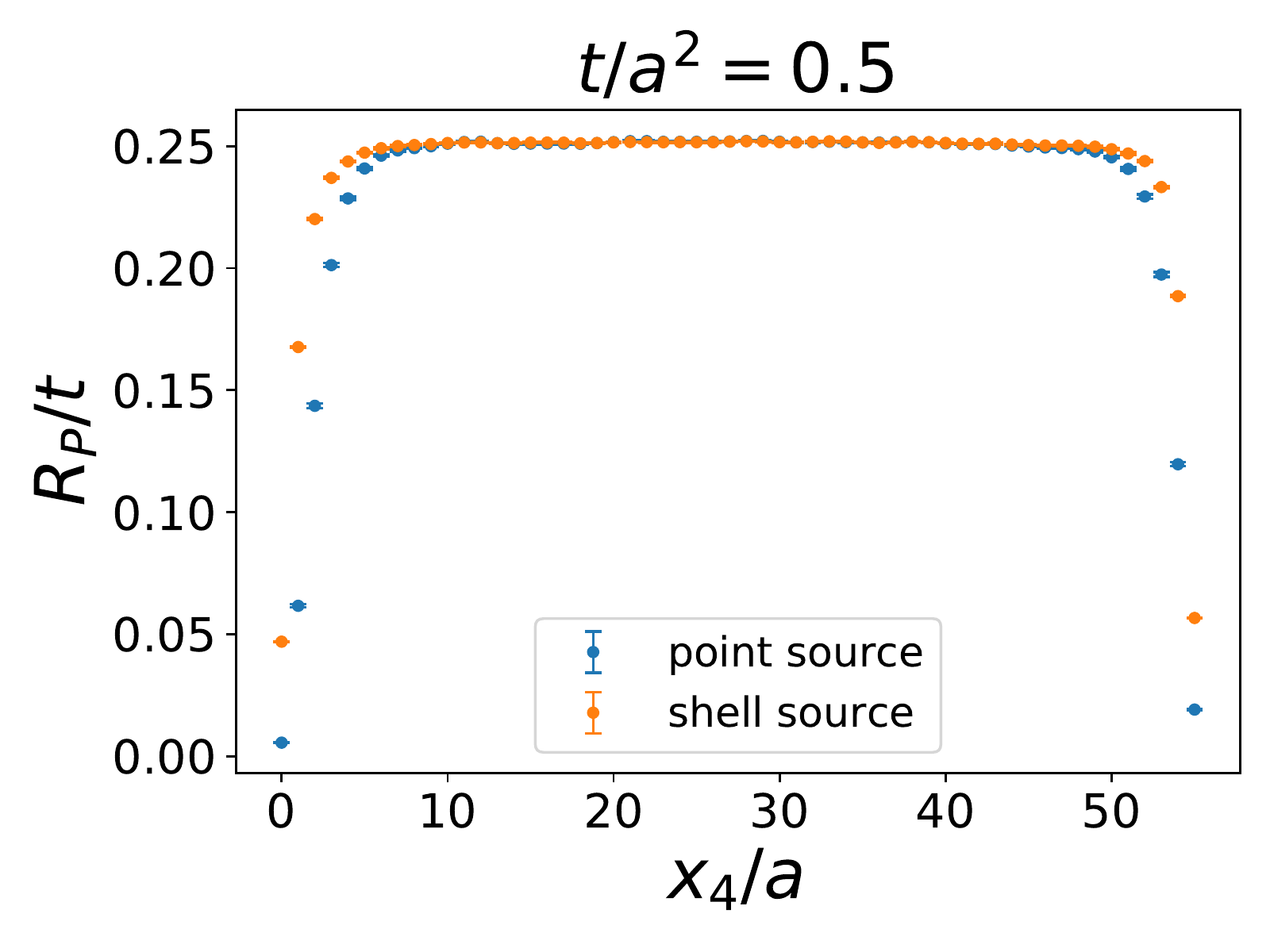}
    \includegraphics[width=0.32\textwidth]{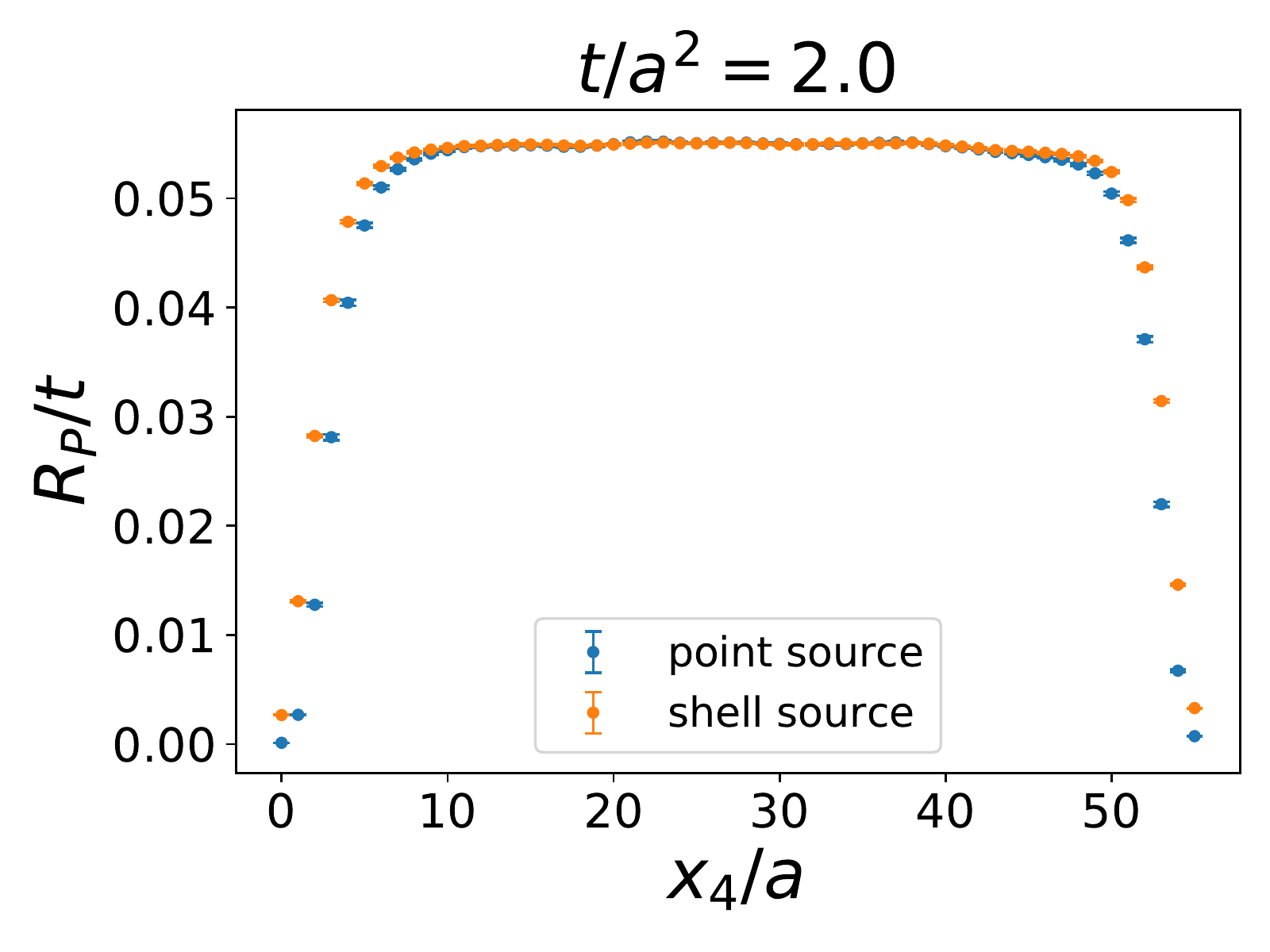}
		\caption{The ratio $R_P/t$ defined in Eqs.~(\ref{eq:RP}, \ref{eq:RP_hadron})
		and determine on the ensemble $A_3$ 
		(see Table~\ref{tab:latpar1}) for several values of the flow time $t/a^2 = 0, 0.5, 2.0$,
    corresponding to a flow time radius $r_f = \sqrt{8t} = 0, 2a, 4a$. Different colors correspond
	to different sources.}
    \label{fig:R_plateau}
\end{figure}

In Fig.~\ref{fig:R_plateau}, we show the source-sink separation, $x_4$,
dependence of $R_P(x_4;t)/t$ for several values of the flow time $t/a^2 = 0,
0.5, 2.0$, corresponding to values of the flow time radius $r_f = \sqrt{8t} =
0, 2a, 4a$.
We observe that the asymptotic plateau value
is reached with no particular problem 
for every value of the flow time we adopt in this work.
In Fig.~\ref{fig:R_plateau} we also show a comparison between a point and a
gauge-invariant Gaussian smeared source~\cite{Gusken:1989qx, Alexandrou:1990dq}
with $64$ iterations of the smearing algorithm and, using the definition of
Ref.~\cite{Gusken:1989qx}, a smearing parameter of $\alpha=0.39$.
The determination of the plateau is fairly straightforward so we perform a
simple constant fit, where the fit range is determined in a standard way
minimizing the corresponding $\chi^2$.

\begin{figure}
    \centering
    \includegraphics[width=0.49\textwidth]{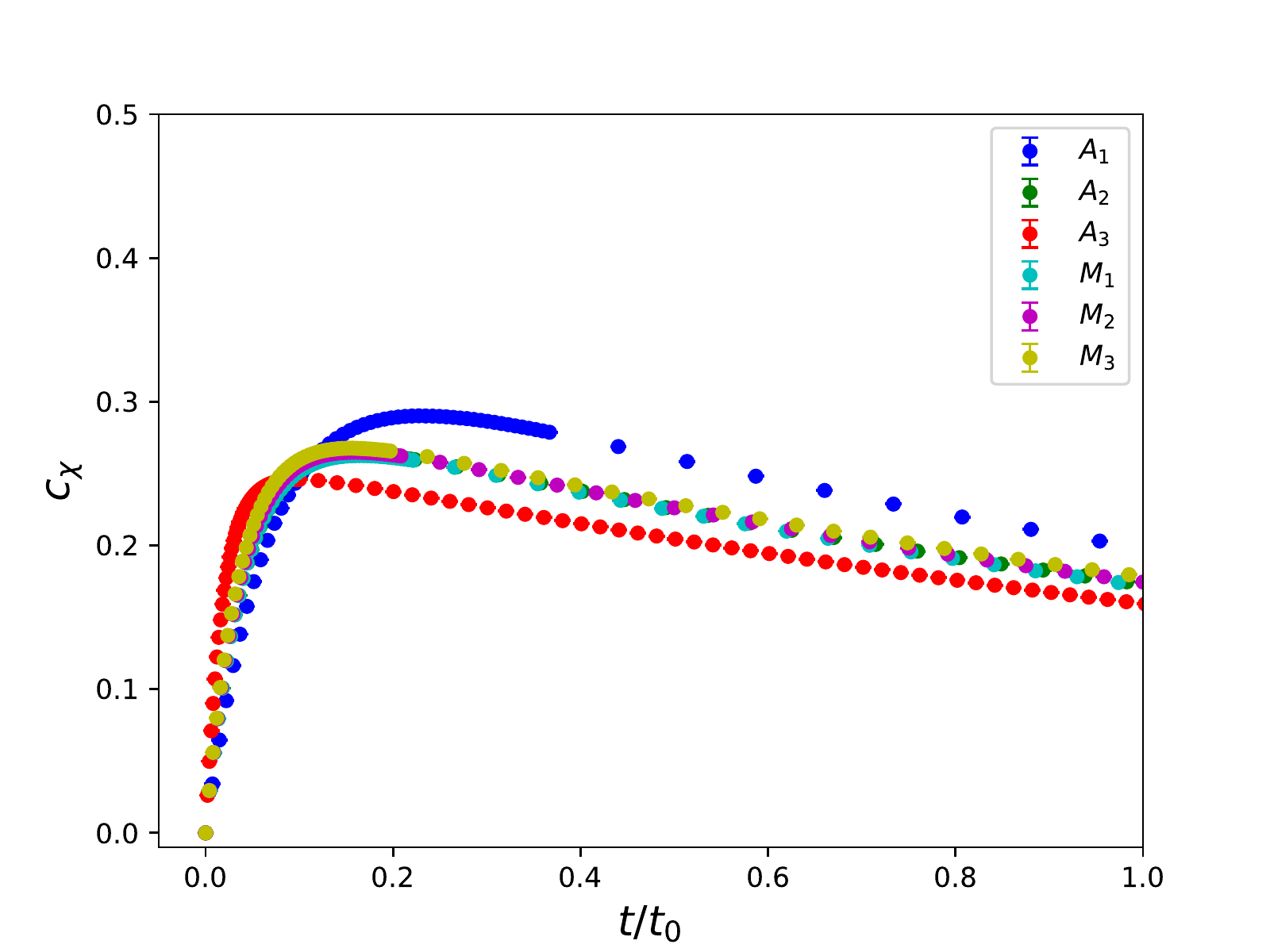}
    \includegraphics[width=0.49\textwidth]{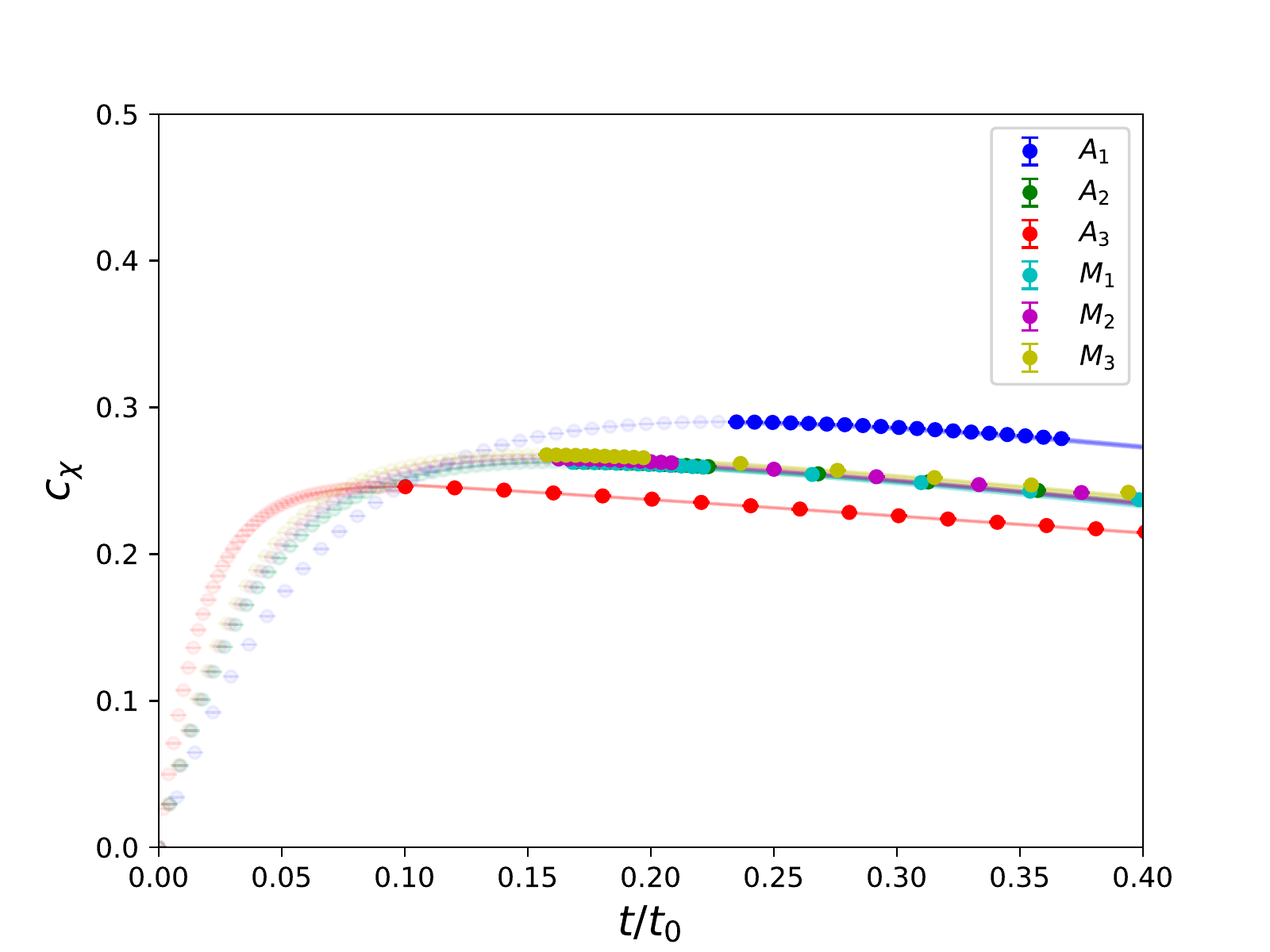}
		\caption{Ratio $R_P$ in Eqs.~(\ref{eq:RP}, \ref{eq:RP_hadron}) 
        as a function of $t/t_0$. 
        In the right plot, error bands are reconstructed based on the 
        final fitting results and solid data points are belonged 
        in the selected fitting ranges in Table~\ref{tab:fit_range}.}
    \label{fig:t_R_flow_dep}
\end{figure}

The result of the fit is the ratio $R_P(t)$ of Eq.~\eqref{eq:RP_hadron} and it
is shown, as a function of $t/t_0$, in the left plot of Fig.~\ref{fig:t_R_flow_dep}. 
We determine $t_0/a^2$ in a standard way~\cite{Luscher:2010iy} and the values 
for all our ensembles are given in Appendix~\ref{app:data} in 
Table~\ref{tab:fit_range}. 
The flow time dependence of $R_P(t)$ in Fig.~\ref{fig:t_R_flow_dep} 
can be explained as follows: at short distances $r_f \lesssim 2 a$ 
the ratio $R_P$ is dominated by cutoff effects, while at large flow times, aside from the 
expansion coefficient we want to determine, $R_P$ contains contributions from higher dimensional
operators linear in $t$. For this reason we decided to perform a simple fit using the fit function
\be
R_{\textrm{fit}}(t) = B_{-1}\frac{t_0}{t} + B_0 + B_1 \frac{t}{t_0}\,,
\label{eq:R_fit}
\ee
where $B_{-1}$ parametrizes O($a^2$/t) effects, while $B_1$ parametrizes collectively 
effects from higher dimensional operators. A similar analysis has been done 
in Refs.~\cite{Kitazawa:2016dsl,Iritani:2018idk,Suzuki:2020zue} 
to analyze finite temperature quantities and renormalized 4-fermion operators.
The fit coefficient $B_0$ provides the value of $c_\chi$ in Eq.~\eqref{eq:c_chi} 
for each ensemble.
\begin{figure}
    \centering
    \includegraphics[width=0.49\textwidth]{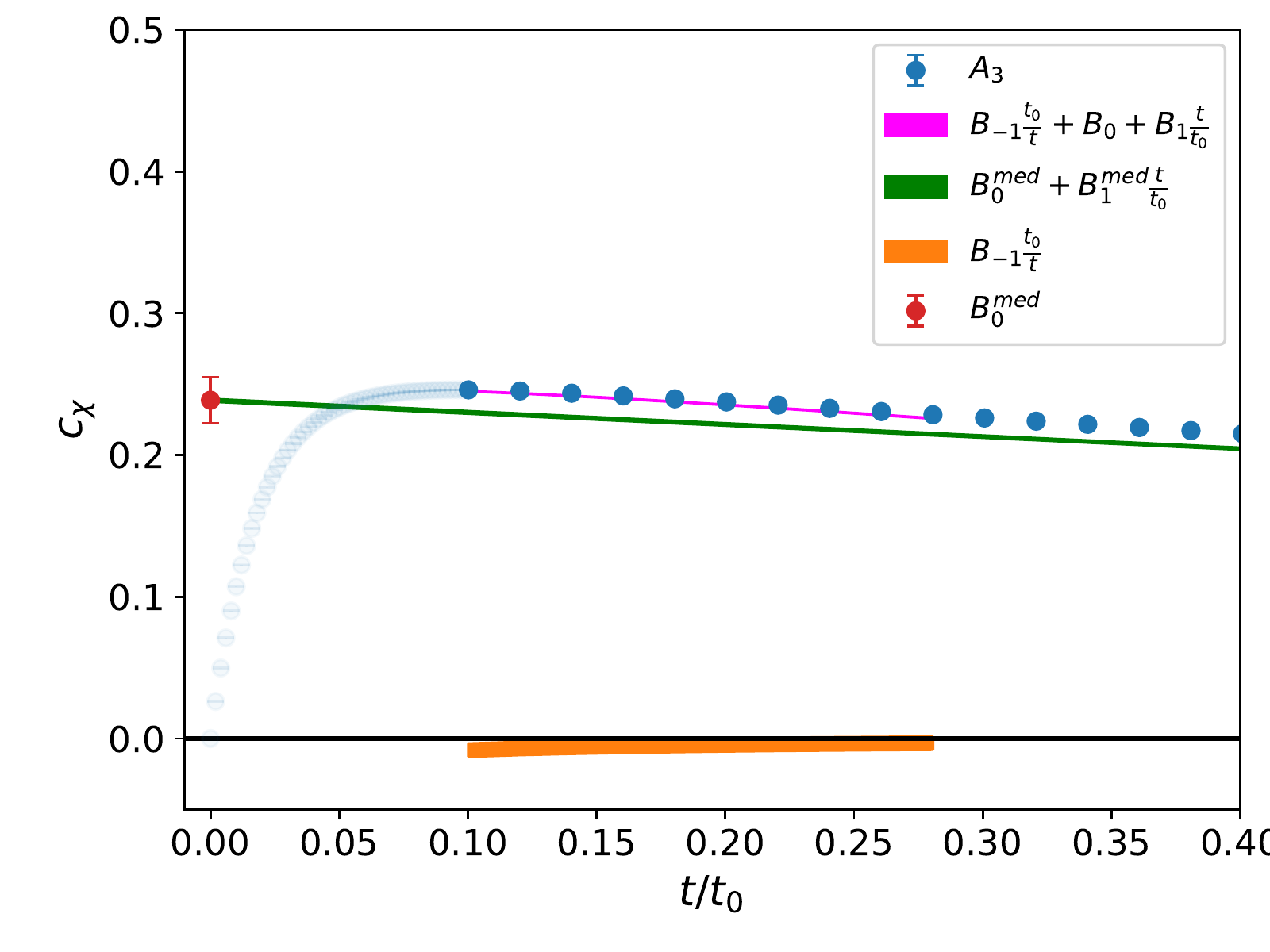}
    \includegraphics[width=0.49\textwidth]{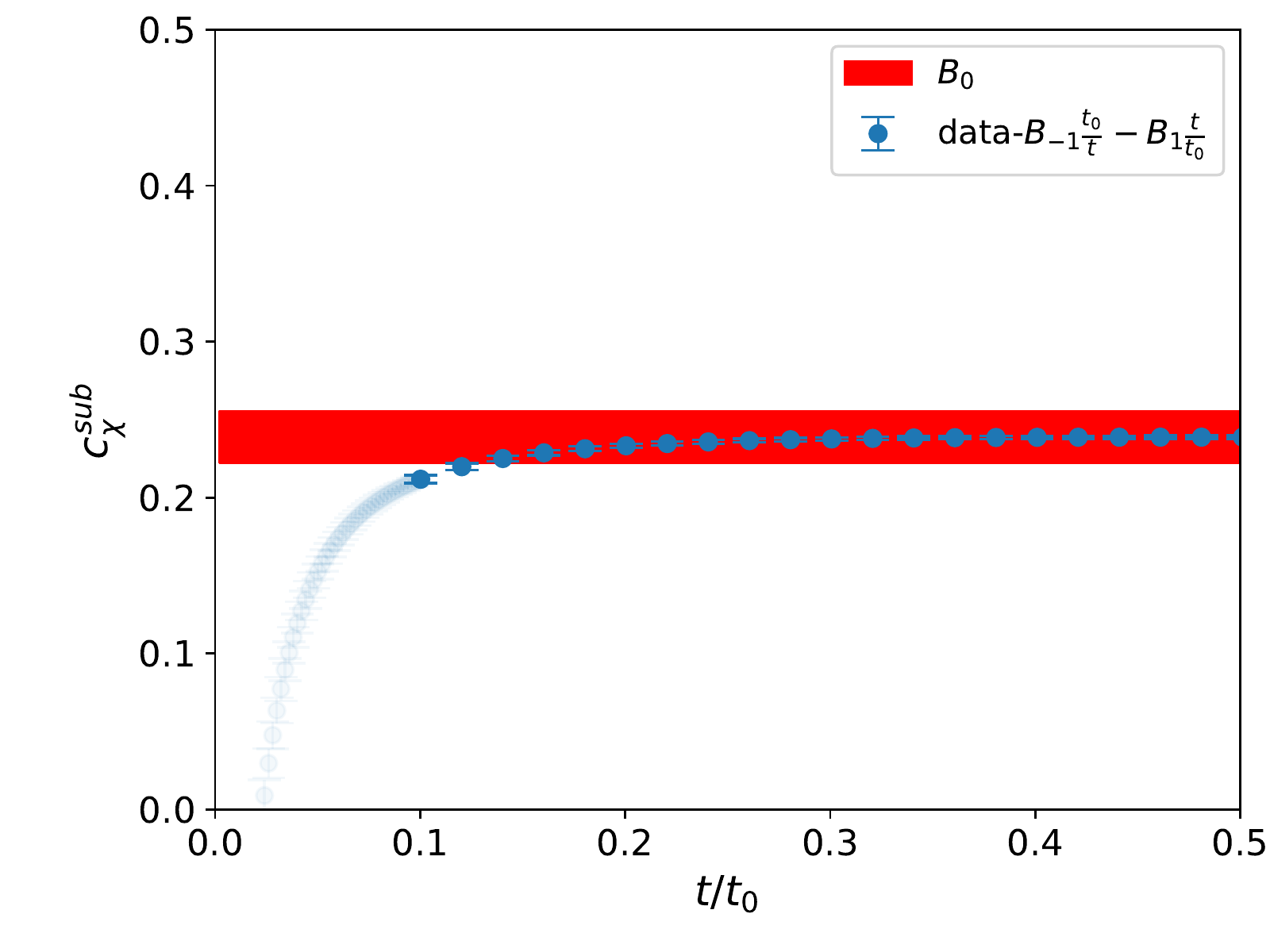}
		\caption{Left plot: flow-time dependence of $c_\chi$ for the ensemble $A_3$. 
        The different colored lines show the following contributions. 
        Magenta: example of a single fit satisfying the p-value condition. The magenta band represents the 
        statistical uncertainty for the particular fit range chosen.
        Orange: the contribution of the cutoff effect, parametrized by $B_{-1}$, 
        to the single fit chosen in the plot. 
        The orange band represents the associated statistical and
        systematic uncertainties stemming from the variation of the fit ranges. 
        Green: the fit result obtained removing cutoff effects. The central value this time represents the 
        median of the distribution obtained varying the fit ranges and satisfying our p-value condition 
        (see Appendix~\ref{app:data}). The width of the band represents the associated statistical uncertainty.
        Red data point: the central value is $B_0$ obtained as the median of the distribution of values of 
        $B_0$ varying the fit ranges and the error represents the sum in quadrature 
        of the statistical and systematic uncertainties on $B_0$.
		Right plot: the blue data points
		represent the raw data after subtracting the cutoff effects and the higher dimensional operator
        contributions, linear in $t/t_0$, determined from the fit.
		The red band represents our estimate of $c_\chi$ including the statistical and systematic 
        uncertainties added in quadrature.}
    \label{fig:subtraction}
\end{figure}

The efficacy of the method depends on the robustness of the 
determination of $B_0$ with respect to the other contributions.
To include all possible systematic effects in the determination of $B_0$, we 
scan many possible fit ranges in $t/t_0$ determining both the $\chi^2$ and the 
p-values.

All the details of the analysis are deferred to Appendix~\ref{app:data}.
The value of $B_0$, for each ensemble, is determined taking into account the statistical uncertainty
and the systematic one stemming from the choice of the fitting range in $t/t_0$ 
(see Appendix~\ref{app:data}).
In the right plot of Fig.~\ref{fig:t_R_flow_dep} we show the effective coefficient $c_\chi$
as a function of $t/t_0$ for all our ensembles together with the fit functions, where
with the thick symbols we show the data included in the fit. 
We note that the data are described well by the fit function~\eqref{eq:R_fit}. 
We also note that closer to the continuum limit we are able to describe the data
at smaller values of $t/t_0$. This is consistent with the expectation
that at smaller lattice spacing short distance effects are milder.

In Fig.~\ref{fig:subtraction} we show an example of the study of the flow time 
dependence of $c_{\chi}$. 
In the left plot with the magenta band we indicate a single fit using Eq.~\eqref{eq:R_fit}
with the associated statistical uncertainty.
With the orange band we show only the contribution from the cutoff effects, proportional to $B_{-1}$,
this time including the systematic uncertainties stemming from varying the fit ranges.
The green line is the fit result obtained removing cutoff effects.
The central value this time represents the 
median of the distribution obtained varying the fit ranges and satisfying our p-value condition 
(see Appendix~\ref{app:data}). The width of the band represents the associated statistical uncertainty.
The red data point has a central value representing $B_0$ obtained as the median 
of the distribution of values of $B_0$ varying the fit ranges 
and the error represents the sum in quadrature 
of the statistical and systematic uncertainties.
We observe that cutoff effects become important at $t/t_0 <0.1$ 
while the fit function describes the data 
over a large range of flow times, $0.1 < t/t_0 < 0.3$. 
The analysis described in Appendix~\ref{app:data}
also shows that more fit ranges are statistical acceptable and our final error budget
include the systematic error induced by varying the fit ranges.
A similar behavior is observed 
for the other ensembles with different fit ranges. 

In the right plot of Fig.~\ref{fig:subtraction} we show the data for $c_{\chi}$ after subtracting 
the contributions from cutoff effects and higher dimensional operators
\be
c_\chi^{\textrm{sub}}(t) = c_\chi(t) - B_{-1}\frac{t_0}{t} - B_1\frac{t}{t_0}\,.
\label{eq:R_sub}
\ee
The red band now represents our estimate of $c_\chi$ including systematic
uncertainties and clearly covers any possible ambiguity
coming from the choice of the fit range and it coincides 
with the red point in the left plot.
The systematic uncertainties we associate
to $B_0$ come from the different choices of fit ranges. 
Again details are given in Appendix~\ref{app:data}.
\begin{figure}
    \centering
		\includegraphics[width=0.55\textwidth]{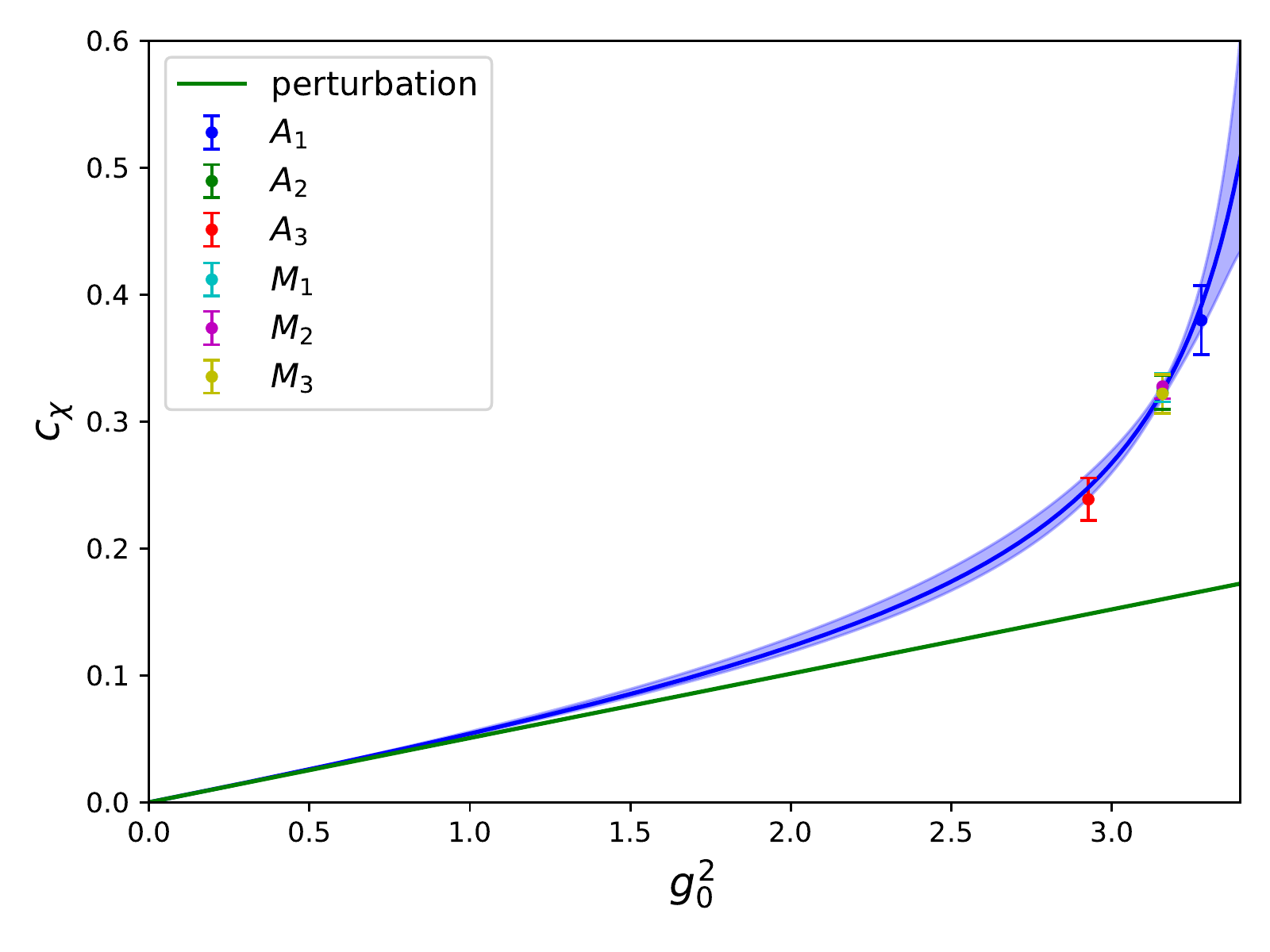}
		\caption{Nonperturbative dependence of $c_\chi$ as a function of the bare coupling $g_0$.
        The blue band represents the Pad\'e approximant in Eq.~\eqref{eq:pade}. The width of the band
        represents the statistical and systematic uncertainties added in quadrature.
        \label{fig:beta_dep}}
\end{figure}
While for the coefficient $B_0$ it is not possible to perform the continuum
limit because the value of $Z_\chi$ is not known, we can nevertheless
parametrize the dependence of $c_\chi$ on the bare coupling $g_0$ with a Pad\'e
approximant.

To make sure that the behavior at small coupling is reproduced, we have
calculated in perturbation theory the same ratio $R_P$ in Eq.~\eqref{eq:RP}.
The result of the calculation is described in Appendix~\ref{app:PT}.
At one loop in perturbation theory there is no contribution to the expansion coefficient
$\cCP$ coming from the renormalization of the flowed fermion field or the pseudoscalar density.
For completeness we quote here the result
\be
c_\chi = c_{\chi}^{(1)} g^2 + O(g^4)\,, \quad  c_{\chi}^{(1)}(t)=\frac{1}{2\pi^2}\,.
\ee
In Fig.~\ref{fig:beta_dep} the data points represent the  nonperturbative determination of $c_\chi$
with error bars including statistical and systematic uncertainties.
We have parametrized the dependence on the bare coupling of $c_\chi$ with a Pad\'e approximant
\be
c_\chi(g_0^2) = \frac{\frac{1}{2\pi^2} g_0^2+c_2 g_0^4}{1+c_4 g_0^2}\,,
\label{eq:pade}
\ee
where we have constrained the leading order in $g_0^2$ to be consistent with perturbation theory.
The green straight line in Fig.~\ref{fig:beta_dep} represents the perturbative result described
in Appendix~\ref{app:PT}, consistent with the result of Ref.~\cite{Rizik:2020naq}.
The blue curve represents the Pad\'e approximant obtained from fitting 
to data we obtained all our ensembles.
We note that in some ensembles, for example $M_3$,  
the data became bimodal in distribution and because of this we 
restricted our fitting to a smaller range in $t/t_0$.
Otherwise we use the full possible fitting range, when possible, consistent with 
the p-values chosen. See Appendix~\ref{app:data} for details.
Our final result is summarized by Eq.~\eqref{eq:pade} with the values
\be
c_2 =  -0.01115(63)\,,\qquad c_4 = -0.2690(61)\,.
\label{eq:fit_par}
\ee

The parametrization of Eq.~\eqref{eq:pade} with the fit parameters in Eq.~\eqref{eq:fit_par}
provides the coefficient of the 
power divergence for each value of the bare coupling for the particular choice 
of the lattice action in this paper.
With a different lattice action $c_\chi(g_0^2)$ changes, but this paper provides
a general method that can be adapted to any lattice action.

\section{O($a$) improvement at finite flow time}
\label{sec:Oaflow}

In Sec.~\ref{sec:Oaimpro} we concluded that to nonperturbatively remove O($a$) cutoff effects
in the flowed correlation functions we use, beside improving the action and the
local operators, we need to add nonstandard O($a$) terms like
$\wGamma_{PP}(x_4;t)$ and $\wGamma_{CP}(x_4;t)$ defined in
Eqs.~(\ref{eq:wC_PP}, \ref{eq:wC_OP}). 
The numerical determination of these types of correlation functions requires,
in addition to the calculation of flowed propagators, the calculation of ``kernel"
lines, where the Lagrange multipliers $\lambda$ and $\lambdabar$ are
contracted with flowed fermion fields. 
Details on how to determine ``kernel" lines are given
in Appendix~\ref{app:fermion_flow}.
\begin{figure}
    \centering
    \includegraphics[width=0.32\textwidth]{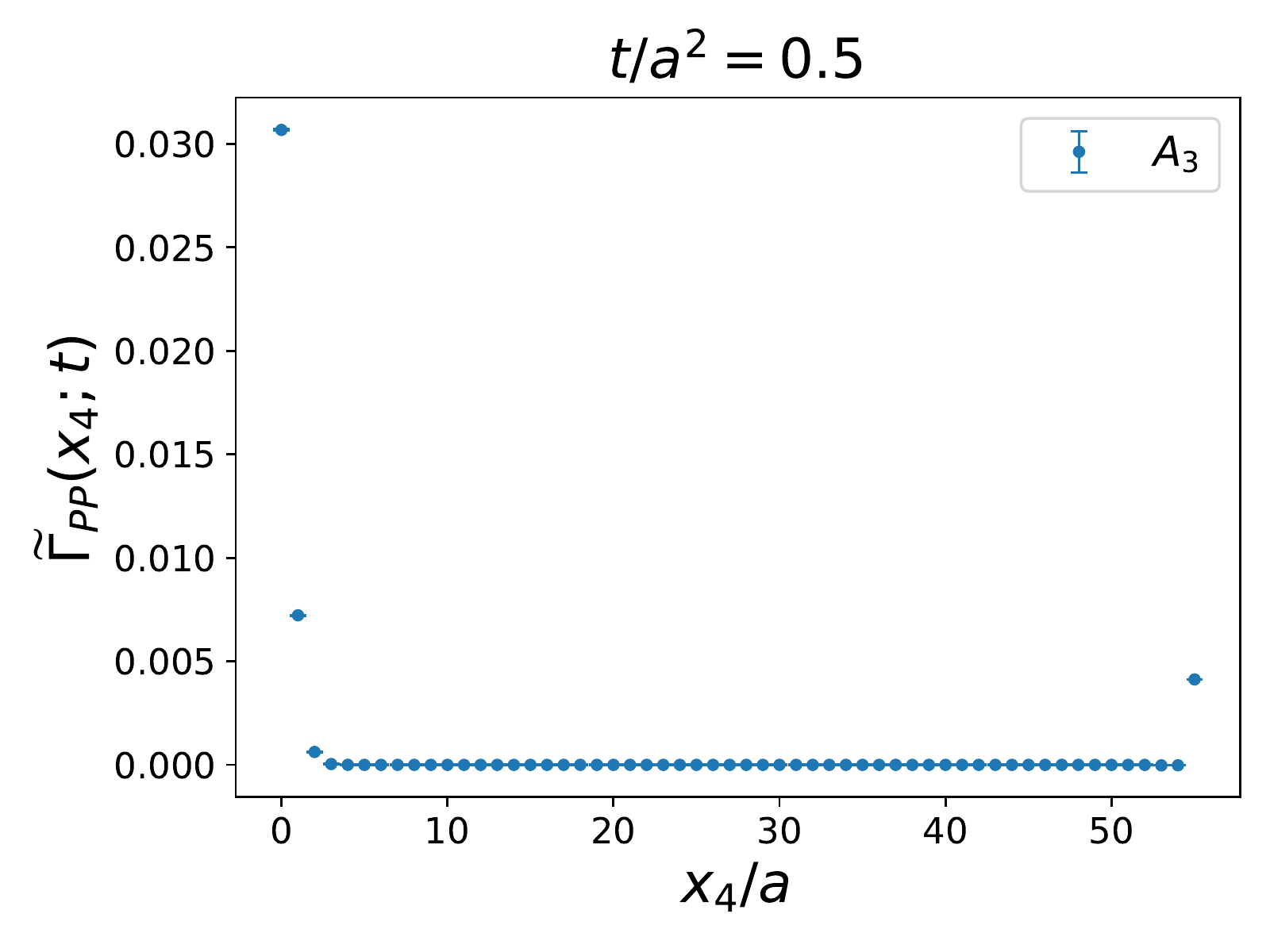}
    \includegraphics[width=0.32\textwidth]{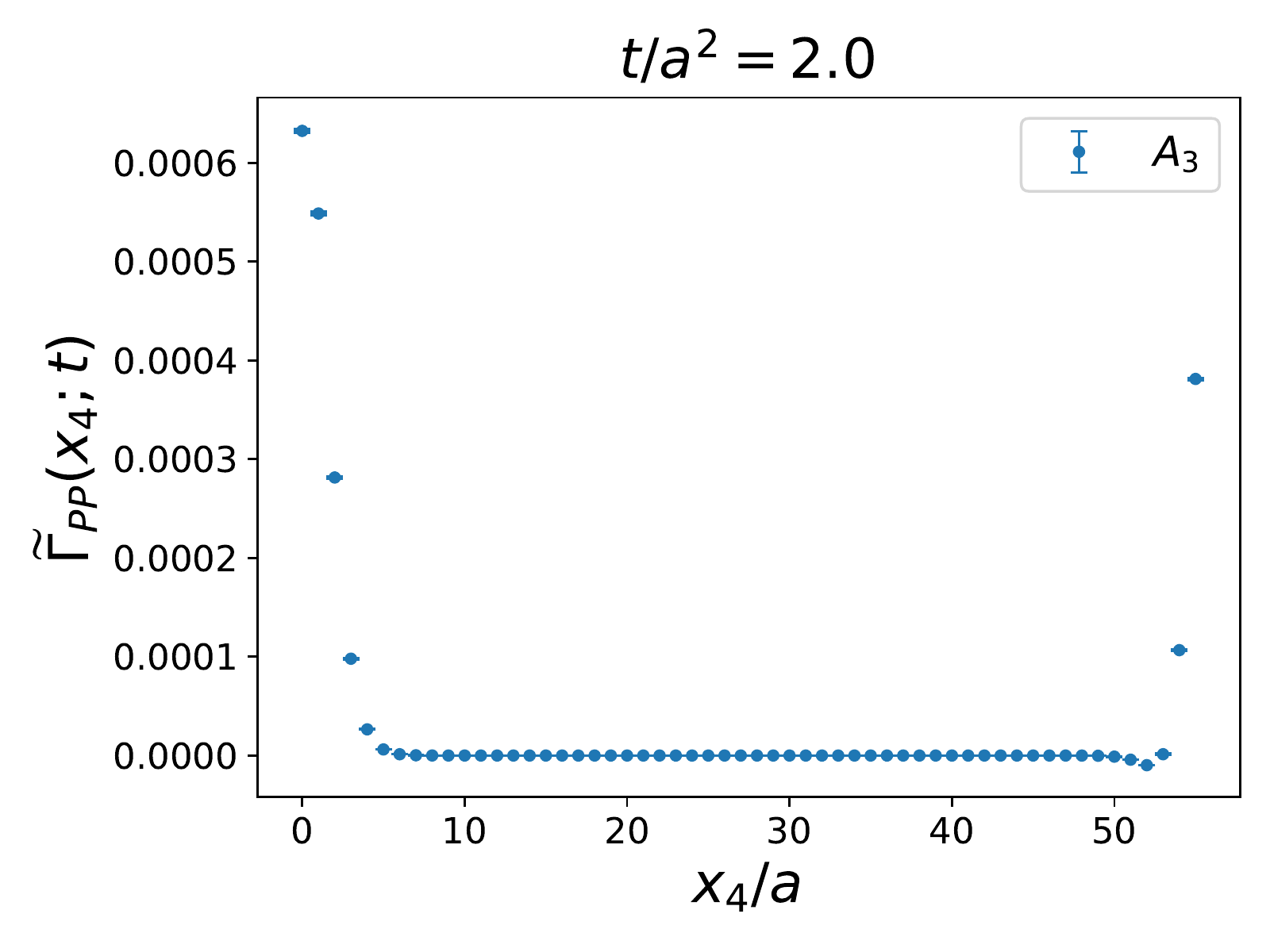}
    \includegraphics[width=0.32\textwidth]{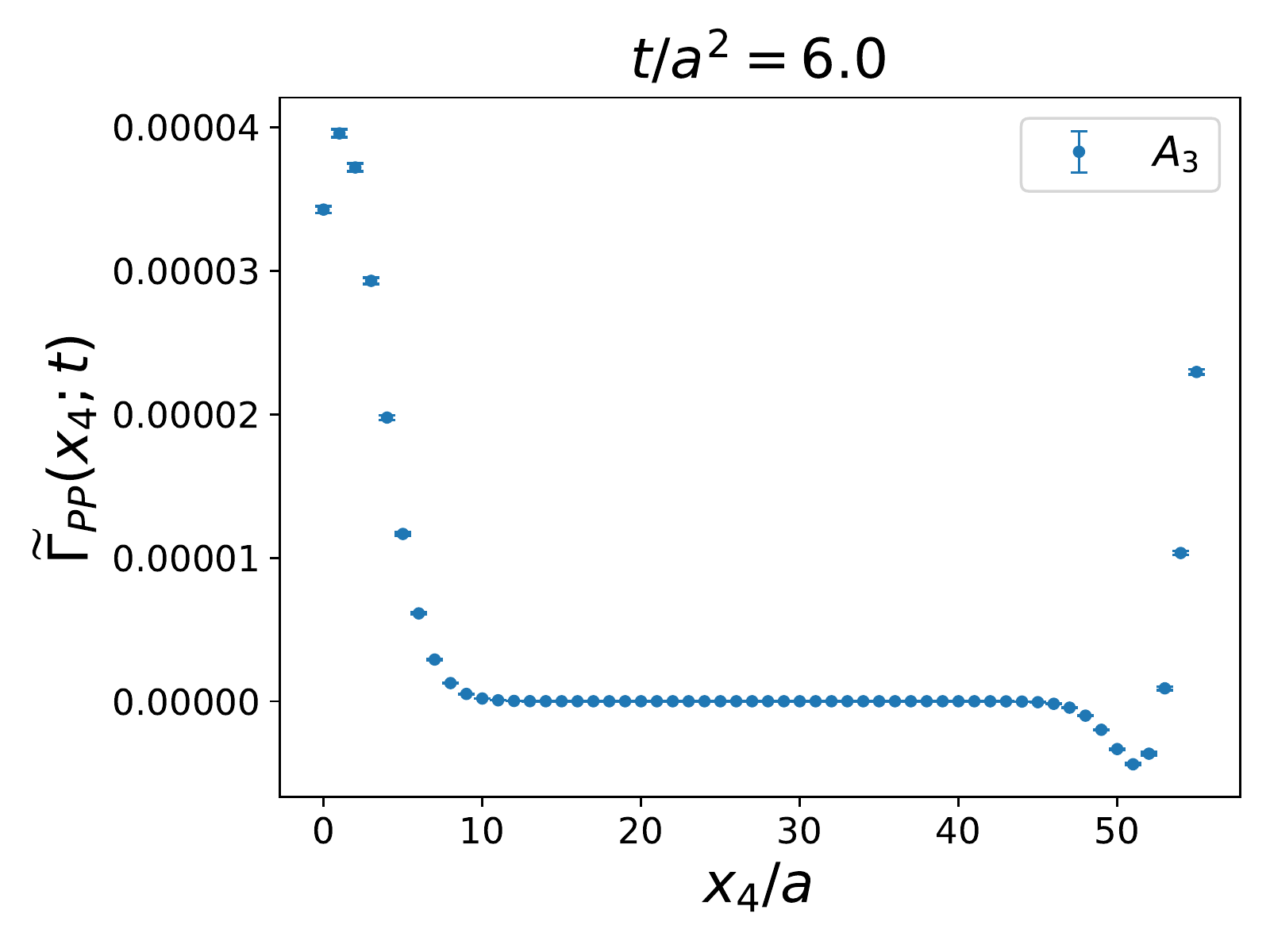}
    \caption{\label{fig:C_PP_tilde_A3}
Euclidean time, $x_4$, dependence of $\wGamma_{PP}(x_4;t)$ for the ensemble $A_3$ at $3$ values of the flow time.
	}
\end{figure}

\begin{figure}
    \centering
    \includegraphics[width=0.32\textwidth]{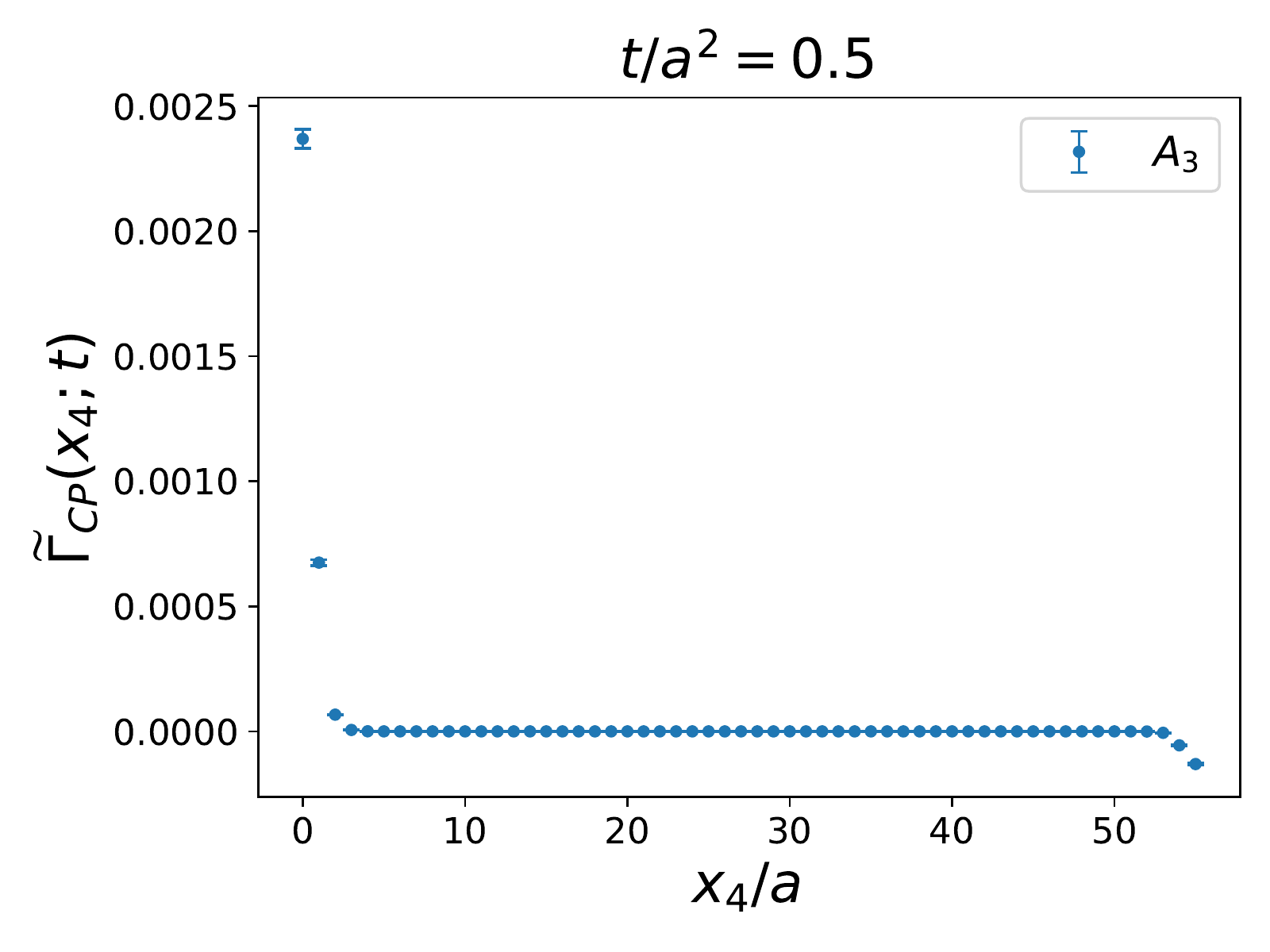}
    \includegraphics[width=0.32\textwidth]{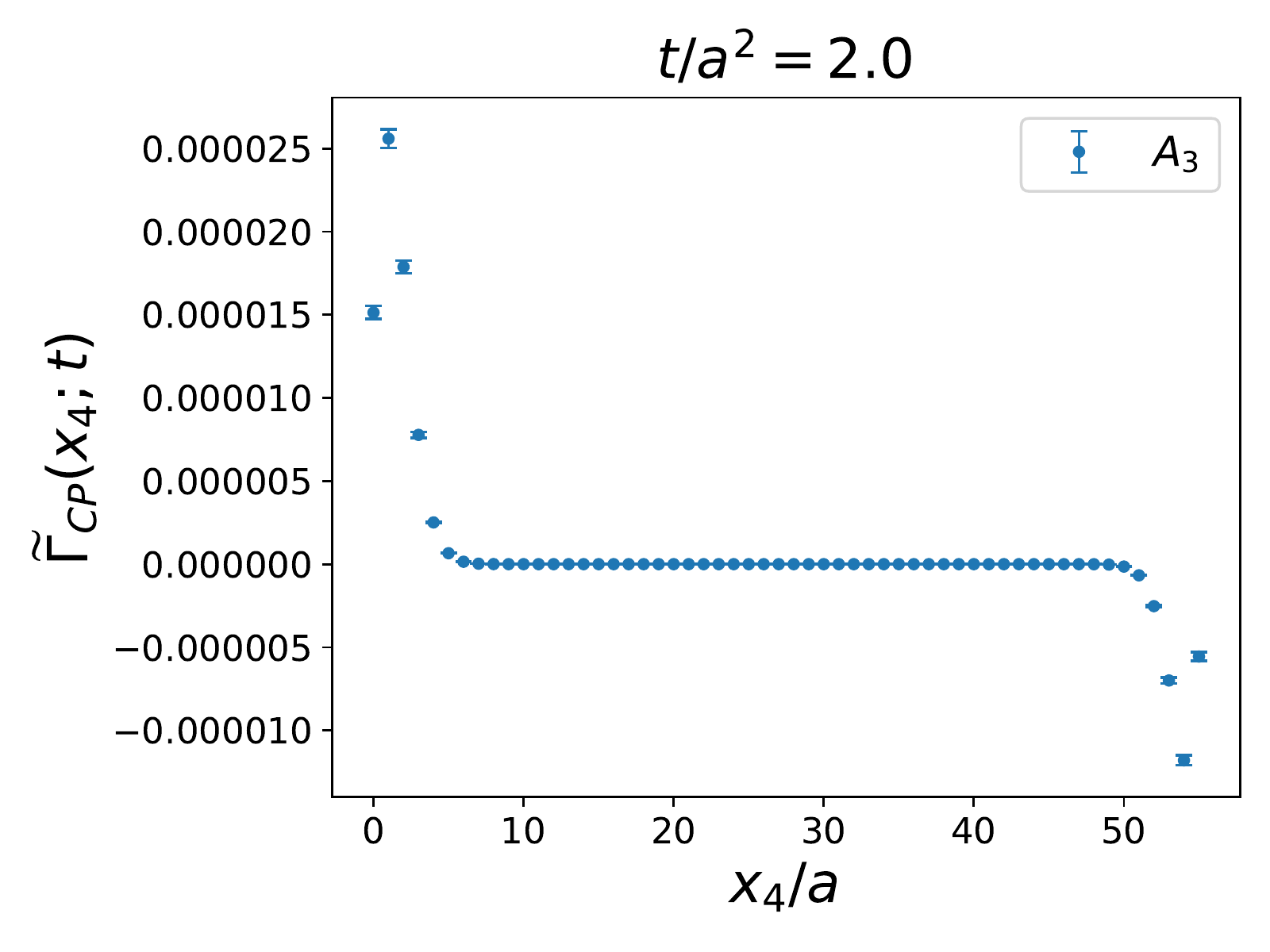}
    \includegraphics[width=0.32\textwidth]{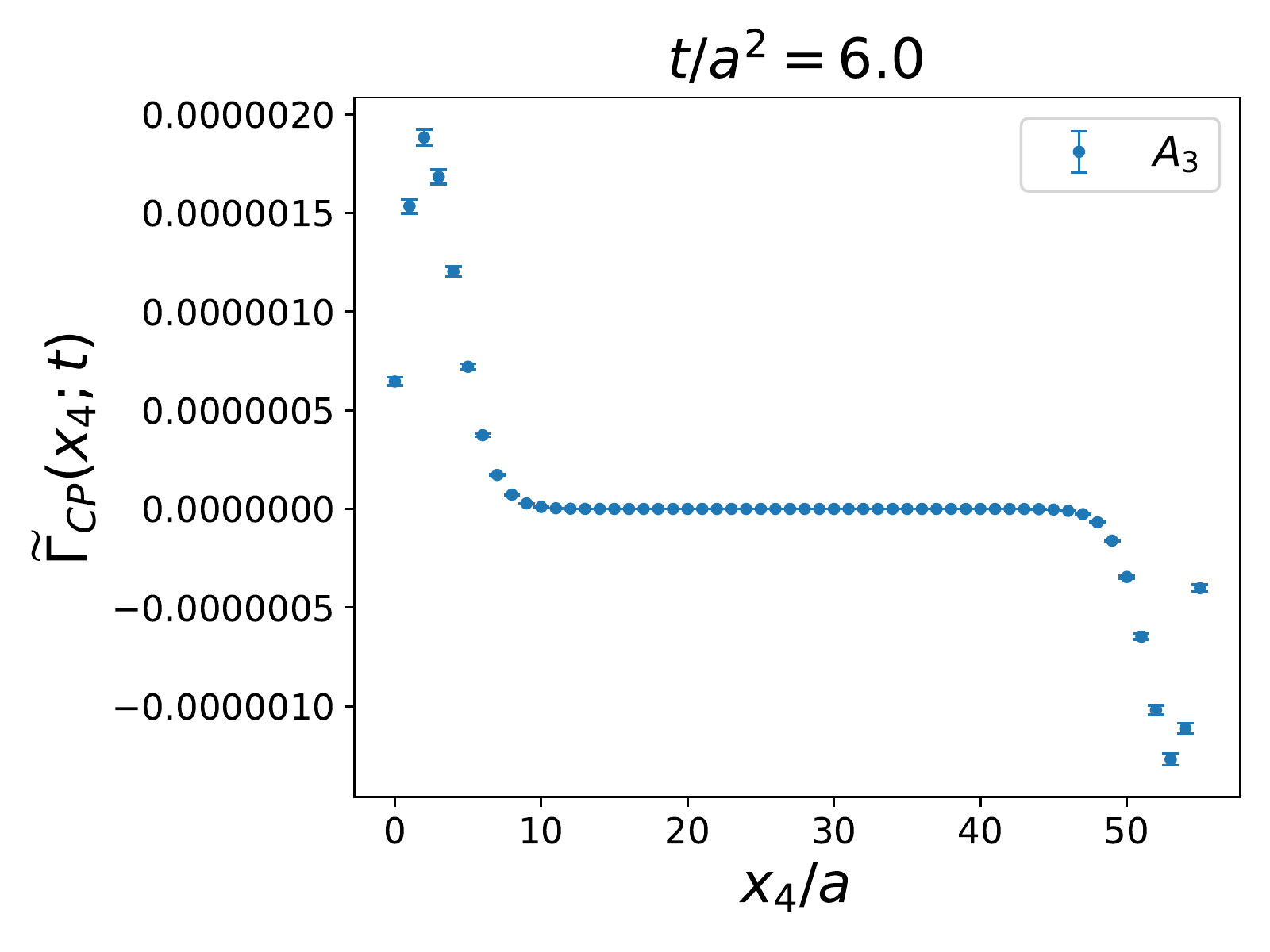}
    \caption{\label{fig:C_OP_tilde}
	Euclidean time, $x_4$, dependence of $\wGamma_{CP}(x_4;t)$ for the ensemble $A_3$ at $3$ values of the flow time.
	}
\end{figure}
In Figs.~(\ref{fig:C_PP_tilde_A3}, \ref{fig:C_OP_tilde}) we show the Euclidean time dependence, $x_4$, 
respectively of $\wGamma_{PP}(x_4;t)$ and $\wGamma_{CP}(x_4;t)$ for the ensemble 
$A_3$ and $3$ values of the flow time $t/a^2=0.5, 2.0, 6.0$.
It is clear that any nonzero contribution is localized in the region $x_4 \lesssim \sqrt{8t}$
while for $x_4 \gg \sqrt{8t}$ the correlation functions vanish.
We conclude that as far as $x_4 \gg \sqrt{8t}$ our determination 
is nonperturbatively O($a$) improved up to small O($amg^2$) terms. 
This result, obtained with minimal numerical effort,
is one of the great advantages of using the gradient flow to renormalize higher
dimensional operators.

\section{Summary and Outlook }
\label{sec:summary} 

When a $CP$-violating signal is measured in electric dipole moment experiments in the future, 
it will be imperative that theory provides guidance on the origins of this measured violation.   
As the sources themselves can come from various BSM scenarios, 
it is important to understand the systematics of each source 
and its ensuing impact on observables.

To that end we have analyzed the quark chromo-EDM operator, 
for the first time using the gradient flow method to provide control 
on the power divergences that occur due to mixing 
during renormalization when using a discrete spacetime regulator.
In essence, our gradient flow analysis trades induced power divergences 
with cutoff dependencies, the latter being much more amenable to a continuum limit extrapolation.  

Our most important result is shown in Fig.~\ref{fig:Rbar_poly}
and the corresponding description in Eqs.~\eqref{eq:Rbar_poly}.
In the plots we show the nonperturbative determination of the finite renormalization
connecting, for a wide range of renormalized coupling values, 
the qCEDM operator with the pseudoscalar
density at finite flow time. The calculation of this finite renormalization
reduces the power divergence problem 
to the determination of the nonperturbative evolution 
of the pseudoscalar density at finite flow time. 
Once a nonperturbative determination of the expansion coefficient of the pseudoscalar
density is available, it is possible to determine, nonperturbatively and in the continuum,
not only of the leading power divergences but also the subleading logarithmic corrections
to the power divergences. In view of the increased precision of lattice data
it becomes critical to control also this subleading corrections 
and work in this direction is in progress.

The scheme defined in this work, even if technically difficult, can also be used in 
perturbation theory allowing a matching at high-energy and in Appendix \ref{app:PT} we show 
the detail of the calculation.
For completeness we also perform an analysis in terms of the bare coupling, where
the dependence of the power divergence coefficient is reconstructed 
using a Pad\'e approximant. We emphasize that the analysis in terms of the bare coupling,
and the corresponding Pad\'e approximant, depends on the lattice action used. 

We have also discussed the O($a$) improvement of the qCEDM and the appropriate 
modifications at finite flow time. We conclude that the improvement is greatly simplified
using the gradient flow definition of the qCEDM.

We consider this work as a first nonperturbative solution of the problem of the power divergences
for the qCEDM. The method we develop in this paper can be adopted for any local 
operator mixing with lower dimensional operators and with any lattice action. 
\begin{acknowledgments}
We thank the members of the SymLat collaboration, Jack Dragos, Chirstopher J. Monahan,
Giovanni Pederiva, and Jordy de Vries for very useful discussions and 
a most enjoyable collaboration. M.D.R. and A.S. acknowledge very useful and informative discussions
with A. Hasenfratz, C. J. Monahan, and O. Witzel on the gradient flow as a renormalization 
tool for local operators
and J. de Vries, E. Mereghetti, and P. Stoffer on the renormalization of the qCEDM operator. 
J. K. acknowledges support by the Deutsche Forschungsgemeinschaft (DFG, German Research Foundation) 
through the CRC-TR 211 ``Strong-interaction matter under extreme conditions''– project number 315477589 – TRR 211.
T. L. was partly supported by the Deutsche Forschungsgemeinschaft (DFG, German Research
Foundation) through the funds provided to the Sino-German Collaborative
Research Center TRR110 “Symmetries and the Emergence of Structure in QCD”
 (DFG Project-ID 196253076 - TRR 110).
M.D.R. and A.S. acknowledge funding support under the National 
Science Foundation grant PHY-1913287.
We would like to thank the Center for Scientific Computing, University of
Frankfurt for making their High Performance Computing facilities available. 
The authors also gratefully acknowledge the computing time granted through JARA-HPC 
on the supercomputer JURECA \cite{jureca} at Forschungszentrum J\"ulich.
This work used the Extreme Science and Engineering Discovery Environment 
(XSEDE)~\cite{Towns:2014qtb} 
Stampede2 system at the Texas Advanced Computing Center (TACC) 
through allocation TG-PHY200015.

\end{acknowledgments}

\appendix
\section{FLOWING FERMION FIELDS}
\label{app:fermion_flow}

Here we elaborate on the fermion and Lagrange multiplier fields we use in our work.  
The standard lattice QCD quark propagator
\be
\left[ \psi(x) \psibar(y) \right]_F = S(x,y)\,,
\ee
where $\left[ \cdot \right]_F $ denotes a fermionic field contractions, represents the usual inverse of
the lattice QCD action with a given source $\eta(x)$. 
The propagators are flowed as described, for example, in~\cite{Luscher:2010iy}:
\be
\left[\chi(x;t) \psibar(y)\right]_F = a^4 \sum_v K(x,t;v,0) S(v,y)\,,
\ee
\be
\left[\psi(x)\chibar(y;s)\right]_F = a^4 \sum_v  S(x,v) K(y,s;v,0)^\dagger\,,
\ee
where the kernel $K$ is the solution of the equation
\bea
&& \left(\partial_t - D^x_\mu D^x_\mu\right) K(x,t;y,s) =0\,,\nonumber \\
&& K(x,y;t,t) = \frac{1}{a^4}\delta_{xy}\,.
\eea
The flowed propagators are used to determine the correlators 
in Eqs.~(\ref{eq:COP}, \ref{eq:CPPt}).

In order to compute the correlation functions in Eqs.~(\ref{eq:wC_OP}, \ref{eq:wC_PP}),
parametrizing specific O($a$) terms for flowed correlation functions, 
we need to flow the Lagrange fields $\lambda(x;t)$ in the following manner,
\be
\left[\chi(x;t)\lambdabar(y;0)\right]_F = a^4 \sum_v  \delta_{xv} K(v,t;y,\epsilon)\,,
\ee
\be
\left[\lambda(x;0) \chibar(y;s)\right]_F = a^4 \sum_v K(v,s;x,\epsilon)^\dagger \delta_{vy}\,,
\ee
where $\delta_{xy}$ is a Kronecker delta in spacetime, color, and spin indices. 
The numerical calculation proceeds similarly as in the case of flowed propagators,
with a point source as initial condition of the flow equation.

As an example we show explicitly the contractions 
for the correlation function in Eq.~\eqref{eq:wC_OP} 
\bea
\wGamma_{CP}(x_4;t) &=& a^3\sum_{\bx}\left\langle\Oc^{ij}(x_4,\bx;t) \wP^{ji}(0,\bzero;0)\right\rangle = \nonumber \\
&=& a^3\sum_{\bx} \left\langle\bigg[ \chibar(x_4,\bx;t) \gamma_{5}\sigma_{\mu\nu}G_{\mu\nu}(x_4,\bx;t) \chi(x_0,\bx;t) \right. \nonumber \\
& \times & \left.\big(\lambdabar(0,\bzero;0) \gamma_{5} \psi(0,\bzero;0) + \psibar(0,\bzero;0) \gamma_{5} \lambda(0,\bzero;0)\big) \bigg]\right\rangle  \nonumber\\
&=& - a^3\sum_{\bx}\Tr \bigg[ [\psi(0,\bzero;0) \chibar(x_4,\bx;t)]_F \gamma_{5}\sigma_{\mu\nu}G_{\mu\nu}(x_4,\bx;t) [\chi(x_4,\bx;t) \lambdabar(0,\bzero;0)]_F \gamma_{5} \bigg] \nonumber \\
&-& \Tr \bigg[ [\lambda(0,\bzero;0) \chibar(x_4,\bx;t)]_F \gamma_{5}\sigma_{\mu\nu}G_{\mu\nu}(x_4,\bx;t) [\chi(x_4,\bx;t) \psibar(0,\bzero;0)]_F \gamma_{5} \bigg]\,,
\eea
where with $[ \cdot]_F$ we denote a fermionic contraction.
\section{DETAILS ON DATA ANALYSIS}
\label{app:data}

In this Appendix we discuss in more details the analysis presented in Sec.~\ref{sec:analysis2}.
The data in Fig.~\ref{fig:t_R_flow_dep} and the fit functional form 
in Eq.~\eqref{eq:R_fit} suggest to restrict the fit ranges using $t/t_0$ larger than 
the value corresponding to the maximum value of $R_P/t$. We then fit 
in all ranges between $(t/t_0)_{\textrm{min}}$ and $(t/t_0)_{\textrm{max}}$
given in Table~\ref{tab:fit_range}, keeping always a minimum number of $5$ data points.
\begin{table}
	\caption{\label{tab:fit_range} $t_0/a^2$ and fit ranges for each ensemble. 
	In the 3rd and 4th column
	we show the complete fit ranges, $(t/t_0)_{\text{min/max}}$, 
	while in the 5th and 6th 
	we show the fit ranges after analyzing the p-values of the fits (see main text in this Appendix).}
	\centering\begin{tabular}{|c|c|c|c|c|c|}
		\hline
		Designation & $t_0/a^2$ & range of $(t/t_0)_\text{min}$ & range of $(t/t_0)_\text{max}$ & range of $(t/t_0)^{\text{sel}}_\text{min}$ & range of $(t/t_0)^{\text{sel}}_\text{max}$  \\
		\hline
		\hline
		$M_1$       & 2.2586(12) & (0.1682, 0.7525) & (0.1903, 0.9738) & (0.1682, 0.2656) & (0.1903, 0.4869) \\
		\hline
		$M_2$       & 2.3993(12) & (0.1624, 0.7500) & (0.1833, 0.9583) & (0.1624, 0.2500) & (0.1833, 0.4167) \\
		\hline
		$M_3$       & 2.5371(15) & (0.1576, 0.7881) & (0.1773, 0.9851) & (0.1576, 0.2758) & (0.1773, 0.5123) \\
		\hline\hline 
		$A_1$       & 1.3627(15) & (0.2348, 0.5870) & (0.2715, 0.9539) & (0.2348, 0.5870) & (0.2715, 0.9539)\\
		\hline
		$A_2$       & 2.2378(24) & (0.1697, 0.7594) & (0.1921, 0.9827) & (0.1697, 0.2680) & (0.1921, 0.4914) \\
		\hline
		$A_3$       & 4.9879(65) & (0.1002, 0.8820) & (0.2005, 0.9822) & (0.1002, 0.8820) &  (0.2005, 0.9822)\\
		\hline
	\end{tabular}
\end{table}

\begin{figure}
    \centering
    \subfigure[$A_3$ ensemble]{\includegraphics[width=0.49\textwidth]{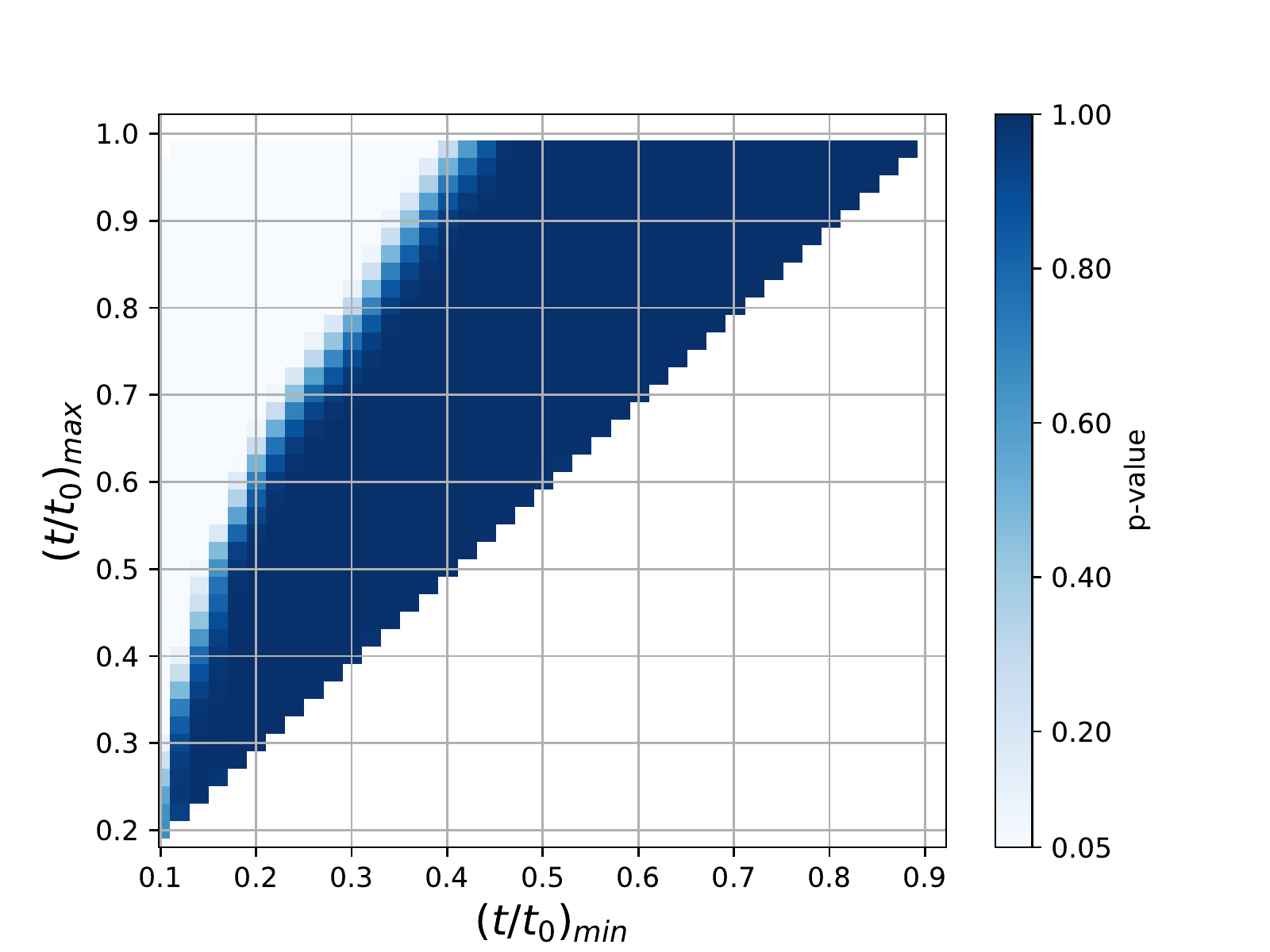}}
    \subfigure[$M_3$ ensemble]{\includegraphics[width=0.49\textwidth]{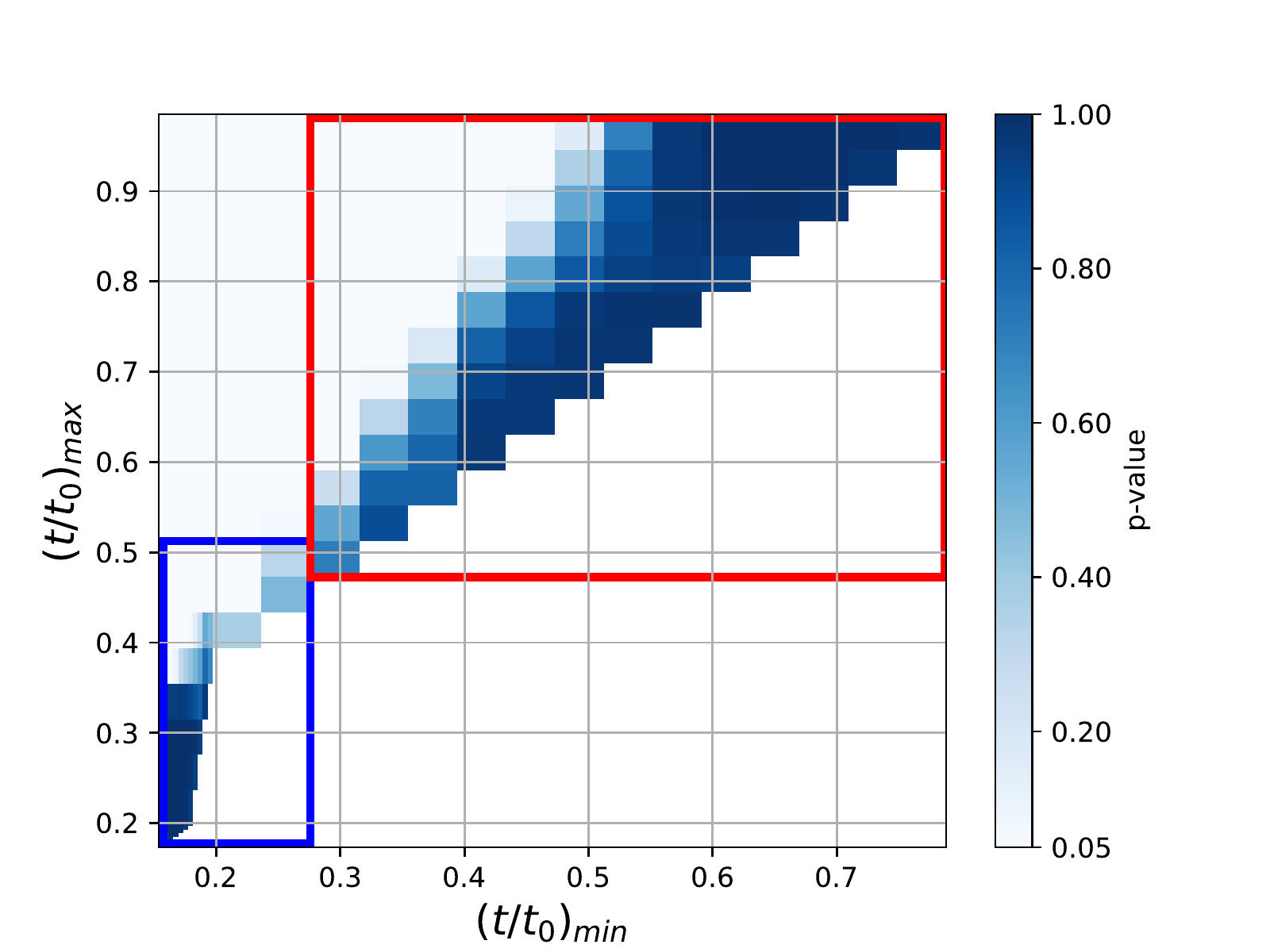}}
		\caption{p-values obtained from the $\chi^2/\text{d.o.f}$. 
		The darker regions correspond to the acceptable fit ranges.
		In the right plot, the blue and red rectangles correspond to blue and red peak 
		in Fig.~\ref{fig:histogram_c}. See main text in this Appendix.}
    \label{fig:p-value}
\end{figure}

\begin{figure}
    \centering
    \subfigure[$A_3$ ensemble]{\includegraphics[width=0.49\textwidth]{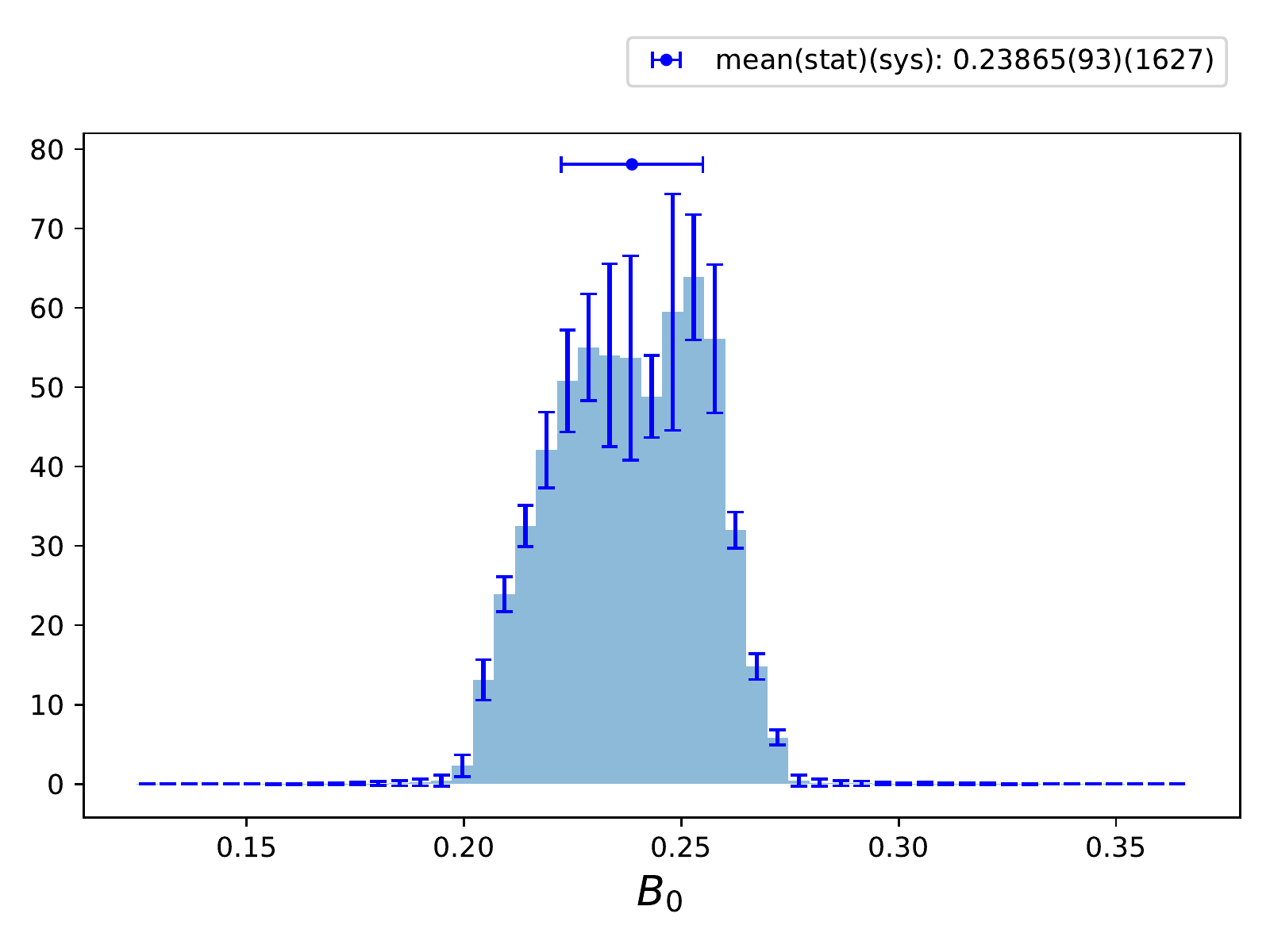}}
    \subfigure[$M_3$ ensemble]{\includegraphics[width=0.49\textwidth]{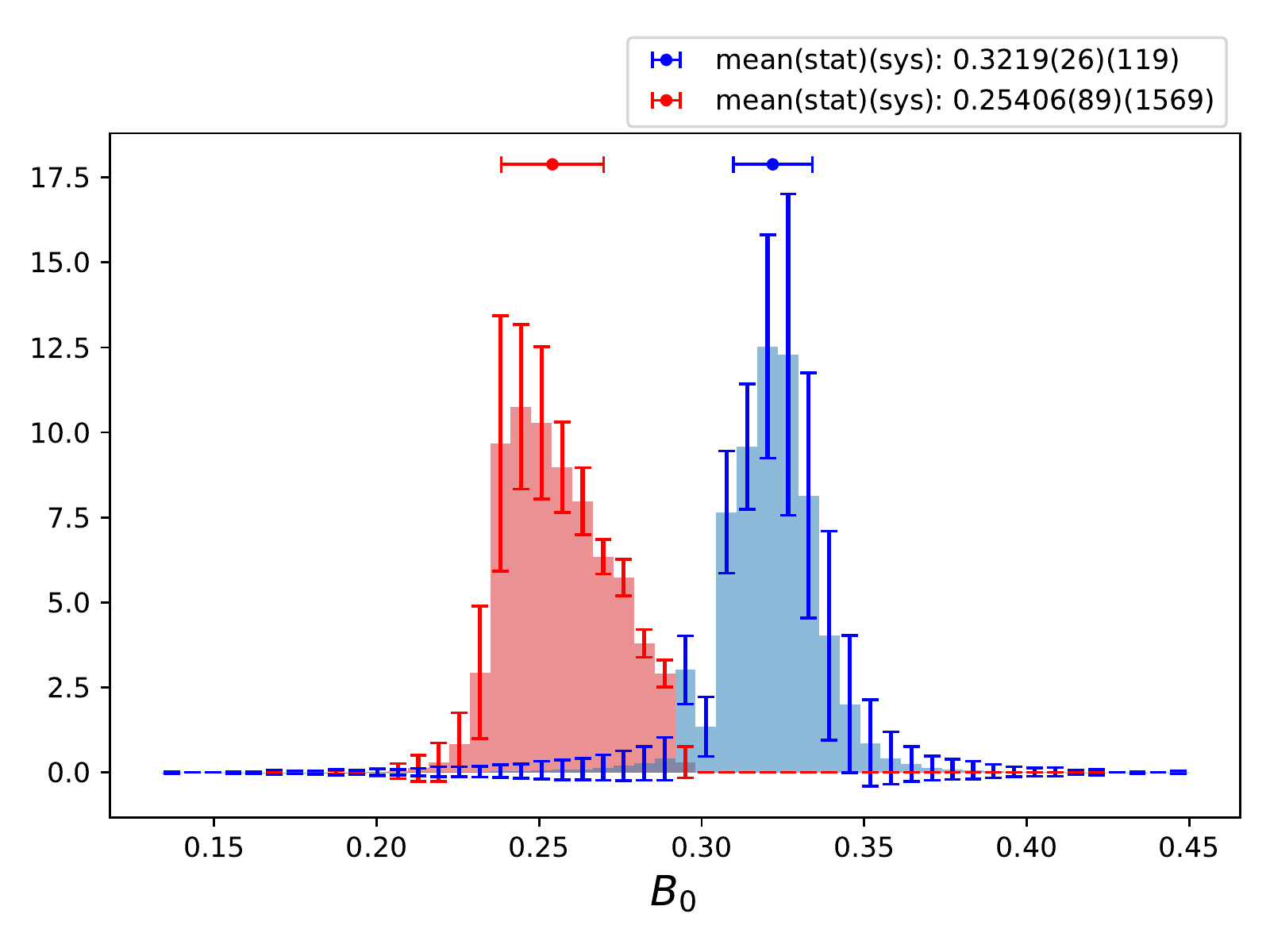}}
    \caption{Distribution of the fit parameter $B_0$ for all the fit ranges satisfying $p > 0.05$.}
		\label{fig:histogram_c}
\end{figure}

To select the acceptable fit ranges and the corresponding values 
of the fit parameters we scan the $\chi^2$ and the p-values for each fit,
using central values and statistical errors obtained by a standard bootstrap analysis 
of the raw data.
The analysis shows $2$ typical behaviors for the fit parameters, that can be 
exemplified plotting a heat map of the p-values for different fit ranges
in the $2$ representative ensembles $A_3$ and $M_3$ (see Fig.~\ref{fig:p-value}).
While for $A_3$, the lattice spacing closer to the continuum limit, we observe 
a stable values for the fit parameters for a wide choice of fit ranges,
for the ensemble $M_3$ we observe $2$ well separated regions where 
the null hypothesis is not rejected: a fit range region
at small flow time (blue box) and one at larger flow time (red box).

To analyze the distributions of the fit parameters obtained,
we first generate $N_B=1000$ bootstrap samples of the raw data
and perform $N_B$ fits for each fit range selected by the p-value condition.
In this way we have the values of $B_{0,-1,1}$ with the corresponding bootstrap
uncertainties for each fit range. 
We then take each value of the fit parameters and plot
a histogram with the distribution obtained changing the fit ranges.
This is equivalent to generating $N_B=1000$ slightly different histograms,
showing the distribution of the fit parameters with varying fit ranges.
We then take each bin of each histogram and perform a standard bootstrap analysis obtaining
for each bin a central value with a statistical fluctuation.
These histograms are shown in Figs.~\ref{fig:histogram_c} for the ensembles 
$A_3$ (left plot) and $M_3$ (right plot).
They are a summary of the 
statistical and systematic uncertainties
giving a visual representation
of the systematic uncertainty stemming from the choice 
of the fit ranges, and for each bin the statistical uncertainty.

We then proceed to determine the median of each 
histogram obtained for each bootstrap sample. 
This gives us $N_B$ bootstrap values for the median: 
a standard statistical analysis gives us the central value
and the associate statistical error. 
To determine the systematic uncertainty we take the median 
of the summary histograms, like the ones in Figs.~\ref{fig:histogram_c},
and determine symmetrically around the median the region with $68\%$ of the area 
of the normalized distribution.
The statistical and systematic errors are then summed in quadrature
and are shown on top of the histograms in Fig.~\ref{fig:histogram_c}. 

We notice that in the right plot of Fig.~\ref{fig:histogram_c} 
we have clearly a two-peak structure.
A deeper investigation of the origin of the $2$ peaks is revealed
when we separate $2$ very distinct regions in the flow time dependence 
of $c_\chi$. This is transparent when we analyze the heat map in the right plot
of Fig.~\ref{fig:p-value}. We clearly notice that the p-value prefers 
$2$ very distinct regions in the fit ranges. It turns out that the 
$2$ distinct regions clearly correspond to the $2$ peaks in 
the right plot of Fig.~\ref{fig:histogram_c}.
For the ensemble $M_3$, right plots, 
we draw in blue and red the values of the $B_0$ obtained in 
the region isolated with the heat maps, i.e. red for fit ranges at 
``large" values of $t/t_0$ and blue for  
fit ranges at ``small" values of $t/t_0$.
Instead for the ensemble $A_3$ we observe that all fit ranges, satisfying the p-value condition,
give values of $B_0$ all within a seemingly well defined distribution (see left plot
of Fig.~\ref{fig:histogram_c}).

The values of $c_\chi$ we plot in Fig.~\ref{fig:beta_dep} correspond to the
fit ranges selected by the p-values condition, and the selection of
small flow time fit ranges explained above.
For the ensembles $A_1$ and $A_3$ we take every fit range selected by the p-value condition.
While choosing the same fit ranges for all the ensembles
will give us a smaller total uncertainties in the fit parameters,
we consider for the ensemble $A_1$
and $A_3$, that do not show a double peak structure, 
the larger range of fit intervals providing us with a more conservative estimate
of the total uncertainty.

\begin{figure}
    \centering
		\subfigure[Fit parameter $B_{-1}$]{\includegraphics[width=0.49\textwidth]{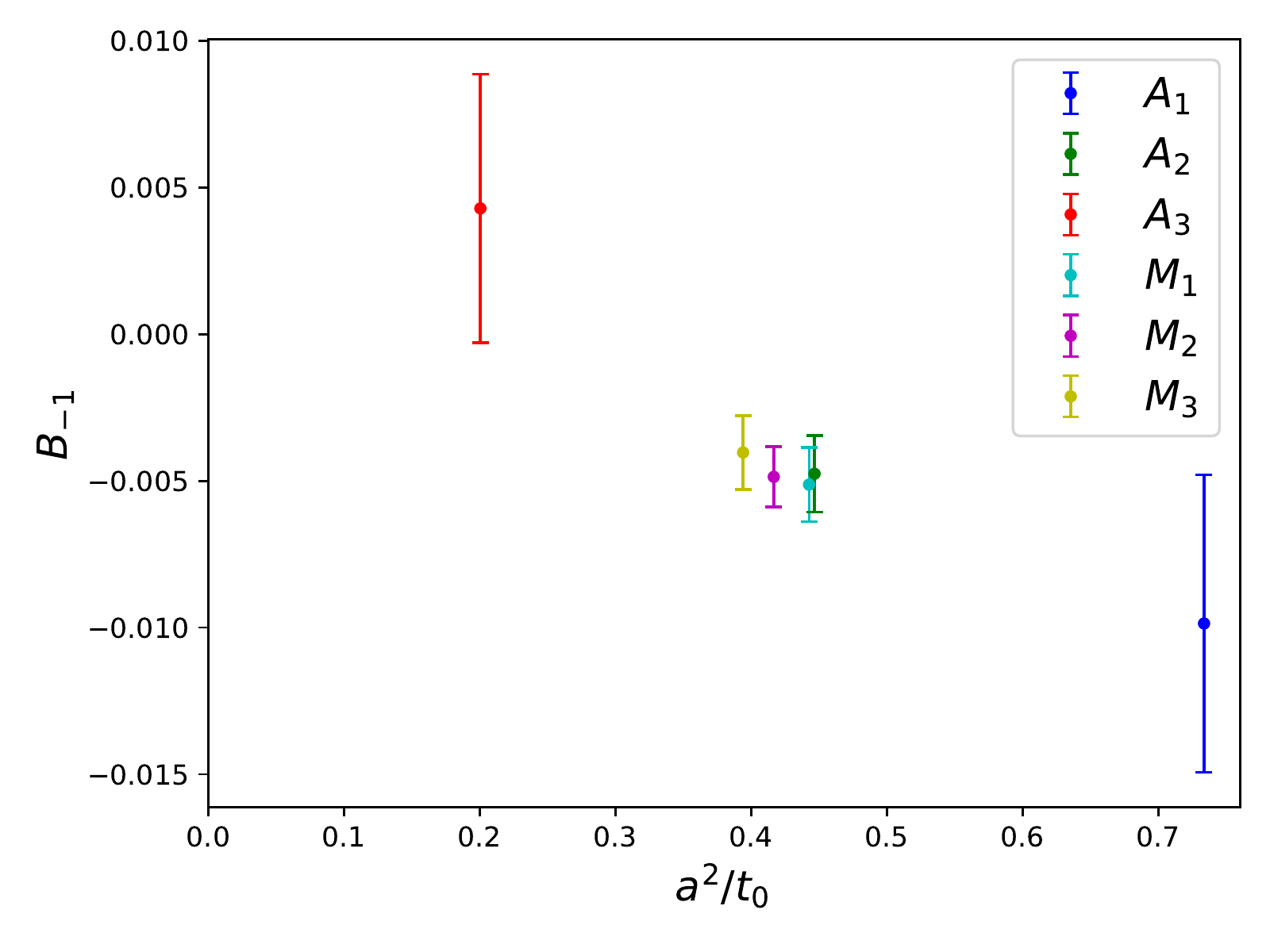}}
		\subfigure[Fit parameter $B_{1}$]{\includegraphics[width=0.49\textwidth]{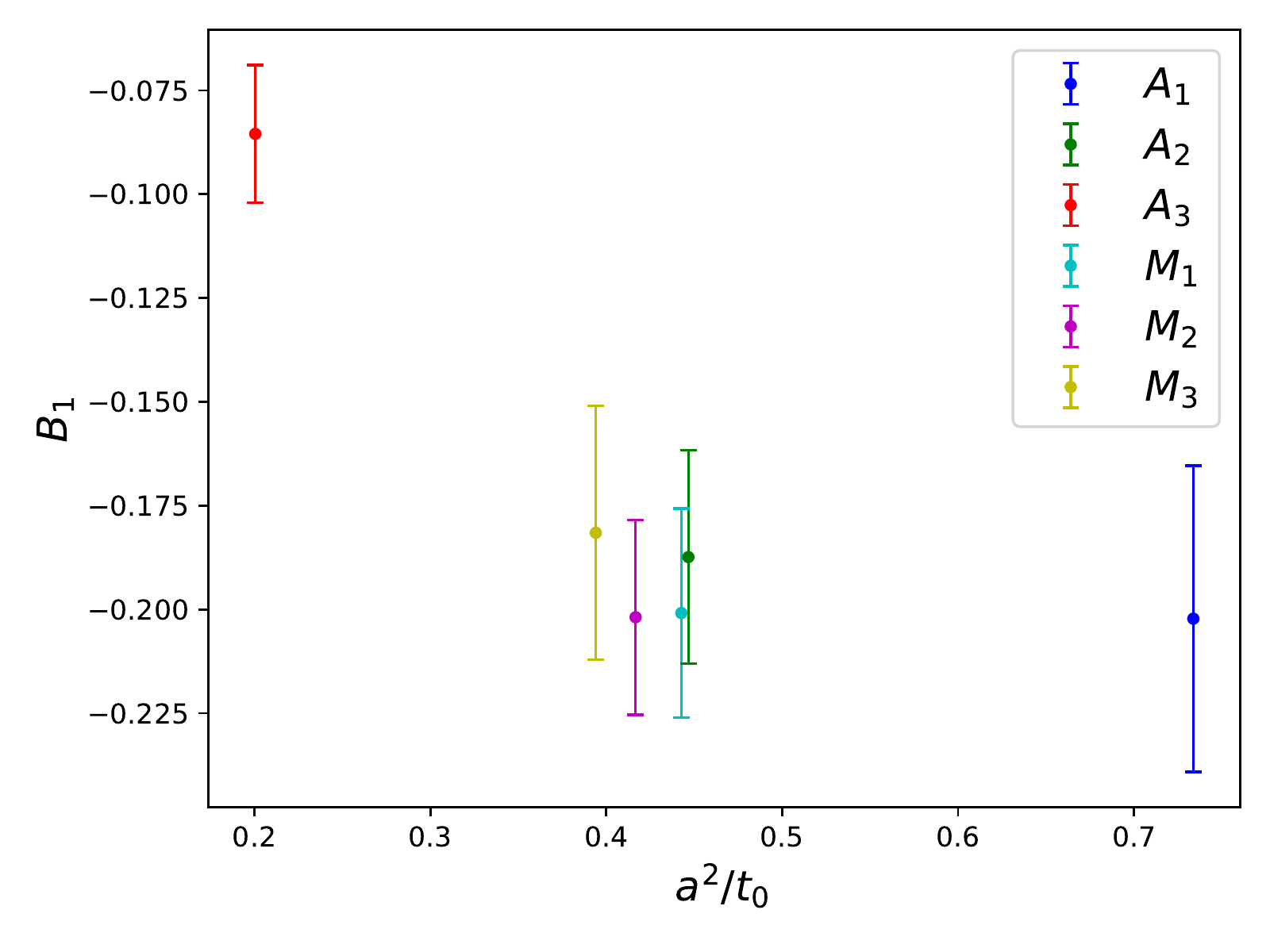}}
		\caption{These results are $t_0/a^2$ dependence of each fitting parameters $B_{-1}$, $B_{1}$. 
		With the exception of the $A_1$ and $A_3$ ensembles, the data points for each ensemble were computed from the fitting ranges from peak and splitting positions given in Table~\ref{tab:fit_range}.}
    \label{fig:t_0_dependence}
\end{figure}

The results for $B_0$ are used to estimate $c_\chi$ of Fig.~\ref{fig:beta_dep}.
For the other $2$ fit parameters in Fig.~\ref{fig:t_0_dependence},
we plot the lattice spacing dependence of the fit parameter $B_{-1}$ and $B_1$.
It is reassuring to notice that $B_{-1}$ practically 
vanishes at a lattice spacing of $a \simeq 0.65$ fm,
indicating small discretization errors.
We also find that the contributions of the higher dimensional operators,
proportional to $B_1$, 
are not negligible, and will be the focus of forthcoming publications.

\section{PERTURBATIVE CALCULATION OF $\cCP$}
\label{app:PT}

In this Appendix we detail the perturbative calculation
of the power divergent coefficient in perturbation theory.
We refer to~\cite{Rizik:2020naq} for the details of the short flow time expansion (SFTE)
of the qCEDM with $2$ external quarks.
For now, we simply summarize the relevant results. 
Near the $t=0$ boundary, we may reconstruct the flowed qCEDM in a basis $\mathcal{A}$ 
of gauge-invariant, $CP$-violating local operators with a modified operator-product expansion:
\be
	O^R_C(x;t)\stackrel{t\rightarrow0}{\sim}\sum_\mathcal{A}c_{Ci}(t)O^R_i(x;0),
\ee
	where all of the flow time dependence of the expanded operator is 
	encoded strictly by the Wilson coefficients $c_{Ci}$. 
	For the qCEDM, for which $[O_C]=d+1$, 
	the leading order contribution comes from the pseudoscalar density,
\be
	O_P(x;t)=P(x;t)=\chibar(x;t)\gamma_5\chi(x;t),
\ee
	which has canonical dimension $[O_P]=d-1$. s
	(In what follows, note that with respect to~\cite{Rizik:2020naq} 
	we have fixed the normalization of the pseudoscalar density and 
	qCEDM operators such that $k_P=1$ and $k_C=-i$.)
Consequently, we expect a linear divergence in the flow time. We may then write
\be
	O^R_C(x;t)=c_{CP}(t)P^R(x;0)+\cdots,
\ee
	where the ellipsis signifies contributions from higher-dimensional operators. 
	The leading contribution to the mixing coefficient appears at one-loop order 
	and is easily extracted by studying the correlation function 
	$G_C(x,y;t)=\int_{z\in\mathbb{R}^d}\langle\psibar(x)O_C(z;t)\psi(y)\rangle$ at $\mcO(g^2)$. 
	We quote our previous result~\cite{Rizik:2020naq}:
\be
	c^{(1)}_{CP}(t)=6\frac{C_2(F)}{(4\pi)^2}\frac{1}{t}.
\ee
	We may, however, extract the same coefficient by studying Eq.~\eqref{eq:RP} 
	in perturbation theory, where the nature of the SFTE demands 
	that the coupling be arbitrarily small:
\be
	c_{CP}(t)=\frac{1}{t}\lim_{x_4\rightarrow\infty}[R_P(x_4;t)]_R,
	\label{eq:cCP_rat}
\ee	
	where
\be	
\left[R_P(x_4;t)\right]_R = t \frac{\left[\Gamma_{CP}(x_4;t)\right]_R}{\left[\Gamma_{PP}(x_4)\right]_R}\,,
\label{eq:RPpt}
\ee
where we adopt for continuum correlation functions
the same notation as for the lattice ones (cf. Eqs.~(\ref{eq:COP}, \ref{eq:CPP})).

This is a manifestly gauge-invariant scheme,
and the separation of the two operators in Euclidean time ensures 
that the correlation function is ground-state dominated.
Thus it is naturally more amenable to a lattice implementation.
\begin{figure}[h!]
	\centering
	\includegraphics[width=95pt,height=74pt]{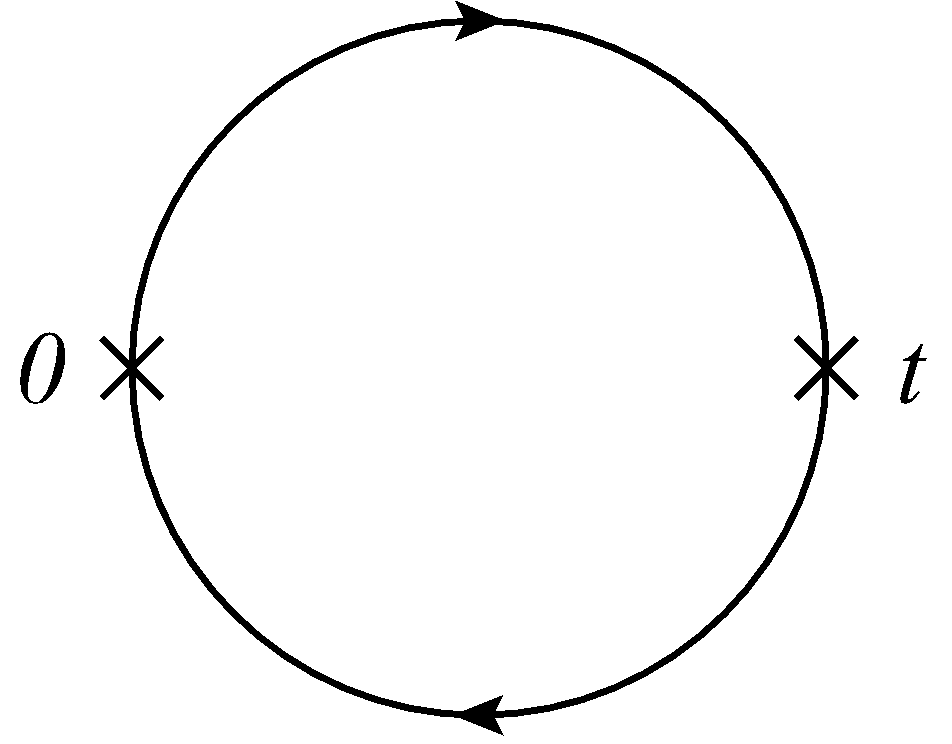}
	\caption{The only contribution to $\Gamma_{PP}(x_4;t)$ at tree level.}
	\label{fig:PP_Graph}
\end{figure}
We work in dimensionally regularized Euclidean space at leading order. 
The denominator of Eq.~\eqref{eq:RPpt} generates a single 
one-loop diagram (Fig.~\ref{fig:PP_Graph}),
which is easily evaluated directly 	
\be
	\Gamma_{PP}^{(0)}(x_4;0)=-4\frac{\dim(F)}{(4\pi)^2}\frac{1}{x_4^3}.
\ee
The spacing $x_d$ provides an ultraviolet cutoff,
so that the above correlator converges as $d\rightarrow4^\pm$ (hence $x_d\rightarrow x_4$).
	
\begin{figure}[h!]
	\centering
	\includegraphics[width=97pt,height=73pt]{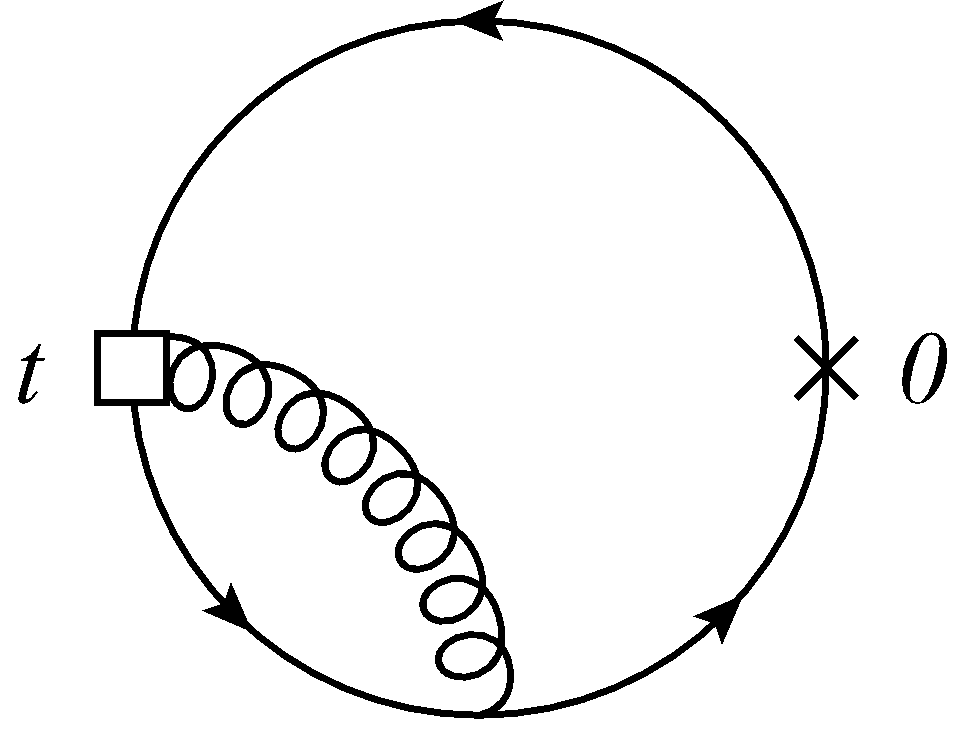}
	\qquad
	\includegraphics[width=97pt,height=76pt]{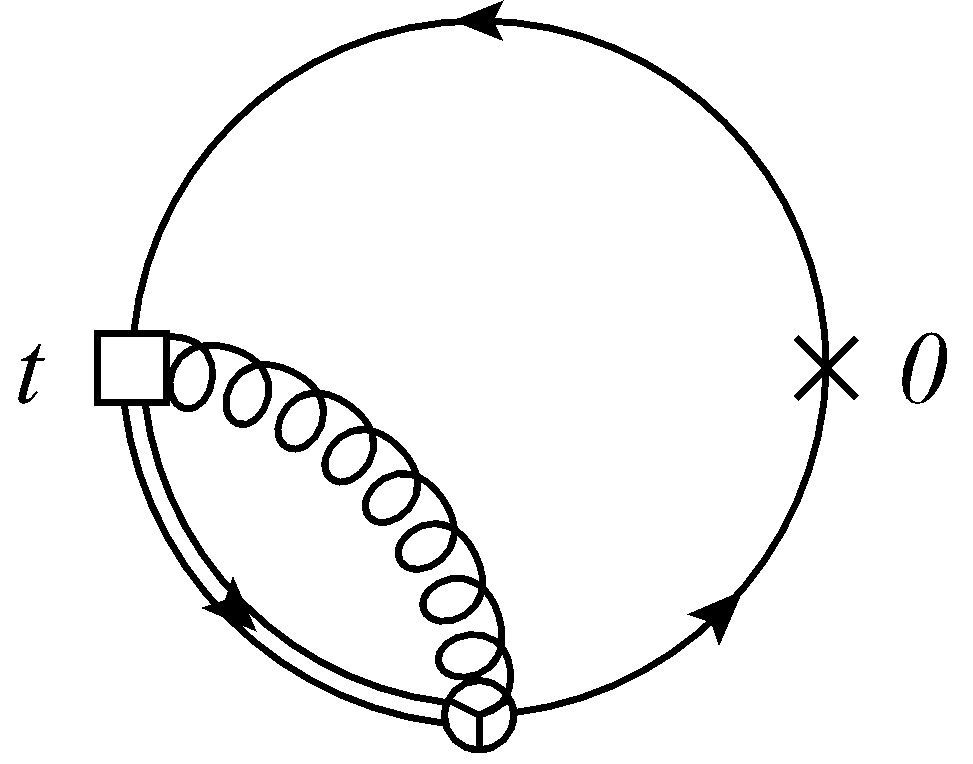}
	\qquad
	\includegraphics[width=97pt,height=75pt]{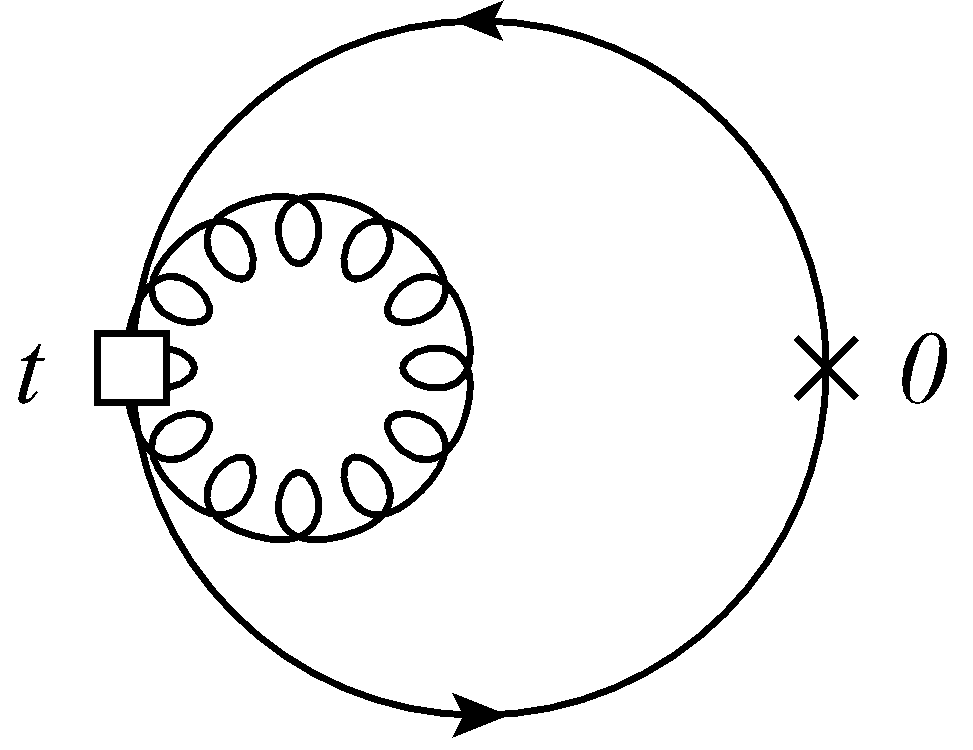}
	\qquad
	\caption{Leading-order contributions to $\Gamma_{CP}(x_4;t)$.}
	\label{fig:CP_Graphs}
\end{figure}

Turning our attention to the numerator,
we study $\int_{\textbf{x}\in\mathbb{R}^{d-1}}\langle O_C(\textbf{x},x_d;t)P(0;0)\rangle$, 
where the pseudoscalar density is fixed at the origin as allowed by translational symmetry. 
The leading contributions in this case are three two-loop ``vacuum bubbles,'' pictured in Fig.~\ref{fig:CP_Graphs}. 
	Since the Euclidean time coordinate is not integrated,
	it is tempting to proceed directly in position space for that component,
	transforming only the spatial coordinates to momentum space. 
	Instead, we have found that the simplest method is to inject 
	some momentum $q$ into both operators,
	proceed as usual in momentum space, and pass back to real 
	space only at the end, projecting the spatial momentum $\textbf{q}$ to zero. 
	Summarily:
\be
	\Gamma_{CP}^{(1)}(x_d;t)=\lim_{p\rightarrow0}
	\int_{\mathbb{R}^{d-1}}d^{d-1}x\ e^{-ipx}\int_qe^{iqx}\tilde{\Gamma}_{CP}^{(1)}(q;t),
\label{eq:gammaCP_PT}
\ee
	where $\tilde{\Gamma}_{CP}(q;t)$ is the Fourier transform of $\Gamma_{CP}(x;t)$. As it turns out, the relevant integrals in this case all converge in four dimensions, so we implicitly take the $d\rightarrow4$ limit in what follows. The middle and right diagrams in Fig.~\ref{fig:CP_Graphs} trivially and identically vanish. 
	For the left diagram in Fig.~\ref{fig:CP_Graphs}, we insert the one-loop result of the analogous calculation with two external quarks 
	(Eq.~(34a) from~\cite{Rizik:2020naq}), which we denote here with $\widetilde{G}_C^{(1)}(p;t)$.
	Without expanding in $t$ we obtain
\be
	\widetilde{G}_C^{(1)}(p;t) = 3\frac{C_2(F)}{(4\pi)^2}\frac{1}{t}\cdot\tilde{f}(p^2t) \gamma_5,
\ee
	where
\be
	\tilde{f}(z)=2\left\{\frac{z}{2}\text{E}_1\left(\frac{z}{2}\right)-z\text{E}_1\left(z\right)+\frac{1-z}{z}\left(e^{-z/2}-e^{-z}\right)\right\}
	\label{eq:PTf}
\ee
	is a common function in flowed perturbation theory, and $E_n(z)$ is the generalized exponential integral. Then, closing the gauge-invariant
	vacuum diagram around a pseudoscalar density operator at the origin, we have
\be
	\tilde{\Gamma}_{CP}^{(1)}(q;t)=2\cdot\Tr\int_p\frac{\slashed{p}}{p^2}\tilde{G}_C^{(1)}(p;t)\frac{\slashed{p}-\slashed{q}}{(p-q)^2}e^{-(p-q)^2t}+\mcO(g^4).
\ee
	Using standard techniques, we arrive at
\be
	\tilde{\Gamma}_{CP}^{(1)}(q;t)=-12\frac{C_2(F)\dim(F)}{(4\pi)^4t^2}\tilde{g}(q^2t),
\ee
	where
\be
	\tilde{g}(z)=
	\begin{aligned}[t]
		\frac{1}{3z}\Bigg\{&(3z^2+12z-3)e^{-z}-(6z^2+30z-12)e^{-\frac{z}{2}}+(3z^2+18z-9)e^{-\frac{z}{3}}\\
		&-(3z^3+15z^2+16z)E_1(z)+(3z^3+21z^2+12z)E_1\left(\frac{z}{2}\right)\\
		&-(z^3+9z^2+6z)E_1\left(\frac{z}{3}\right)\Bigg\}.
	\end{aligned}
\ee
	Finally, we Fourier transform this result to find
\be
	\Gamma_{CP}^{(1)}(x_4;t)=-12\frac{C_2(F)\dim(F)}{(4\pi)^4t^2}\cdot t^{-1/2}g(t^{-1/2}x_4),
\ee
	where
\be
	g(\epsilon)=\frac{1}{2\epsilon^5}
		\begin{aligned}[t]
			\Bigg\{&(\epsilon^6-4\epsilon^4+20\epsilon^2-48)\text{erf}\left(\frac{1}{2}\epsilon\right)-(4\epsilon^6-8\epsilon^4+28\epsilon^2-48)\text{erf}\left(\frac{\sqrt{2}}{2}\epsilon\right)\\
			&+(3\epsilon^6-4\epsilon^4+12\epsilon^2-16)\text{erf}\left(\frac{\sqrt{3}}{2}\epsilon\right)+\frac{2}{\sqrt{\pi}}\Big[e^{-\frac{1}{4}\epsilon^2}(\epsilon^5-6\epsilon^3+24\epsilon)\\
			&-\sqrt{2}e^{-\frac{1}{4}(\sqrt{2}\epsilon)^2}(2\epsilon^5-6\epsilon^3+24\epsilon)+\sqrt{3}e^{-\frac{1}{4}(\sqrt{3}\epsilon)^2}(\epsilon^5-2\epsilon^3+8\epsilon)\Big]\Bigg\}.
		\end{aligned}
\ee
	For large argument, the function $g$ goes as
\be
	g(\epsilon)\stackrel{\epsilon\gg0}{\sim}\frac{2}{\epsilon^3}.
\ee	
	and we find that
\be
	c_{CP}^{(1)}(t)=\lim_{x_4\rightarrow\infty}\frac{\Gamma^{(1)}_{CP}(x_4;t)}{\Gamma^{(0)}_{PP}(x_4)}=6\frac{C_2(F)}{(4\pi)^2}\frac{1}{t},
\ee
	as desired. Indeed, for $N_c=3$, we have
\be
	c_{CP}^{(1)}(t)=\frac{1}{2\pi^2t}
\ee
	as in Eq.~\eqref{eq:cP_PT}. 
	For completeness, we remark that
	the same leading-order result may be obtained with the ratio 
\be	
\Delta(\bar{g}^2) = \lim_{x_4\rightarrow\infty}
t\ \frac{\Gamma_{CP}(x_4;t)}{\Gamma_{PP}(x_4;t)}\Big|_{t=\textrm{constant}}\,,
\label{eq:RPpt_1l}
\ee	
in which now one of the pseudoscalar density operators in the denominator is fixed at $t$, and the renormalized coupling $\bar{g}$ is taken to be an implicit function of $t$. Since the self-mixing of the flowed pseudoscalar density is unity at leading order, we may for now safely ignore the $\textrm{log}\ t$ corrections, which are suppressed by the coupling.
We find that
\be
	\Gamma_{PP}^{(0)}(x_4;t)=-2\frac{\dim(F)}{(4\pi)^2t}\cdot t^{-1/2}f(t^{-1/2}x_4),
\ee
	where
\be
	t^{-1/2}f(\epsilon)=
		\begin{aligned}[t]
			t^{-1/2}\cdot\frac{1}{\epsilon^3}\bigg\{
			&\left(\epsilon ^4+4\right) \text{erf}\left(\frac{\epsilon}{2}\right)-\left(\epsilon^4+2\right) \text{erf}\left(\frac{\epsilon}{\sqrt{2}}\right)\\
			&-\frac{1}{\sqrt{\pi}}\left[\left(\epsilon^3-2\epsilon\right) \left(\sqrt{2} e^{-\frac{1}{4}(\sqrt{2}\epsilon)^2}-2 e^{-\frac{1}{4}(\epsilon)^2}\right)\right]\bigg\}
		\end{aligned}
\ee
	is the inverse Fourier transform of Eq.~\eqref{eq:PTf}. Again, we have
\be
	f(\epsilon)\stackrel{\epsilon\gg0}{\sim}\frac{2}{\epsilon^3},
\ee
	so that
\be
\Delta^{(1)}=\lim_{x_4\rightarrow\infty}t\ \frac{\Gamma_{CP}^{(1)}(x_4;t)}{\Gamma_{PP}^{(0)}(x_4;t)}\Big|_{t=\textrm{constant}}=
	6\frac{C_2(F)}{(4\pi)^2}\cdot\lim_{\epsilon\rightarrow\infty}\frac{g(\epsilon)}{f(\epsilon)}=
	6\frac{C_2(F)}{(4\pi)^2}\xrightarrow{N_c\rightarrow3}\frac{1}{2\pi^2}
\ee
	as in Eq.~\eqref{eq:Rbar_PT}.
		
It is clear from the calculation described above that the leading-order 
contribution to the power divergence is universal whether we probe the local operator
with external quarks (cf. Eq.~\eqref{eq:cP_PT} and Ref.~\cite{Rizik:2020naq})
or with a pseudoscalar density. This should be expected, since the SFTE is an operator-level relation, and fluctuations are neglected at leading order. Moreover, this confirms $c_P^{(0)}(t)=1+\mcO(t)$, which follows trivially from the perturbative solution to the fermionic flow equations. The universality property, however,
strongly depends on the condition that $x_4$ is much larger than $\sqrt{8t}$.
This is somehow expected as well, because no additional contact terms are present when both $t\rightarrow 0$ and $x_4 \rightarrow 0$ with this kinematical choice; \textit{viz}, the qCEDM and the pseudoscalar density do not form a local operator product so long as the physical separation is well larger than the smearing radius. We have indeed repeated the calculation presented in this appendix integrating Eq.~\eqref{eq:gammaCP_PT} over the whole spacetime volume, thus including also the region at $x_4\sim 0$, obtaining a different result. 

\bibliography{ref} 

\end{document}